\documentclass[10pt,journal,compsoc]{IEEEtran}

\ifCLASSOPTIONcompsoc
\usepackage[nocompress]{cite}
\else
\usepackage{cite}
\fi

\usepackage[caption=false,font=footnotesize]{subfig}

\usepackage{amsmath}
\usepackage{todonotes}
\usepackage{booktabs}
\usepackage{url}
\usepackage{array}
\usepackage{acronym}
\usepackage{textcomp}

\acrodef{GDPR}[GDPR]{General Data Protection Regulation}
\acrodef{CCPA}[CCPA]{California Consumer Privacy Act}
\acrodef{ECJ}[ECJ]{European Court of Justice}

\begin{document}

\title{Privacy Policies Across the Ages: Content and Readability of Privacy Policies 1996--2021}

\author{Isabel Wagner%
	\IEEEcompsocitemizethanks{\IEEEcompsocthanksitem I. Wagner is with the Cyber Technology Institute, De Montfort University, Leicester,
		LE1 9BH, United Kingdom.\protect\\
		E-mail: isabel.wagner@dmu.ac.uk
	}%
}

\IEEEtitleabstractindextext{%
\begin{abstract}
It is well-known that most users do not read privacy policies, but almost all users tick the box to agree with them.
In this paper, we analyze the 25-year history of privacy policies using methods from transparency research, machine learning, and natural language processing.
Specifically, we collect a large-scale longitudinal corpus of privacy policies from 1996 to 2021 and analyze the length and readability of privacy policies as well as their content in terms of the data practices they describe, the rights they grant to users, and the rights they reserve for their organizations.
We pay particular attention to changes in response to recent privacy regulations such as the GDPR and CCPA.
Our results show that policies are getting longer and harder to read, especially after new regulations take effect, and we find a range of concerning data practices.
Our results allow us to speculate why privacy policies are rarely read and propose changes that would make privacy policies serve their readers instead of their writers.
\end{abstract}

\begin{IEEEkeywords}
	privacy policy, longitudinal study, natural language processing, machine learning, neural networks, readability formulas
\end{IEEEkeywords}
}

\maketitle

\IEEEdisplaynontitleabstractindextext

\IEEEpeerreviewmaketitle

\IEEEraisesectionheading{\section{Introduction}}

A website's privacy policy is a legal document that explains what data the site collects from its users, how and for what purpose it processes the data, and with what other parties it shares the data. In addition, privacy policies can explain the users' rights regarding opting in or out of data collection, data correction, and data deletion.
Privacy policies are notorious for being lengthy documents that are hard to understand \cite{libert2018automated,linden2020privacy,mcdonald2008cost}. They are rarely read by users, but website owners assert that by visiting their site users agree to their privacy policy.
This gap in understanding between users and website owners deserves closer study: why are policies not read more frequently, what obstacles are there, and which data practices do users unwittingly agree to?

In the 25-year history of website privacy policies, the privacy and data protection rules have changed several times and in different jurisdictions, and privacy policies were updated to accommodate new requirements.
For example, when the \ac{GDPR} came into force in 2018, privacy policies became longer and many websites introduced new privacy policies \cite{degeling2019we,kretschmer2021cookie}.

However, research into privacy policies has been limited to studying single points in time or short periods, e.g., across the GDPR introduction. In contrast, this paper analyzes \textit{longitudinal} changes in privacy policies over the last 25 years, in terms of length, readability, and content.
Studying longitudinal changes in privacy policies is important
to understand whether new privacy regulations lead to increased user privacy or not;
to understand whether privacy regulations lead to substantive changes in data practices or whether they merely result in corporate box-ticking exercises;
to understand the state of privacy on the web and the extent to which websites respect user privacy;
to guide the design of new privacy protections;
and to guide the design of next-generation privacy regulations.

We leverage recent advances in machine learning and natural language processing \cite{harkous2018polisis} to automate the analysis of privacy policies at scale.
In this paper, we answer the following research questions:
\begin{itemize}
	\item How user-friendly are privacy policies in terms of their length and readability?
	\item What was the effect of the \ac{GDPR} and \ac{CCPA} on the privacy policy landscape?
	\item How have the terminology and jargon in privacy policies evolved, for example in terms of the use of key terms?
	\item How has the content of privacy policies evolved in terms of covered privacy practices and user rights?
\end{itemize}
To answer these research questions, we collect a corpus of more than 50,000 unique privacy policy texts spanning 25 years, from 1996 to 2021.
To the best of our knowledge, we are the first to apply BERT to classify the attributes of privacy policy segments and the first large-scale longitudinal study of the contents of privacy policies.
We find that the length of the average privacy policy has approximately doubled in the last ten years and quadrupled since 2000. Both \ac{GDPR} and \ac{CCPA} have led to significant increases in policy length.
Similarly, privacy policies have become less readable over time, particularly around the times when new privacy regulations were introduced. In 2021, the average Flesch Reading Ease of privacy policies in our corpus was 31, which is similar to academic articles (e.g., the Harvard Law Review has scores in the low 30s \cite{mcdonald2009comparative}), compared to 34 in 2011 and 37 in 2001.

Analyzing the content of privacy policies, we identify several concerning trends, including the increasing use of location data, increasing use of implicitly collected data, lack of meaningful choice, lack of effective notification of privacy policy changes, increasing data sharing with unnamed third parties, and lack of specific information about security and privacy measures.
Overall, our results show that recent privacy regulations have not substantially improved the privacy of users online, but rather led to more bloated privacy policies that describe more and more invasive data practices.
Our policy corpus, including readability data and policy content labels, is available as an open access dataset \cite{wagner2022longitudinal}.

The remainder of this paper is structured as follows.
Section \ref{sec:background} discusses background information and related work.
Section \ref{sec:methodology} explains our methodology for collecting and analyzing privacy policies.
Our results on the readability and content of privacy policies are described in Sections \ref{sec:readability} and \ref{sec:content-analysis}, respectively.
Limitations of our approach are discussed in Section \ref{sec:limitations}, and 
Section \ref{sec:conclusion} concludes.

\section{Background and related work}
\label{sec:background}

\begin{figure*}[!t]
	\centering
	\includegraphics[width=\linewidth]{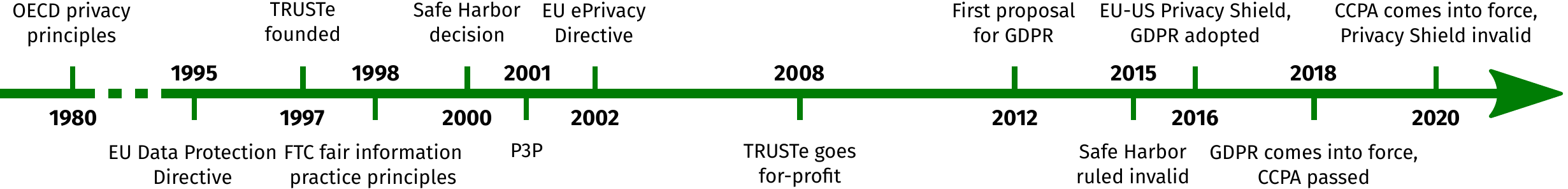}
	\caption{Timeline of privacy-relevant events, regulations, and principles since 1980.}
	\label{fig:timeline}
\end{figure*}

In the last two decades, the landscape of privacy regulations has seen significant changes (see Figure \ref{fig:timeline}), including the introduction of Europe's \ac{GDPR} in 2018 and California's \ac{CCPA} in 2020. 

\subsection{Privacy regulations}

The OECD privacy principles, introduced in 1980, encompass collection limitation, data quality, purpose specification, use limitation, security safeguards, openness, individual participation, and accountability \cite{organisationforeconomicco-operationanddevelopmentoecd2013oecd}.
These principles were intended to be implemented into national law by member countries. For example, European privacy legislation--most recently the \ac{GDPR}--closely follows these principles.
In contrast, privacy rules in the United States are more closely aligned to the Federal Trade Commission's fair information practice principles (FIPPs), which are based on notice/awareness and choice/consent. 
The FIPPs are seen as less comprehensive than the European privacy regime.

The International Safe Harbor Privacy Principles were developed between 1998 and 2000 as a way to reconcile the stricter European privacy regime with the FIPPs.
The Safe Harbor decision by the European Commission in 2000 ruled that the Safe Harbor principles complied with the EU Data Protection Directive, and as a result companies who complied with the principles could register their (self-)certification and transfer data from EU to US.
However, the Safe Harbor decision was invalidated by the \ac{ECJ} in October 2015.
In its place, the EU-US Privacy Shield was agreed in February 2016, introducing stronger obligations on US companies, stronger monitoring, and stronger enforcement.
However, in 2020 the Privacy Shield was again declared invalid by the \ac{ECJ}.
As a result, EU-US data transfers are now governed for example by Article 49 of the \ac{GDPR} which allows data transfers subject to explicit informed consent, or by contracts between EU data subjects and US data controllers.
The \ac{CCPA}, passed in 2018 and taking effect in 2020, was the first US state law that introduced comprehensive privacy rules in the European sense \cite{stallings2020handling}.

\subsection{Length and readability of privacy policies}
Privacy policies have been studied for more than a decade. For example, in 2008, the annual economic opportunity cost for reading privacy policies was estimated at \$781 billion for US internet users~\cite{mcdonald2008cost}.
In 2009, privacy policies were found to have Flesch Reading Ease (FRE) scores between 32 and 46 \cite{mcdonald2009comparative}, which is consistent with our finding that 50\% of policies from 2009 have scores between 29 and 39.
In 2018, privacy policies from the top 1 million websites
had an average Flesch Reading Ease of 39.8 \cite{libert2018automated}, which indicates better readability than the policies from 2018 in our corpus (FRE=32.2). This may be due to a different FRE implementation (we evaluate the influence of implementations on FRE scores in Appendix \ref{sec:readability-implementations}).
In addition, higher-ranked policies were found to be longer (2,000+ words) than lower-ranked policies (1,400 words on average) \cite{libert2018automated}. In comparison, policies from 2018 in our corpus had an average length of 3,200 words, but almost all were higher-ranked policies.
An analysis of a longitudinal corpus of policies from 130k websites \cite{amos2021privacy} shows a steady increase in policy length with the word count doubling between 2009 and 2019, which is similar to our findings.
However, the average length of policies is shorter than in our corpus (e.g., for March 2019: 2178 words vs. 3599 words), which could be due to our effort in retrieving \textit{full} policy texts (see Section \ref{sec:methodology}) or to their inclusion of many lower-ranked policies which tend to be shorter.
Similar to our findings, \cite{amos2021privacy} found a decrease in readability over the years.

After the GDPR came into effect, more than 84\% of European websites had a privacy policy and 62\% displayed a cookie consent notice, an increase of 4.9\% and 16\%, respectively \cite{degeling2019we}.
Policies became significantly longer, increasing from 2,145 words on average in 2016 to 3,603 words on average in May 2018 \cite{degeling2019we}.
In addition, the use of GDPR-related terms, such as \textit{complaint}, \textit{data portability}, or \textit{erasure}, increased between 6 and 12\% between January and May 2018. 
In comparison, our longitudinal approach allowed us to observe increases for the same terms, but finding larger increases (15--25\%) if the period beyond May 2018 is taken into account.
Even though some studies find improvements in readability of privacy policies after introduction of the GDPR, \cite{kretschmer2021cookie} conclude that privacy policies are still not understandable by the general public. 

\subsection{Machine-readable privacy policies}

The Platform for Privacy Preferences (P3P) was introduced in 2001 as a W3C recommendation.
P3P aimed at making privacy policies machine-readable, opening complex privacy policies to automated analysis and matching against user preferences.
P3P relied on two components: the server-side component provides a policy following P3P's XML schema, and the user-side component retrieves and analyzes the policy, matches it against user preferences, and displays the result to the user.
However, P3P was never mandated and 
as a result was never widely adopted \cite{reay2007survey}.
In addition, many P3P policies were erroneous and seldom corrected or updated \cite{reay2007survey}, and there were no mechanisms to ensure that the natural-language and P3P policy versions were equivalent, or that a website's actual practices conformed to the stated policy \cite{mazmudar2020mitigator}.

Standardized presentation of privacy policies has also been proposed. For example, a tabular presentation can significantly increase comprehension and usability \cite{kelley2010standardizing}.
However, in 2021 natural-language policies are still the norm and neither standardized presentations nor machine-readable policies have been adopted.

The application of machine learning to label the contents of privacy policies may be a way to realize the intended benefits of P3P purely on the client-side.
The first machine learning classifiers for privacy policies were based on support vector machines and hidden Markov models \cite{wilson2016creation}, presented together with the OPP-115 corpus of labeled privacy policies.
Classifiers based on convolutional neural networks \cite{harkous2018polisis} and BERT models \cite{mousavinejad2020establishing} improved the classification performance. 
These classifiers were used to build a user interface that maps privacy icons to policy statements~\cite{harkous2018polisis} and to analyze GDPR-related changes in the policy landscape using queries that assess the specificity and compliance of privacy policies \cite{linden2020privacy}.
However, to the best of our knowledge, these classifiers have not yet been applied to analyze the content of privacy policies on a large scale.

Natural language processing (NLP) has been used to identify contradictory statements in privacy policies of mobile apps \cite{andow2019policylint}, where 14\% of policies contained misleading statements, including redefinitions of common understandings of terms and conflicts between terms used in different privacy regulations.
NLP was also used to study the flow-to-policy consistency of mobile apps and their privacy policies, showing that the behavior of 40\% of apps was not consistent with the app's privacy policy \cite{andow2020actions}.

\section{Methodology}
\label{sec:methodology}
Our methodology for collecting and analyzing a longitudinal corpus of privacy policies consists of 4 steps:
(1) crawling websites to find links to their privacy policies, (2) retrieving the policy texts, relying on the Wayback Machine to retrieve historical policy texts going back to 1996, (3) evaluating readability based on length, use of obfuscating words, passive-voice sentences, and readability formulas, and (4) evaluating the data practices described in policies using a hierarchy of machine learning classifiers.

We do not use the corpus published by \cite{amos2021privacy}, because their policy collection ended in 2019 and thus does not reflect the policy updates made after the CCPA came into effect.
In addition, their corpus contains fewer unique policy texts per website than our corpus.

For all crawls, we use computers on our university campus between January 2020 and December 2021. In cases where access to privacy policies was filtered by our university firewall (e.g., for pornographic websites), we used supplementary crawls from a residential internet connection.
We use the Wayback Machine to retrieve policies between 1996 and February 2020, and monthly live crawls between March 2020 and December 2021.

\subsection{Selecting websites and dates to crawl}

To select websites for our longitudinal analysis of privacy policies, we combine two approaches.
First, we select sites from two recent versions of the Tranco list \cite{pochat2019rigging}.
We use the Tranco list from 1 October 2019\footnote{\url{https://tranco-list.eu/list/JL9Y}} 
and select the top 1,000 websites plus an additional 1,000 sites drawn uniformly at random from the top 1k to 10k. In addition, we select the top 1k sites from the Tranco list from 31 March 2021\footnote{\url{https://tranco-list.eu/list/ZLZG}}.

Second, we add sites from the historical Alexa toplist for each year \cite{lerner2016internet}:
for 2010--2021, we use the top 1k sites of the Alexa top one million;
between 2003 and 2009, we scrape the top 500 sites from the Alexa website as archived by the Wayback Machine;
and for 2002, we use the top 100 sites from the archived Alexa website.
This process resulted in a total of 4,997 sites.

For each site, we retrieve the list of available snapshots for the landing page using the Wayback Machine's CDX API.
To pick up when a site's privacy policy moves to a new URL, we select one snapshot per year between 1996 and 2008, quarterly snapshots between 2009 and 2017, and monthly snapshots after that.
In addition, we retrieve the category for each site from Alexa.

\subsection{Finding privacy policy links}

We use Firefox, automated with Selenium, to load the landing page snapshots and parse the HTML with BeautifulSoup 4 \cite{richardson2020beautiful}.
To locate links to privacy policies, we search through link titles and link URLs in reverse order.
Because there is no standard naming scheme for privacy policy links, 
we search for each of the terms \textit{privacy polic}, \textit{privacy}, \textit{terms of service}, \textit{web policies}, \textit{cookie polic}, \textit{data polic} and \textit{legal}.
Across all snapshots, we found 27,329 privacy policy links.

\subsection{Retrieving the policy text}

To retrieve the privacy policy text, we identify available snapshots for each policy link using the Wayback Machine's CDX API and fetch one snapshot per month, as far back as available. For each snapshot, we load the link with Firefox/Selenium, scroll to the bottom, and save the resulting HTML. 
We follow HTTP redirects within the Wayback Machine. If a page has not loaded completely after two minutes, we trigger a timeout and extract the policy text from the partially loaded page. This often succeeds when the page was waiting for embedded resources.
To extract the policy text from HTML pages, we use both Firefox's reader mode and the readability-lxml library to strip navigational elements, page headers, and page footers.
We compare the length of both extracted texts and keep the longer of the two.

Some sites only display a short summary of the policy instead of the full policy when clicking on the landing page's privacy policy link. To catch these cases, we search for links within policies whose titles contain \textit{privacy} or \textit{policy} as well as \textit{full}, \textit{entire}, or \textit{complete}, plus titles that contain \textit{privacy statement}, \textit{privacy polic}, \textit{privacy notice}, or \textit{privacy}, and add these links to our list.

In total, we fetched 1,068,683 documents as potential privacy policies, with 120,265 unique documents (an average of 39.1 policy instances and 4.4 unique policy texts per link).

\subsection{Data cleaning}

Because the readability formulas and machine learning classifiers we use are based on English text,
we remove all non-English policies from our database.
In particular, we use the PYCLD2 package~\cite{al-rfou2019pycld2} for language detection and remove all policies where English was not the language detected with highest confidence.
In addition, we remove short policies with fewer than 100 words because they usually contain brief summaries or error messages, not policy text.

In addition, we implemented the classifier from \cite{linden2020privacy} to identify which of the documents in our corpus are privacy policies.
We trained the classifier with the same corpus of 1,000 privacy policies as \cite{linden2020privacy}, but used our own set of non-policy documents because theirs was not available.
We trained three versions of the classifier with different sets of non-policy documents: (1) a selection of 1,000 landing pages from our crawls, (2) landing pages longer than 5,000 characters, and (3) a selection of 1,000 subsites crawled from the landing pages of the Tranco top 500 sites.
When the non-policy corpus contains legal documents that are not privacy policies but use similar language (such as terms of service), the classifier performs well on the training data, but mis-classifies many true privacy policies. Therefore, we filter the non-policy corpus so that it does not include keywords expected in privacy policies (using the same list of keywords as above).
We remove policy texts that have a low probability to be a privacy policy according to all three classifiers, using empirically determined thresholds of 0.9 for classifier (1), 0.6 for classifier (2), and 0.1 for classifier (3). 

\begin{figure}
	\centering
	\subfloat[Non-unique]{\label{fig:policy-instances}{\includegraphics[width=.47\linewidth]{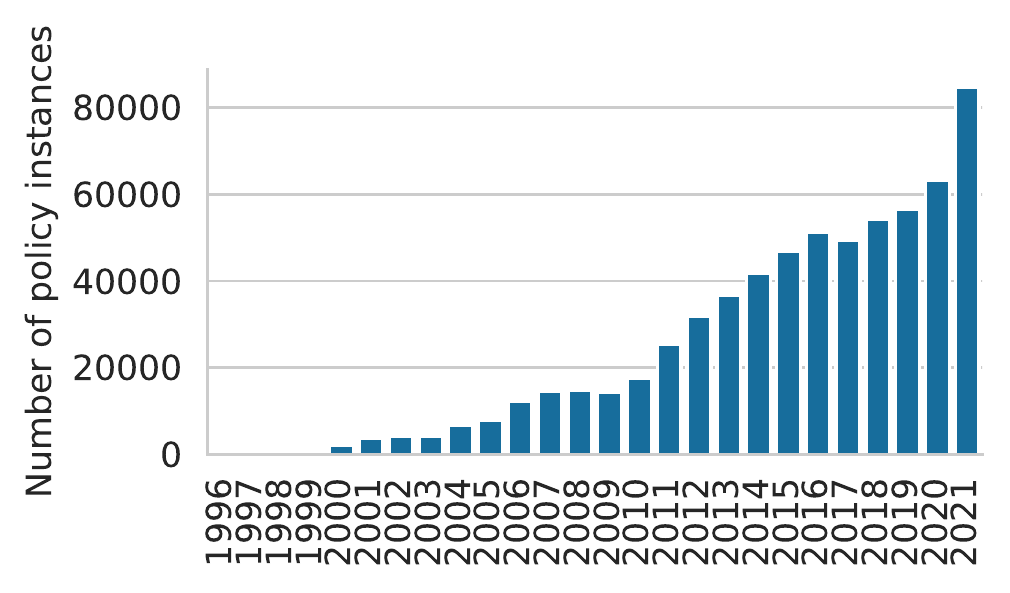}}}
	\subfloat[Unique]{\label{fig:unique-policy-instances}{\includegraphics[width=.47\linewidth]{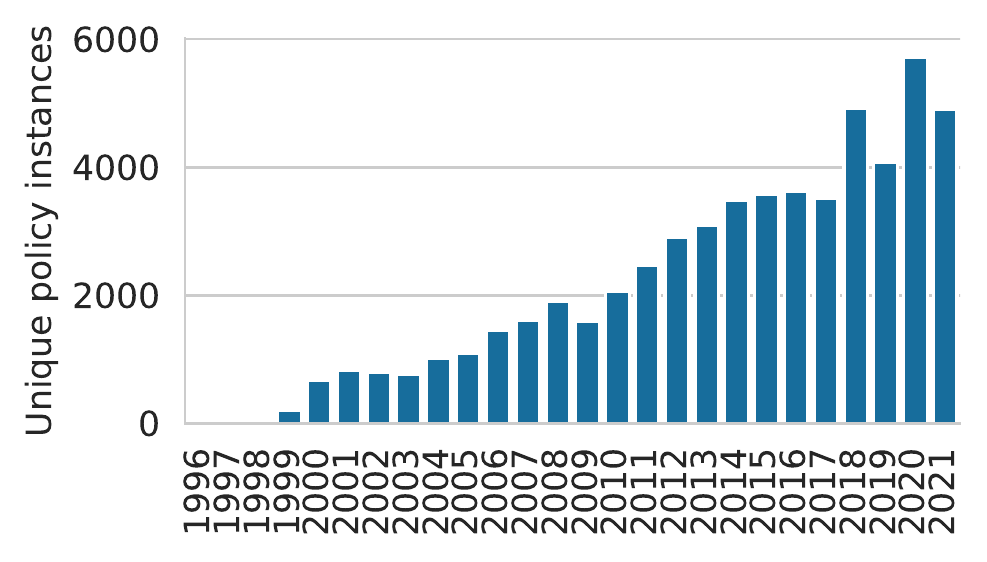}}}
	\caption{Number of policy instances collected for each year. Unique policy texts are counted in the year they first appeared.}
	\label{fig:policies-per-year}
\end{figure}

After filtering, our corpus contains 56,416 unique privacy policy texts.
Figure \ref{fig:policies-per-year} shows how many policy snapshots and unique policies were collected for each year.
The peaks for unique policy texts in 2018 and 2020 (Figure \ref{fig:unique-policy-instances}) show that the introductions of the \ac{GDPR} and \ac{CCPA}, respectively, caused many organizations to update their privacy policies.

\subsection{Readability measures}
We analyze the readability of privacy policies based on length, obfuscating terms, passive-voice sentences, and readability formulas.

\subsubsection{Length}
We measure length as the number of syllables, words, and sentences in a privacy policy. 
The length measure allows us to compare our results with prior work such as \cite{linden2020privacy,amos2021privacy}.
We evaluate open source implementations for length measures in Appendix \ref{sec:readability-implementations}. 

\subsubsection{Obfuscating words}
The use of obfuscating words is a measure for language accessibility \cite{shipp2020how}.
Obfuscating words, such as \textit{acceptable}, \textit{significant}, \textit{mainly}, or \textit{predominantly} make text harder to understand because they reduce the clarity of statements in a privacy policy.
We have used the list of obfuscating adjectives and adverbs published in \cite{shipp2020how}.
In particular, we preprocess both our policy texts and the list of obfuscating words with the \textit{gensim} preprocessor \cite{rehurek2021gensim}, which strips punctuation and numbers, removes whitespace and stop words, converts text to lowercase, and applies the PorterStemmer algorithm \cite{porter1980algorithm}.
We then count instances of obfuscating words in the policy text.

\subsubsection{Passive-voice sentences}
The use of passive-voice sentences can make text harder to understand because it makes it difficult to determine who is doing what.
To detect passive-voice sentences we use the same pattern for the spaCy \cite{explosionaigmbh2021spacy} pattern matcher as \cite{linden2020privacy}, i.e. we match a sequence of nominal subject (\textit{nsubjpass}), auxiliary (\textit{aux}), and passive auxiliary (\textit{auxpass}). 

\subsubsection{Readability formulas}
\label{sec:readability-formulas}
Readability formulas estimate the difficulty of texts, such as reading material for schools or health information.
Many different formulas have been developed since the 1940s \cite{dubay2007smart}.
They generally combine a measure for sentence complexity with a measure for word complexity.
Sentence complexity is most often measured as the average sentence length, i.e., the average number of words per sentence.
Word complexity can be measured as the number of syllables per word, the number of characters per word, or the fraction of \textit{complex} or \textit{difficult} words.

We consider seven readability formulas with open source Python implementations. Next, we introduce these formulas and explain our selection of three: the Flesch Reading Ease (FRE), Coleman-Liau score (CL), and Simple Measure Of Gobbledygook (SMOG).
In Appendix \ref{sec:readability-implementations}, we evaluate the Python implementations and find that they have subtle differences in how they compute word and sentence complexity, which can significantly alter the resulting readability scores. It is therefore important to document the choice of readability library and version.

\paragraph{Syllables as word complexity measure}
The Flesch Reading Ease (FRE) measures word complexity based on the average number of syllables in a word.
FRE is perhaps the most commonly used readability formula. FRE scores are commonly between 0 and 100, although negative scores are possible for very difficult texts, and scores above 100 are possible for very easy texts. 
Scores above 60 are interpreted as plain English, while scores below 30 are best understood by university graduates.
\[FRE = 206.835 - 1.015 \left( \frac{\text{words}}{\text{sentences}} \right) - 84.6 \left( \frac{\text{syllables}}{\text{words}} \right)\]

The Flesch-Kincaid Grade Level (FKG) uses the same components as the FRE, but because they are weighted differently the two scores are not comparable. FKG outputs a \textit{grade level}, i.e. an indicator for how many years of education are needed to understand the text. From the readability formulas that use syllables as word complexity measure, we selected FRE for comparability with prior work \cite{libert2018automated,shipp2020how}.
\[FKG = 0.39 \left ( \frac{\mbox{words}}{\mbox{sentences}} \right ) + 11.8 \left ( \frac{\mbox{syllables}}{\mbox{words}} \right ) - 15.59\]

\paragraph{Characters as word complexity measure}
Both the Automated Readability Index (ARI) and the Coleman-Liau score (CL) use the average number of characters per word to measure word complexity, which is simple to automate.
ARI and CL both output a grade level estimate, but they differ in their weighting and in their measure of sentence complexity: ARI uses the average sentence length, whereas CL uses inverse sentence length.
We selected CL because it is the only formula with a different sentence complexity measure.
\[ARI = 4.71 \left (\frac{\mbox{characters}}{\mbox{words}} \right) + 0.5 \left (\frac{\mbox{words}}{\mbox{sentences}} \right)  - 21.43\]

\[CL = 5.88\left(\frac{\mbox{characters}}{\mbox{words}}\right) - 29.6\left(\frac{\mbox{sentences}}{\mbox{words}}\right) - 15.8\,\!\]

\paragraph{Difficulty as word complexity measure}
The Simple Measure Of Gobbledygook (SMOG) formula measures word complexity based on the number of polysyllabic words, i.e. words that have three or more syllables. SMOG is statistically invalid if the text has fewer than 30 sentences, however, this is rarely the case for privacy policies.
\[SMOG = 1.0430 \sqrt{\mbox{polysyllables}\times{\frac{30}{\mbox{sentences}}} } + 3.1291\]

The Dale-Chall and Gunning-Fog formulas both count words that are defined as \textit{difficult} or \textit{complex}.
\textit{Difficult} words, according to the Dale-Chall score (DC), are words that are not known by 80\% of fourth-graders. To make this notion more precise, the authors have published a list of 3000 \textit{familiar} words, and difficult words are all those not on the list. It is important to note that the DC formula and word list have been updated in 1995, but the versions circulating on the Internet refer to the 1948 version\footnote{We make our implementation of the 1995 DC word list available on request.}.
In addition to the word list, DC has a list of rules that govern in which cases words should not be counted as \textit{difficult}.
For example, regular plurals and possessives of words on the word list are not counted as difficult, and neither are words with other regular word endings, e.g., \textit{-d} or \textit{-ing}. In addition, short numbers are not counted as difficult, and proper names and places are only counted once per 100-word sample.

\[DC = 0.1579 \left (\frac{\mbox{difficult words}}{\mbox{words}}\times 100 \right) + 0.0496 \left (\frac{\mbox{words}}{\mbox{sentences}} \right)\]

The Gunning Fog (GF) formula uses the notion of a \textit{complex} word. Complex words are generally words with three or more syllables. However, if a word is a proper noun, familiar jargon, or a compound word, it is not counted as complex. In addition, common suffixes such as \textit{-es}, \textit{-ed}, or \textit{-ing} are not counted as a syllable.
\[GF = 0.4\left[ \left(\frac{\mbox{words}}{\mbox{sentences}}\right) + 100\left(\frac{\mbox{complex words}}{\mbox{words}}\right) \right]\]
For both DC and GF, their additional rules make it difficult and computationally expensive to correctly implement the formulas. We selected SMOG because its word complexity measure can be implemented efficiently.

\paragraph{Calibration of readability formulas}
It is important to note that these formulas do not have a common zero point because they were calibrated differently. 
The calibration of readability formulas is generally based on reading comprehension tests with human users. The participants read a text and are then given either a multiple-choice comprehension test or a test in which they have to fill in the blanks.
The formulas are then calibrated to indicate the level of a reader who can answer a set percentage of questions correctly, e.g., 75\% for Flesch or 100\% for SMOG \cite{dubay2007smart}.
As a result, the formulas are not expected to report exactly equal estimates of the readability level; in fact, it is expected that SMOG consistently estimates higher grade levels than Flesch or Dale-Chall because its calibration assumes a higher level of text comprehension.

\subsection{Classifying content of privacy policies}

To evaluate which data practices are described in privacy policies, we follow Harkous et al. \cite{harkous2018polisis}.
Specifically, we implement a hierarchy of classifiers so that the top-level classifier labels the topic, or category, of each segment of a privacy policy, and the lower-level classifiers label the attributes described in each segment.

For example, consider the segment:
``As you navigate through and interact with our Website, we may use automatic data collection technologies to collect certain usage information about your equipment, browsing actions, and patterns, including: details of your visits to our Website, including traffic data, location data, logs, and other communication data and the resources that you access and use on the Website.''
This segment is labeled as \textit{First Party Collection/Use}, and its 9 attribute-value pairs are 
Does/Does Not=\textit{Does}, 
Collection Mode=\textit{Implicit}, 
Action First-Party=\textit{Collect on website},
Identifiability=\textit{Unspecified},
Personal Information Type=\textit{Location} and \textit{User online activities},
Purpose=\textit{Unspecified},
User Type=\textit{Unspecified},
Choice Type=\textit{Unspecified},
Choice Scope=\textit{Unspecified}.

To train these classifiers, we rely on the OPP-115 corpus \cite{wilson2016creation}, which is a labeled collection of 115 privacy policy texts. 
Each privacy policy segment was labeled with one or more of ten top-level categories, and then further labeled with attribute-value pairs that represent its data practices in detail. 
Each policy in this corpus was annotated independently by three legal experts. The inter-rater consistency as measured by Fleiss' Kappa ranged between 0.91 and 0.49, depending on the top-level category.

\subsubsection{Preprocessing}

To prepare the OPP-115 corpus for training, we ensure consistent spelling of all attribute labels, in particular consistent use of upper-/lower-case (e.g., ``User Profile'' vs. ``User profile'').
We use the full set of annotations (i.e., the \textit{annotations} folder), but apply majority vote consolidation~\cite{mousavinejad2020establishing}, i.e., we only include labels if at least two annotators agree on the label. 
This is applied for top-level category labels as well as for attribute labels.

In addition, we restrict attribute labels to those reported in \cite{harkous2018polisis}.
The omitted labels have very small support and would therefore be difficult to train correctly. Accordingly, restricting attribute labels improves the performance of our classifiers.
The labels for all classifiers are one-hot encoded using a multi-label binarizer, that is, all policy segments can be assigned more than one label. For example, it is possible that a policy segment covers more than one top-level category, and that it describes several lower-level attributes.

\subsubsection{Classifier training}

We train one top-level classifier and 21 attribute-level classifiers using the fast-bert library \cite{trivedi2022fast-bert}, which is based on HuggingFace transformers \cite{wolf2020transformers}. 
We first fine-tune the \textit{bert-base-uncased} language model using all unique policy texts in our corpus (4 epochs, batch size=8). Fine-tuning is a computationally expensive step (33 hours per epoch on our hardware), but improves classifier performance.

For the top-level classifier, we use the train-test-validation split reported in \cite{mousavinejad2020establishing}, which is a random split with a 3:1:1 ratio using the majority-vote version of the OPP-115 dataset.
We train the classifier for 100 epochs with a batch size of 8 using the \textit{train} portion of the dataset. 
We use the \textit{validation} portion of the dataset to evaluate the loss after each epoch.

Table \ref{tab:top-classifier-performance} shows the performance of our top-level classifier based on the \textit{test} portion of the dataset in terms of F1 score, compared with prior work. The performance on average is slightly better than the fine-tuned BERT model reported in \cite{mousavinejad2020establishing} (differences are due to the fine-tuning step, where we used a different corpus, and possible differences in the batch size and number of epochs).
Our results are slightly worse than the results from \cite{harkous2018polisis} which may be due to differences in their train/test split and data augmentation process.
Overall, the classifier performance is at about the same level as the inter-rater consistency reported for the OPP-115 dataset \cite{wilson2016creation}.

For the attribute-level classifiers, the train-test-validation split from \cite{mousavinejad2020establishing} results in imbalanced splits where some attribute labels are missing from some splits. 
Therefore, we create a separate stratified 3:1:1 split for each attribute.
Because we have multi-label data, we apply an algorithm for multi-label stratification instead of the default stratifiers in scikit-learn \cite{bradberry2022iterative-stratification,sechidis2011stratification}.
We tune the number of training epochs by comparing training loss and validation loss and selecting the final epoch as the one just before the two losses start to diverge.

Table \ref{tab:attr-classifier-performance} shows the macro F1 scores of all 21 attribute-level classifiers, compared with prior work. On average, our BERT classifiers outperform the CNN-based prior work by 10\%.
Detailed performance results for all attribute classifiers are in Appendix \ref{appendix:attribute-results}.

\begin{table*}
	\footnotesize
	\caption{F1 score for our top-level classifier vs. prior work. The best F1 scores for each category are in \textbf{bold}.}
	\label{tab:top-classifier-performance}
	\centering
	\begin{tabular}{lp{1.5cm}p{1.5cm}p{1.5cm}p{1.5cm}}
		\toprule
		& CNN, maj. \cite{mousavinejad2020establishing} & BERT, maj. \cite{mousavinejad2020establishing} & CNN, union \cite{harkous2018polisis} & BERT, maj. (here) \\
		\midrule
		First Party Collection/Use 	& 82 & \textbf{91} & 79 & 90 \\
		Third Party Sharing/Collection & 82 & \textbf{90} & 79 & 87 \\
		User Access, Edit \& Deletion & 70 & 73 & 80 & \textbf{85} \\
		Data Retention & 40 & 56 & \textbf{71} & 56 \\
		Data Security & 75 & 80 & \textbf{85} & \textbf{85} \\
		International/Specific Audiences & 82 & 83 & \textbf{95} & 84 \\
		Do Not Track & 100 & \textbf{100} & 95 & \textbf{100} \\
		Policy Change & 88 & \textbf{90} & 88 & 89 \\
		User Choice \& Control & 72 & 81 & 74 & \textbf{82} \\
		Introductory/Generic & 73 & 79 & 70 & \textbf{81} \\
		Practice Not Covered & 13 & 35 & \textbf{70} & 47 \\
		Privacy Contact Information & 84 & 78 & \textbf{87} & 78 \\
		\midrule
		\textit{Micro average} & 78 & \textbf{85} & -- & \textbf{85} \\
		\textit{Macro average} & 71 & 79 & \textbf{81} & 80 \\
		\bottomrule
	\end{tabular}
\end{table*}

\begin{table}
	\footnotesize
	\caption{Macro F1 scores for attribute-level classifiers vs. prior work. Some results, shown as n/a, were not reported in \cite{harkous2018polisis}.}
	\label{tab:attr-classifier-performance}
	\centering
	\begin{tabular}{lp{1.5cm}p{2cm}}
		\toprule
		& CNN \cite{harkous2018polisis} & BERT (here) \\
		\midrule
Access Scope & 	n/a & 	67 \\
Access Type & 	62 & 	90 \\
Action First-Party & 	65 & 	87 \\
Action Third Party & 	n/a & 	74 \\
Audience Type & 	97 & 	97 \\
Change Type & 	76 & 	90 \\
Choice Scope & 	59 & 	63 \\
Choice Type & 	73 & 	78 \\
Collection Mode & 	n/a & 	85 \\
Do Not Track Policy & 	100 & 	100 \\
Does/Does Not & 	86 & 	93 \\
Identifiability & 	77 & 	91 \\
Notification Type & 	71 & 	94 \\
Personal Information Type & 	81 & 	83 \\
Purpose & 	83 & 	84 \\
Retention Period & 	73 & 	89 \\
Retention Purpose & 	n/a & 	84 \\
Security Measure & 	74 & 	82 \\
Third Party Entity & 	73 & 	80 \\
User Choice & 	n/a & 	81 \\
User Type & 	n/a & 	92 \\

		\bottomrule
	\end{tabular}
\end{table}

\subsubsection{Policy segmentation}

To apply these classifiers to policy texts from our corpus, we have to split each policy into semantically coherent segments.
We find that the list aggregation technique proposed in \cite{harkous2018polisis}, which relies on HTML tags, does not work consistently on our policy corpus, in part due to our reliance on reader mode and readability-lxml.
Instead, we use GraphSeg \cite{glavas2016unsupervised}
(relatedness threshold 0.25, minimal segment size 1).
Instead of the default word embeddings in GraphSeg, we use custom word embeddings that are specific to the privacy policy domain.
Specifically, we use an unsupervised fastText model (model type: skipgram, dimensions: 300, minimum word count: 5) based on our corpus of unique policy texts.

\subsubsection{Policy content labeling}

Before labeling the policy segments, we discard non-English segments. This step improves labeling results for policies that include several languages in one document.
Then, we apply the top-level classifier to label the top-level category for each segment.
Finally, for each attribute that is relevant for the labeled top-level category, we apply the corresponding attribute classifier.
In addition to the predicted labels, we record the numeric prediction confidence (i.e., the output of the final sigmoid function).

\section{Readability of privacy policies}
\label{sec:readability}

\begin{figure}
	\begin{minipage}[b]{0.45\textwidth}
		\centering
		\includegraphics[width=\linewidth]{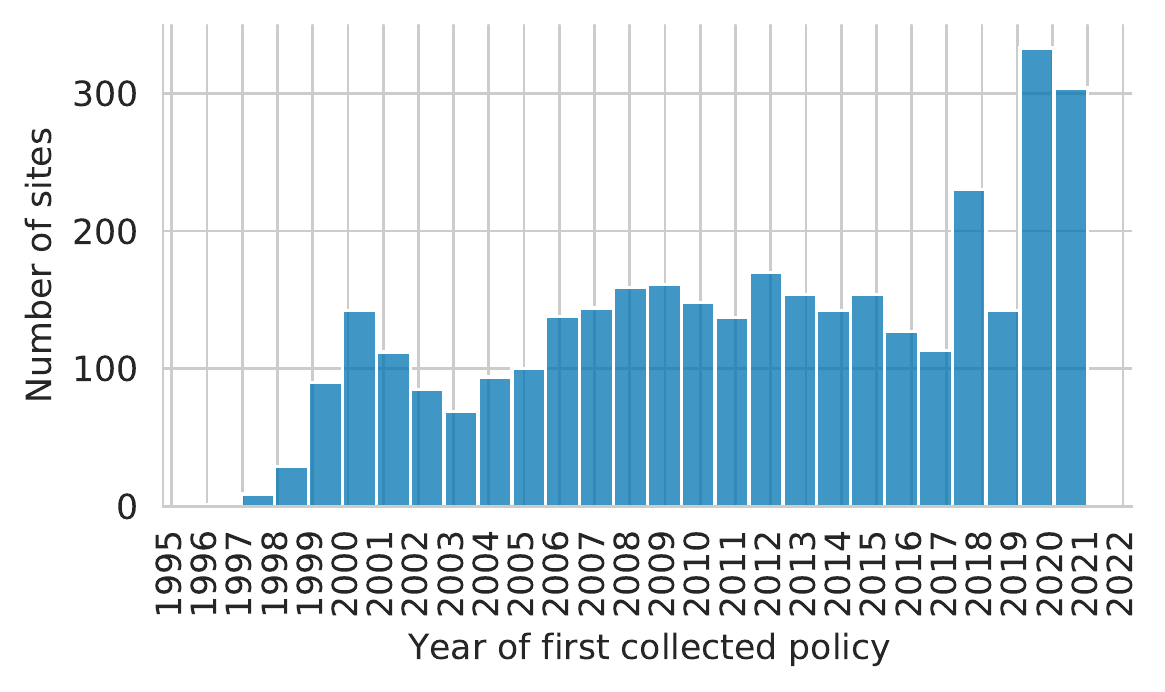}
		\caption{Year of the first policy collected for each URL.}
		\label{fig:year-of-first-policy}
	\end{minipage}\hfill
	\begin{minipage}[b]{0.51\textwidth}
		\centering
		\includegraphics[width=\linewidth]{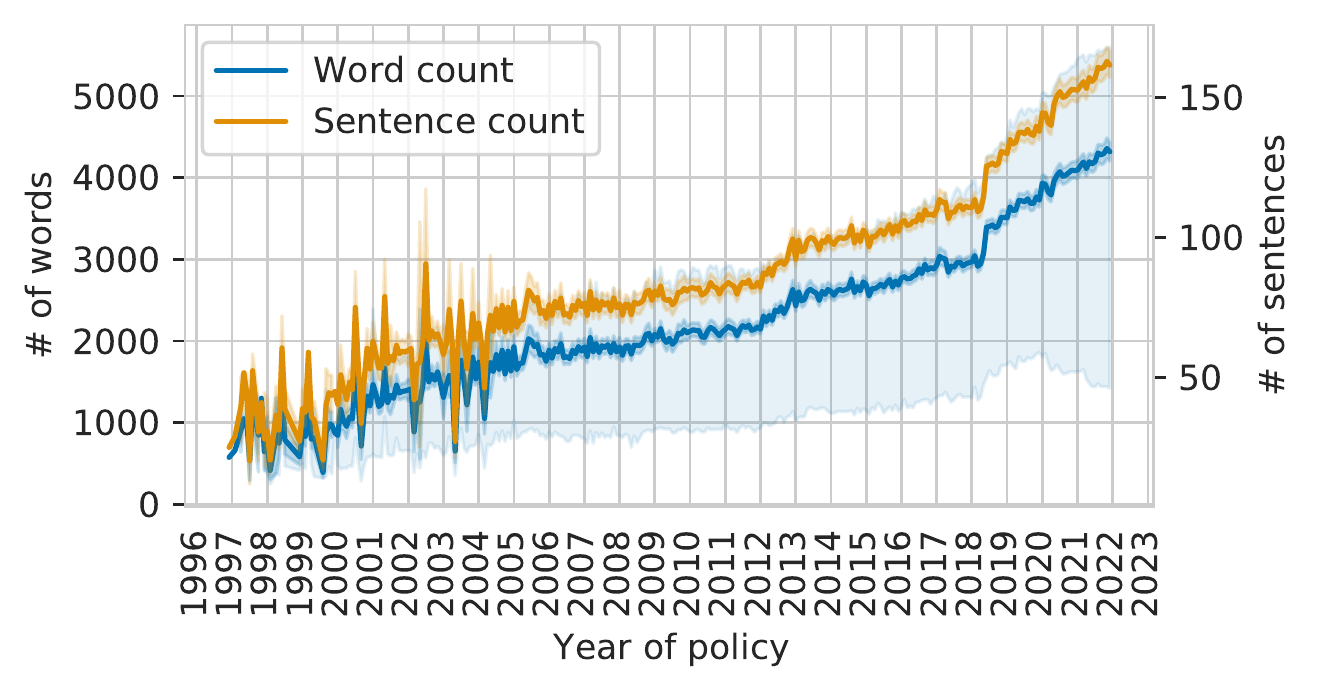}
		\caption{Mean word and sentence count of privacy policies.}
		\label{fig:length-by-month}
	\end{minipage}

\end{figure}

We now present the results of our evaluation of the readability and contents of privacy policies over the last two decades.
Our corpus contains unique privacy policies from 2,181 sites.
Figure \ref{fig:year-of-first-policy} shows in which year a site first introduced their privacy policy, i.e., the first occurrence of a privacy policy for each site.
The years with the largest number of new privacy policies are 2018 and 2020/21, which correspond to the introduction of the GDPR and CCPA.

\subsection{Length of privacy policies}

\subsubsection{Word and sentence counts}
Figure \ref{fig:length-by-month} shows the monthly mean word and sentence counts for privacy policies since 1997. In this and the following figures, the dark shaded areas indicate the 95\% confidence interval for the mean, while the light shaded area (if present) indicates the the area between the 25\% and 75\% quantiles, i.e., length of 50\% of policies.
The average policy length has almost doubled in the last ten years, with 2159 words in March 2011 and 4191 words in March 2021, and almost quadrupled since 2000 (1146 words).
Policy length increased significantly around May 2018 when the GDPR came into force and around the beginning of 2020, when the CCPA came into force, albeit with small effect sizes (Welch's t-test for word counts in March and June 2018: $p<0.001,d=0.18$, for word counts in December 2019 and June 2020: $p<0.001,d=0.08$).

\begin{figure}
	\begin{minipage}[b]{0.47\textwidth}
		\centering
		\includegraphics[width=\textwidth]{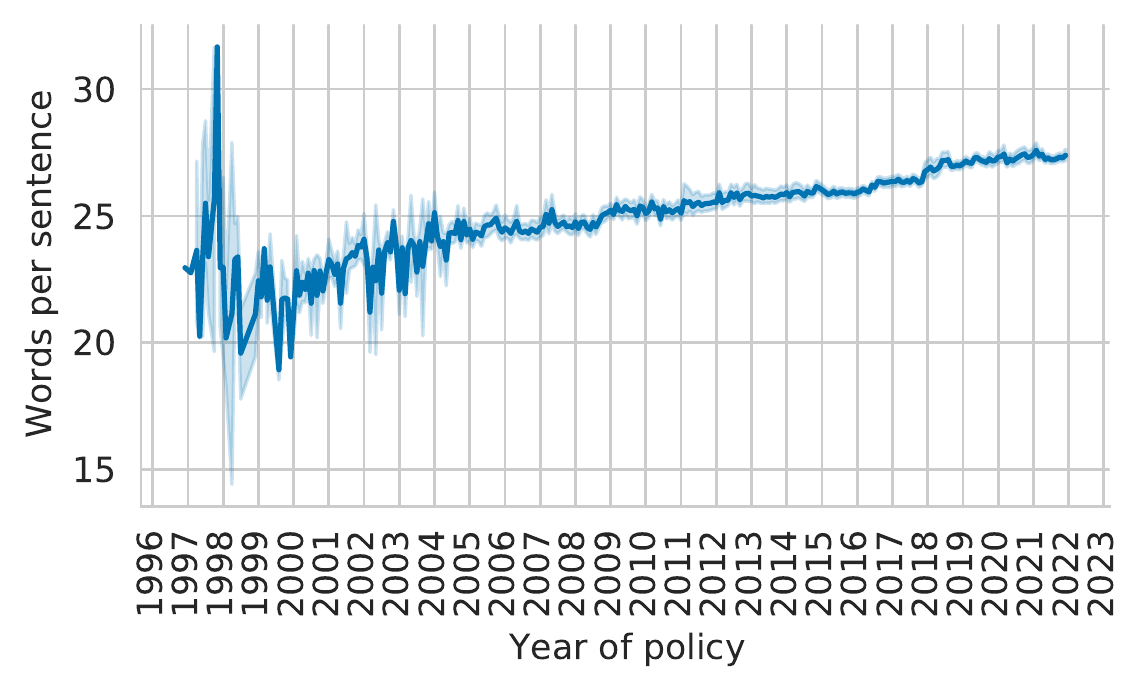}
		\caption{Average number of words per sentence.}
		\label{fig:words-per-sentence}
	\end{minipage}\hfill
	\begin{minipage}[b]{0.47\textwidth}
		\centering
		\includegraphics[width=\textwidth]{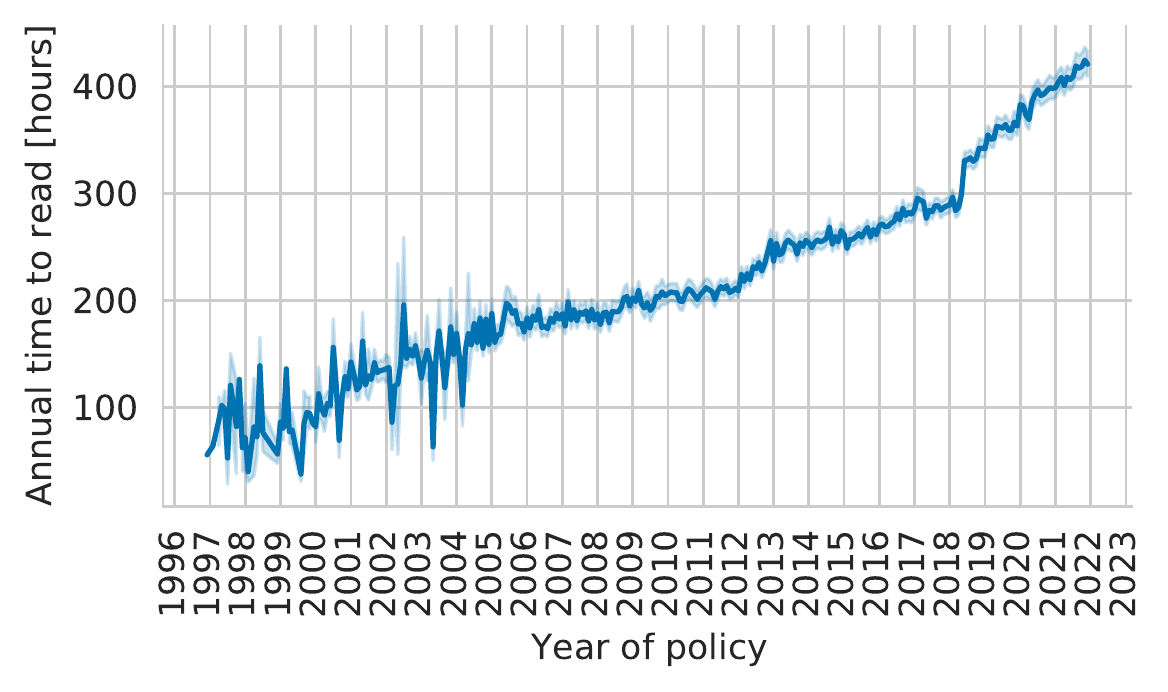}
		\caption{Annual hours needed to read policies.}
		\label{fig:annual-time-to-read} 
	\end{minipage}
\end{figure}

We analyzed whether the length of a privacy policy depends on the rank of its website, based on the tranco rank from 20th July 2020 for all websites. We find that the rank has little influence on policy length: the top 100 policies use 3079 words on average, whereas lower-ranked policies use 2853 words (Pearson's $R=-0.015$).

Figure \ref{fig:words-per-sentence} shows that the average sentence length in privacy policies increases roughly linearly with 0.2 words per sentence per year, except in the year of the GDPR introduction, where it  increased by 0.7 words, from 26.3 words in 10/2017 to 27 words in 9/2018.

\begin{figure*}
	\centering
	\includegraphics[width=\linewidth]{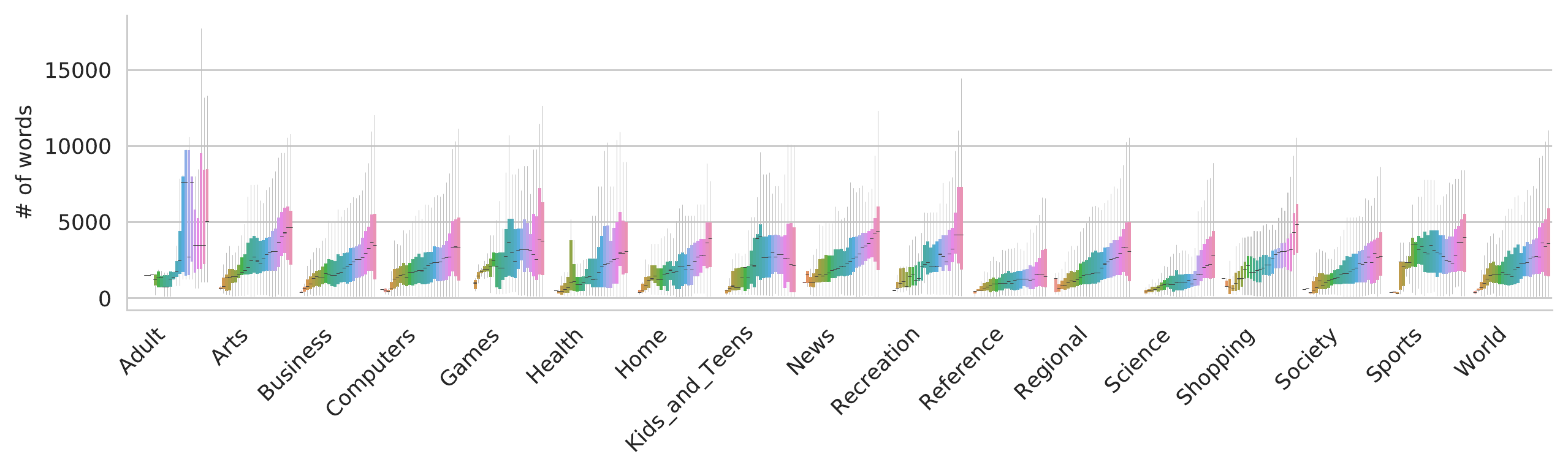}
	\caption{Policy length (word count) by category}
	\label{fig:wordcount-by-category}
\end{figure*}

Figure \ref{fig:wordcount-by-category} shows how policy length has evolved for different categories of websites, as indicated by the second-level category reported by Alexa. We can see that the trend for each individual category is similar to the overall trend shown in Figure \ref{fig:length-by-month}. 
For six categories, the average policy length in 2021 is significantly different ($p<.001$)  from the overall average with at least a small effect size ($d>0.2$): websites from the \textit{reference}, \textit{science}, and \textit{society} categories are 900--1600 words shorter, while policies from the \textit{adult}, \textit{shopping}, and \textit{recreation} categories are 900--2500 words longer.

\subsubsection{Time to read}

To put policy length into the context, we compute the time it would take to read each policy, assuming a reading speed of 250 words per minute \cite{mcdonald2008cost,libert2018automated}.
In 2021, the average privacy policy takes 17 minutes to read, the average top-10 policy takes 23 minutes, and the longest policy (Microsoft) takes 152 minutes.

Assuming that each user visits 1462 unique websites each year \cite{mcdonald2008cost}, Figure \ref{fig:annual-time-to-read} shows an estimate for the annual time spent for reading privacy policies.
Compared with \cite{mcdonald2008cost}, our numbers for the year 2008 (190 hours) are slightly lower than what they report (244 hours), which is likely caused by differences in the policy sample (top 75 websites from the AOL search log in \cite{mcdonald2008cost}, compared to policies from 996 sites for 2008 in our corpus).
It is notable that the annual reading time in 2021 is more than 400 hours, i.e., more than one hour per day.

\subsection{Readability of privacy policies}

\subsubsection{Readability formulas}
\begin{figure}
	\centering
	\includegraphics[width=.9\linewidth]{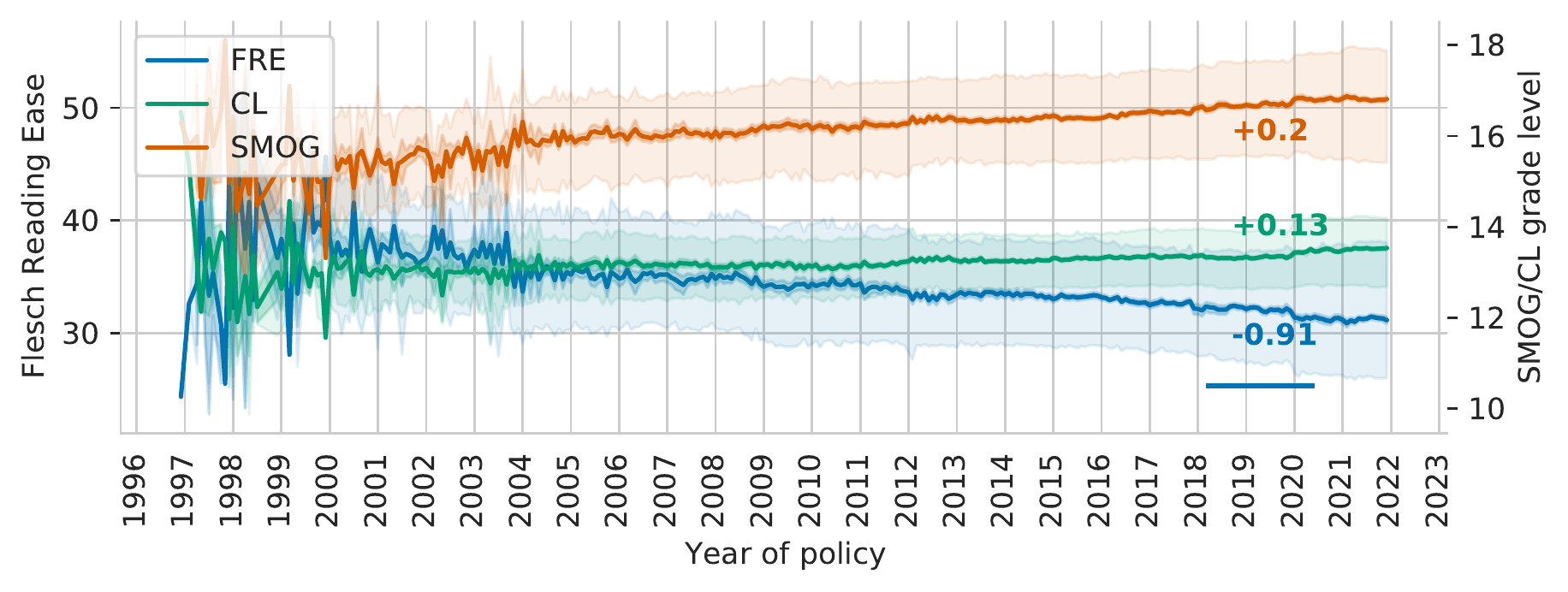}
	\caption{Flesch Reading Ease (FRE), SMOG, and Coleman-Liau (CL) grade levels. Shaded areas indicate 50\% of policies; numbers indicate the change between 3/2018 (before GDPR) and 6/2020 (after CCPA).}
	\label{fig:readability-by-year}
\end{figure}

Figure \ref{fig:readability-by-year} shows the readability of privacy policies indicated by the Flesch Reading Ease, Coleman-Liau, and SMOG measures.
More difficult texts have lower values for FRE and higher values for CL and SMOG. The CL and SMOG grade levels can be interpreted as the years of education needed to understand a text.
All three measures indicate that privacy polices have become increasingly hard to read over the years.
In 2021, 41.6\% of policies had an FRE score below 30. This score indicates a very difficult text that is best understood by university graduates. The median FRE score for 2021 policies was 31.8. 
Only 6.7\% of policies in 2021 had an FRE score above 45, which is the readability standard required for insurance policies in Florida \cite{thefloridasenate1982florida}.

SMOG and CL estimate that the median policy in 2021 requires 16.7 and 13.4 years of education, respectively (the grade levels differ because formulas differ in variables and calibration).
Between March 2018 and June 2020, across the introduction of both GDPR and CCPA, the SMOG grade level increased by 0.20, corresponding to two additional months of education needed to understand the new privacy policies ($p<.001, d=0.09$).

\begin{figure*}
	\centering
	\includegraphics[width=\linewidth]{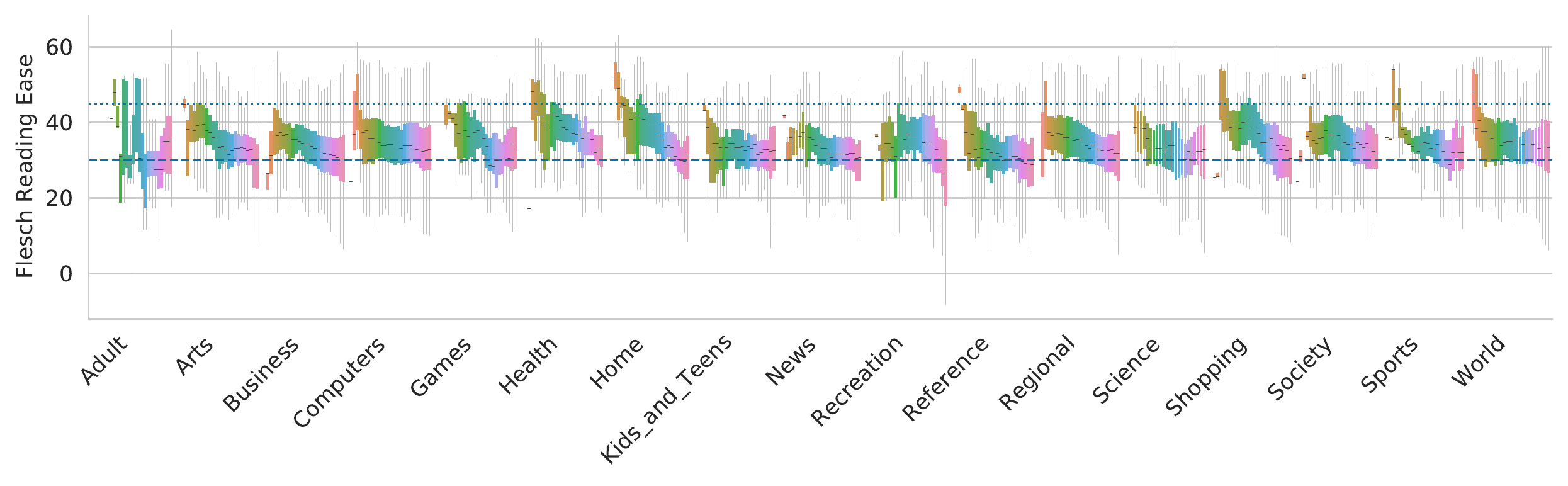}
	\caption{Readability (Flesch Reading Ease) by category. A university degree is needed to understand texts with a score below 30 (dashed line), whereas a score above 45 (dotted line) would comply with Florida's standard for insurance policies.}
	\label{fig:flesch-by-category}
\end{figure*}

To analyze whether some website categories have more readable policies than others, Figure~\ref{fig:flesch-by-category} shows the FRE score for websites grouped by their Alexa category.
The figure shows most categories follow the overall trend of decreasing readability, however, policies from \textit{adult}, \textit{games}, and \textit{science} websites have become noticeably more readable in the last years.
In 2021, four categories have significantly different FRE scores compared to the overall average, with at least a small effect size ($p<.001$ and $d>0.2$): \textit{adult} websites and websites for \textit{kids and teens} have more readable policies (FRE 3--5 points higher than average), whereas \textit{reference} and \textit{recreation} websites have less readable policies (FRE 2--5 points lower than average).

\begin{figure}
	\begin{minipage}[b]{0.47\textwidth}
		\centering
		\includegraphics[width=\linewidth]{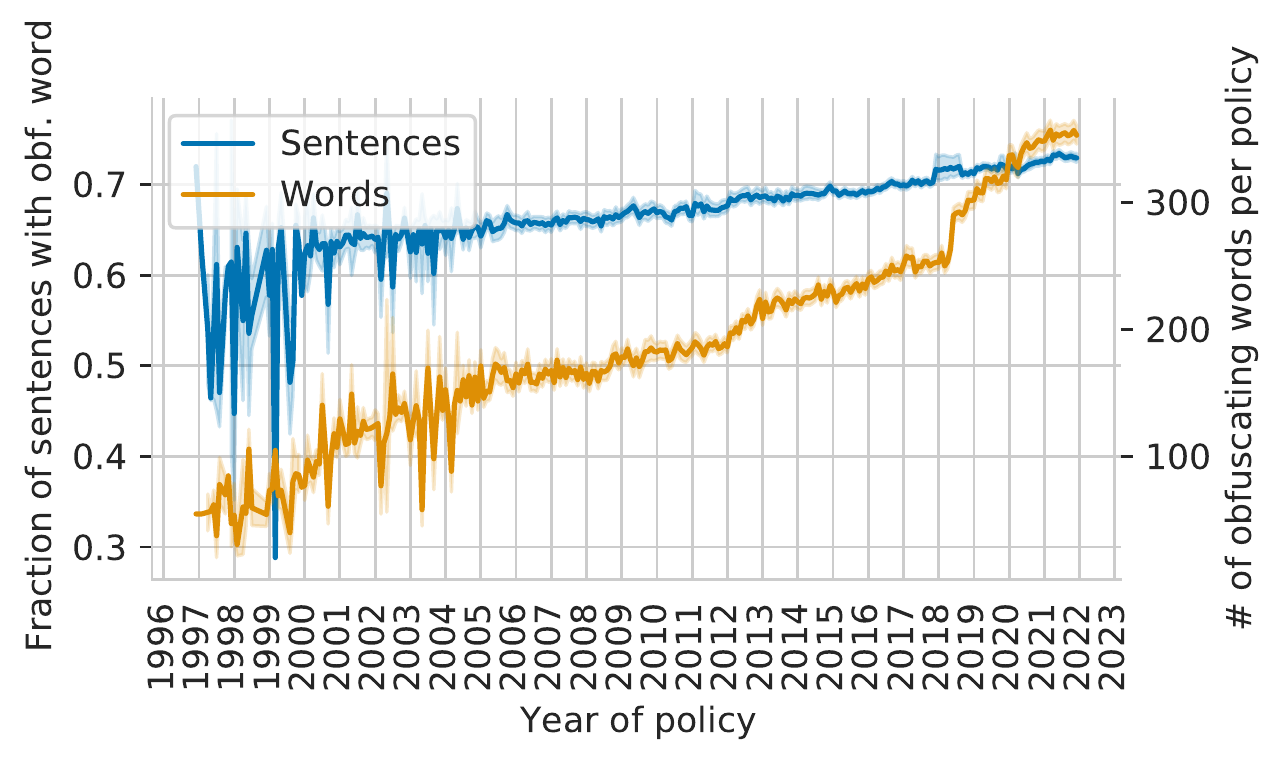}
		\caption{Number of obfuscating words in policies and fraction of sentences with obfuscating words.}
		\label{fig:obfuscating-words-by-year}
	\end{minipage}\hfill
	\begin{minipage}[b]{0.47\textwidth}
		\centering
		\includegraphics[width=\linewidth]{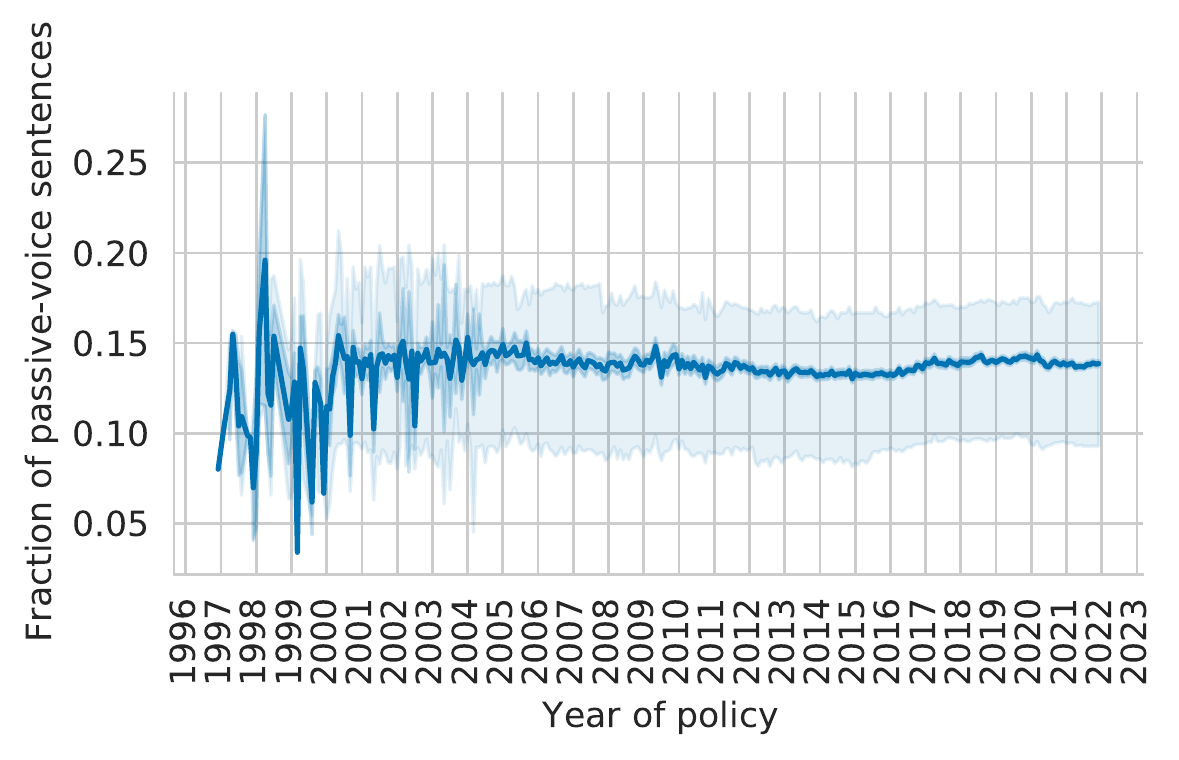}
		\caption{Fraction of sentences in passive voice. The area between the 25\% and 75\% quartiles is shaded.}
		\label{fig:passive-voice}
	\end{minipage}
\end{figure}

\subsubsection{Obfuscating words}
Figure \ref{fig:obfuscating-words-by-year} shows the average number of obfuscating words in privacy policies as well as the fraction of sentences that contain obfuscating words.
We can see that the absolute number of obfuscating words increased steadily before 2018, but then increased rapidly from a median of 227 in March 2018 to 304 in June 2020 ($p<.001, d=0.385$).
In contrast, the fraction of sentences that contain obfuscating words continued its steady increase, indicating that the increase in the absolute number can be explained by the increased policy length post-GDPR.
In 2021, more than two-thirds of sentences in privacy policies contained at least one obfuscating word (72\%).

\subsubsection{Passive-voice sentences}
Passive-voice sentences are harder to understand than active-voice sentences and can obscure who the actor in a sentence is. In a privacy policy, this can indicate that companies do not want to take responsibility for their data practices \cite{mcdonald2009comparative}.
Figure \ref{fig:passive-voice} shows the fraction of passive-voice sentences in privacy policies over the years. There is no significant increase over the years, and across the introduction of the CCPA the fraction of passive-voice sentences even decreased by 0.5\%. 

\subsection{Common terms}

\begin{figure}
	\begin{minipage}[b]{0.47\textwidth}
		\centering
		\includegraphics[width=\linewidth]{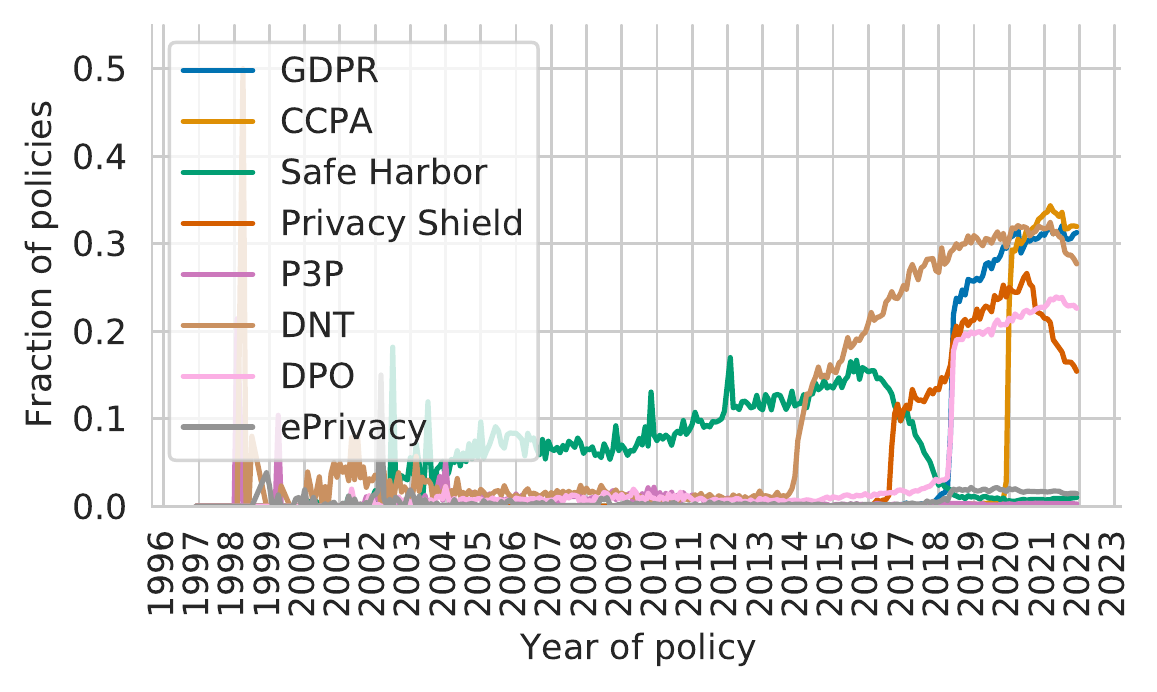}
		\caption{Mentions of eight key terms.}
		\label{fig:perc-terms-by-year}
	\end{minipage}\hfill
	\begin{minipage}[b]{0.47\textwidth}
		\centering
		\includegraphics[width=\linewidth]{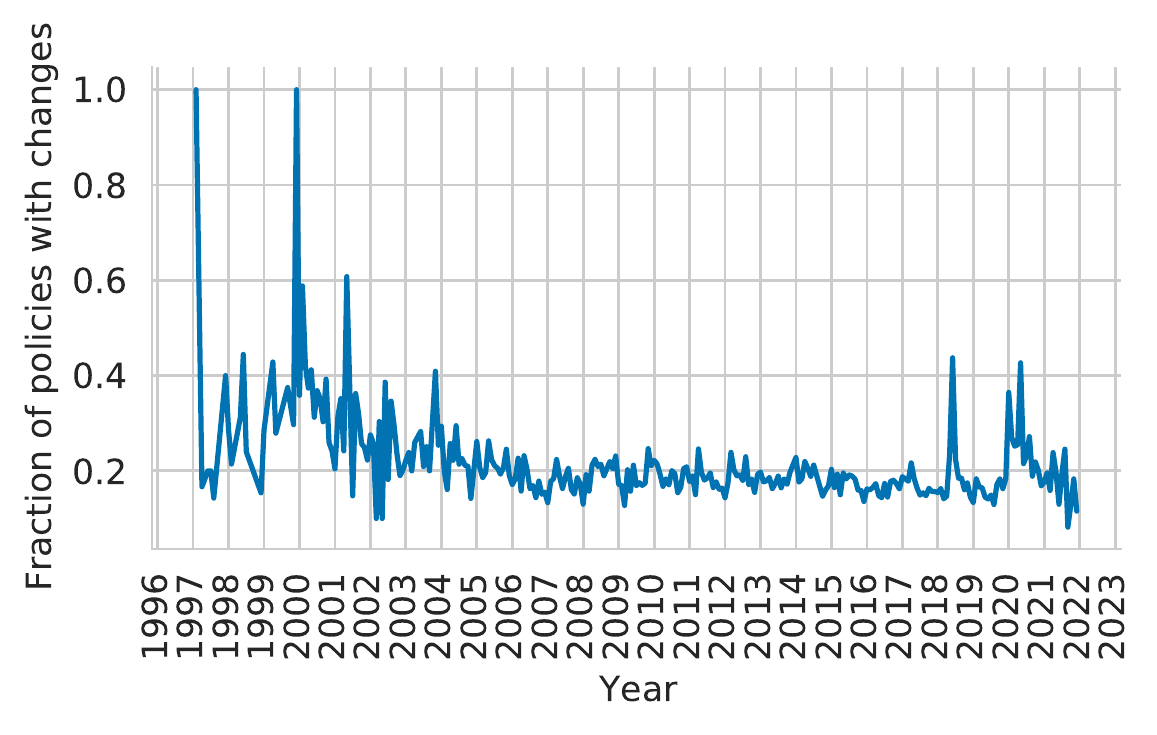}
		\caption{Fraction of URLs with an updated policy.}
		\label{fig:policy-change-by-year}
	\end{minipage}
\end{figure}

Figure \ref{fig:perc-terms-by-year} shows how mentions of eight privacy-related terms have evolved, based on text matches of acronyms and long forms.
The Safe Harbor agreement was invalidated in October 2015. Figure \ref{fig:perc-terms-by-year} shows a drop in policies mentioning Safe Harbor over the following two years from over 15\% in early 2016 to under 2\% in 2018 and 0.7\% in 2021.
Between 2016 and 2018, the increase in mentions of the Privacy Shield mirrors the decrease in Safe Harbor mentions, and keeps climbing to 27\% of policies in mid-2020.
After its invalidation in July 2020, mentions of the Privacy Shield drop to under 20\% in early 2021.

During the first half of 2018, mentions of the GDPR increased steeply, similar to the steep increase in CCPA mentions in early 2020. 
Data Protection Officers (DPOs) started to be mentioned at the same time as the GDPR and at a similar rate, following the GDPR requirement for organizations to name their DPO.

The Platform for Privacy Preferences Project (P3P) was recommended by the W3C in 2002. About 5\% of policies mentioned P3P in 2004, but most mentions stated that the natural language text overrides the machine-readable policy.
As of 2021, only 0.3\% of policies mention P3P.

The Do Not Track (DNT) HTTP header was proposed in 2009, but discontinued in 2018. Beginning in 2014, privacy policies have increasingly mentioned DNT, with almost 32\% of policies mentioning DNT in 2021.
However, as with P3P, almost all policies state that they do not honor DNT (see also Section \ref{sec:content-analysis}).

\subsection{Effect of GDPR and CCPA introduction}

\subsubsection{Update rates}
Figure \ref{fig:policy-change-by-year} shows how many privacy policies were updated each month, measured as a change in at least one sentence.
Between 2005 and 2017, roughly 20\% of sites update their policy every month. In the second half of 2017, the fraction of updated policies dropped slightly, most likely because websites were aware that the GDPR would come into force in 2018, but did not know yet how to update their privacy policies. In May 2018, websites were forced to react because the GDPR came into effect, and more than 40\%  of policies were updated in mid-2018.
A similar effect is observable with the introduction of the CCPA in 2020.

\subsubsection{Policy length}
To further analyze the effect of the introduction of the GDPR and CCPA on the length and readability of privacy policies, we split our policy corpus into those URLs whose policies mention the GDPR or CCPA, and those URLs that do not.

\begin{figure}
	\begin{minipage}[b]{0.47\textwidth}
		\centering
		\includegraphics[width=\linewidth]{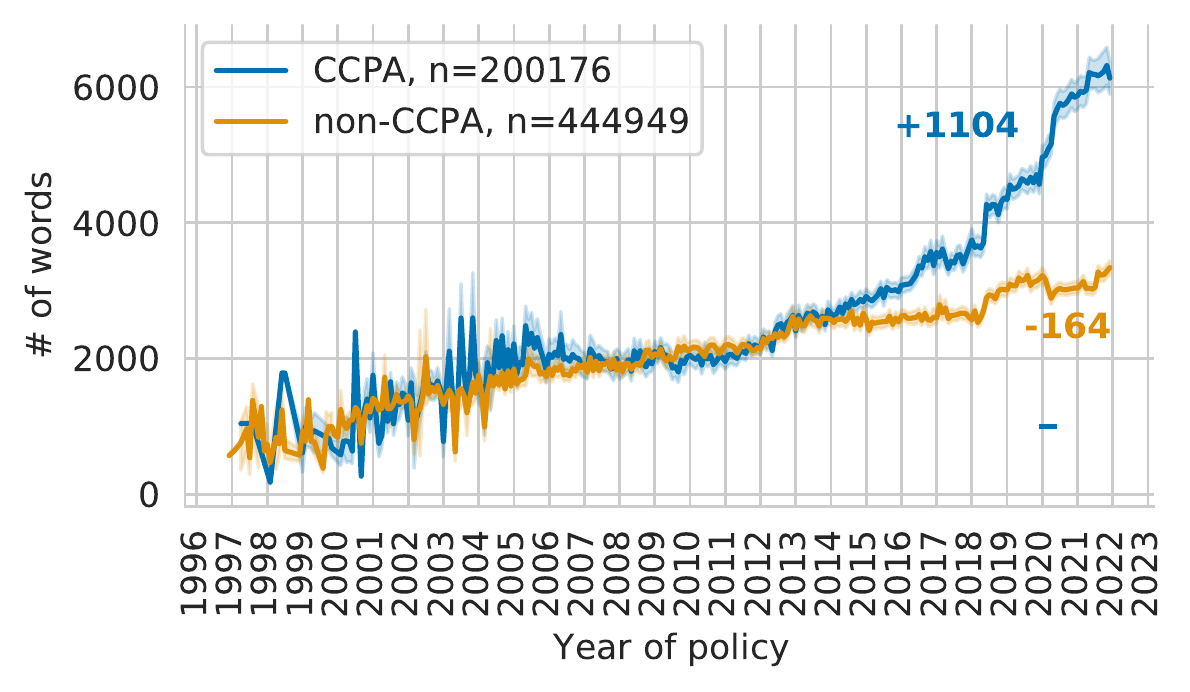}
		\caption{Length of privacy polices over time, split into URLs whose policies mention CCPA and those that do not. Numbers indicate the change between 12/2019 and 6/2020.}
		\label{fig:policy-length-CCPA-vs-non}		
	\end{minipage}\hfill
	\begin{minipage}[b]{0.47\textwidth}
		\centering
		\includegraphics[width=\linewidth]{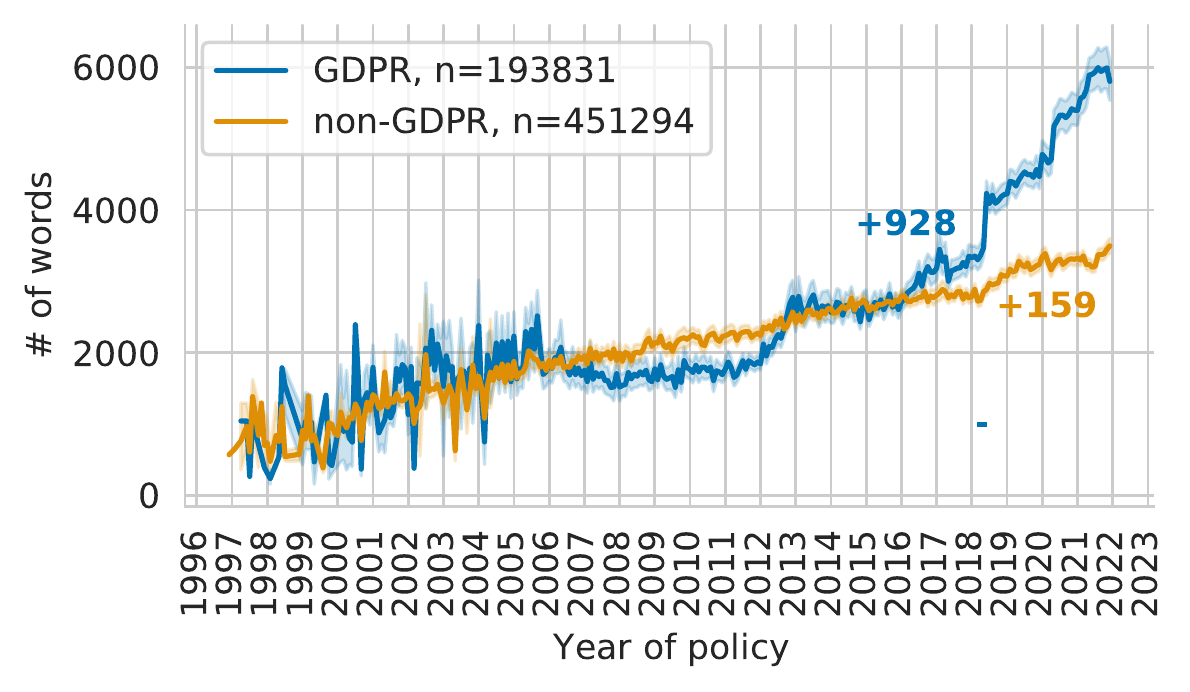}
		\caption{Length of privacy polices over time, split into URLs whose policies mention GDPR and those that do not. Numbers indicate the change between 3/2018 and 6/2018.}
		\label{fig:policy-length-GDPR-vs-non}
	\end{minipage}
\end{figure}

Figure \ref{fig:policy-length-CCPA-vs-non} shows that the length of policies that end up mentioning the CCPA increases significantly by 1104 words between December 2019 and June 2020 ($p<.001, d=0.27$), whereas policies that do not mention the CCPA become 164 words shorter (not statistically significant, $p=0.0197, d=0.061$).
In addition, the rate of increase is higher for policies that mention the CCPA.
Similarly, policies mentioning the GDPR increased by 928 words between March and June 2018 ($p<.001, d=0.28$), compared to 159 words for policies that do not mention the GDPR (not statistically significant, $p=0.00596, d=0.07$, see Figure \ref{fig:policy-length-GDPR-vs-non}).
In addition, the rate of increase post-GDPR is also higher for policies that mention the GDPR.
Possible reasons for this increase in length are that policies needed to add text that addresses the requirements of each regulation (see below), and that policies have used lengthy legalistic language from each regulation's text to ensure compliance.

\begin{figure}
	\begin{minipage}[b]{0.47\textwidth}
		\centering
		\includegraphics[width=\linewidth]{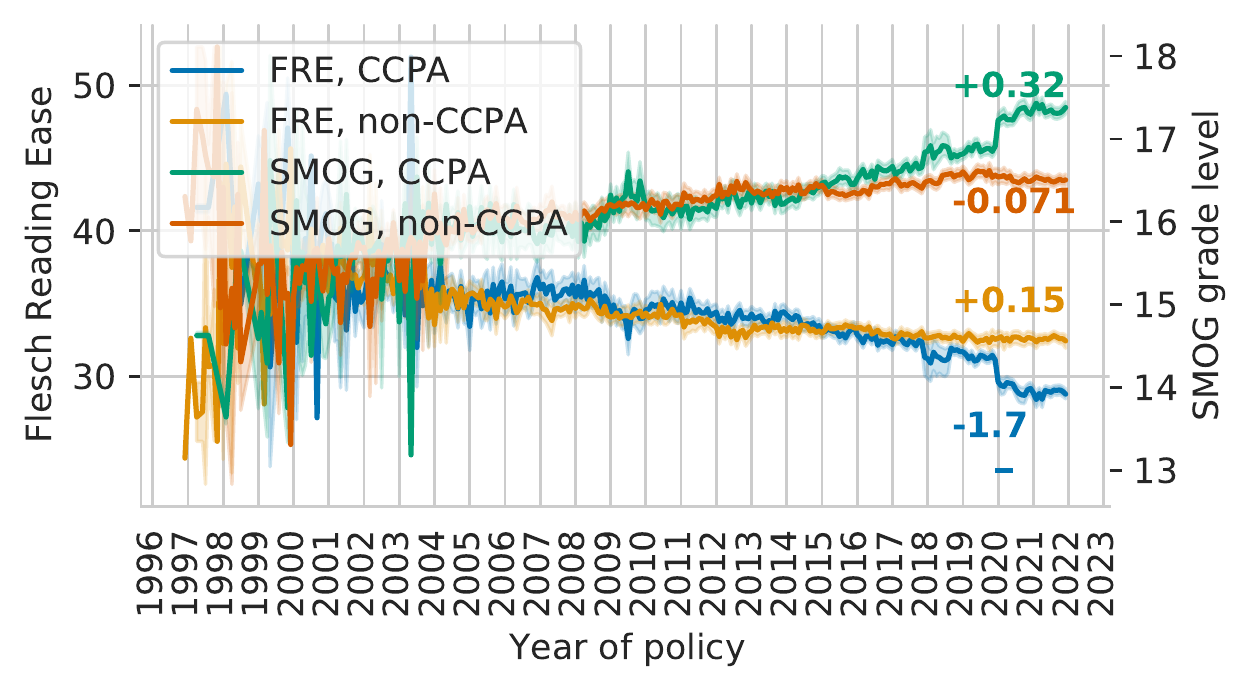}
		\caption{Readability of policies over time, split into URLs whose policies mention CCPA and those that do not. Numbers indicate the change between 12/2019 and 6/2020.}
		\label{fig:readability-CCPA-by-year}
	\end{minipage}\hfill
	\begin{minipage}[b]{0.47\textwidth}
		\centering
		\includegraphics[width=\linewidth]{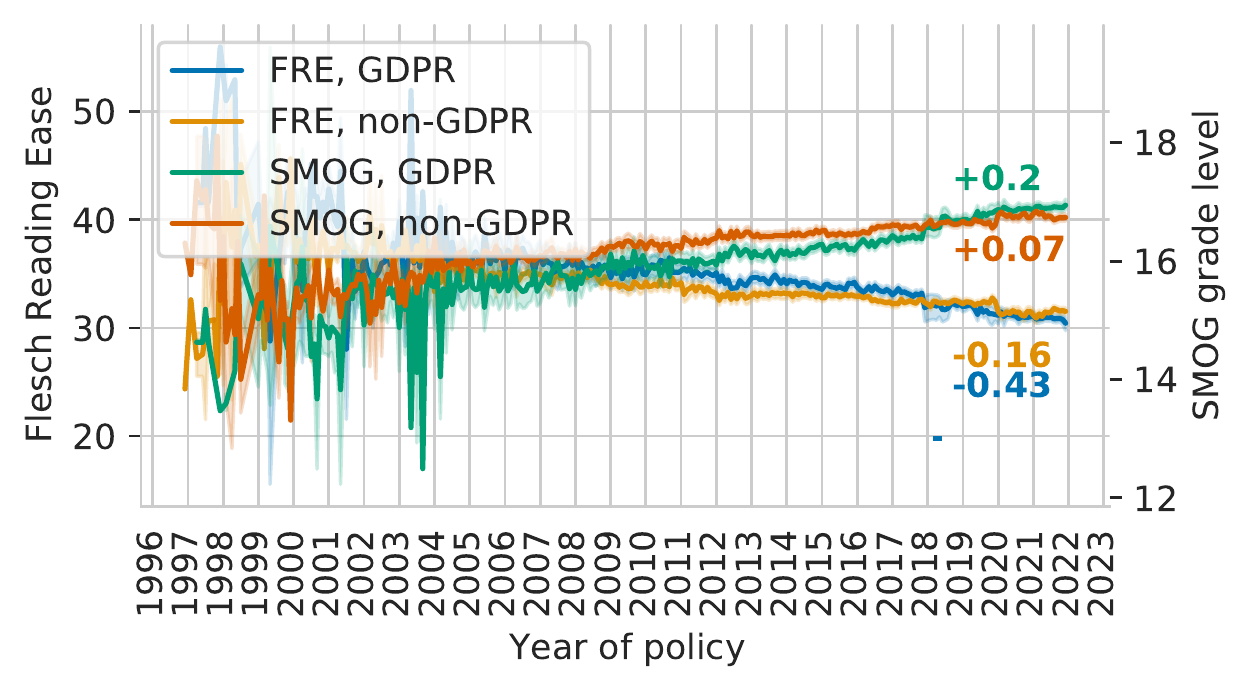}
		\caption{Readability of policies over time, split into URLs whose policies mention GDPR and those that do not. Numbers indicate the change between 3/2018 and 6/2018.}
		\label{fig:readability-GDPR-by-year}
	\end{minipage}
\end{figure}

\subsubsection{Readability}
Figure \ref{fig:readability-CCPA-by-year} compares the readability of policies that end up mentioning the CCPA with those that do not mention it (Figure \ref{fig:readability-GDPR-by-year} for GDPR).
The introduction of the CCPA pushed the average Flesch Reading Ease down by 1.7 points ($p<.001, d=0.21$), below the threshold of 30 points. This means that understanding the average privacy policy that mentions the CCPA requires a university degree.
As a result, half the US population\footnote{Based on OECD tertiary education rates: Australia 47\%, Canada 59\%, New Zealand 39\%, UK 47\%, US 48\%, and Germany 30\% \cite{oecd2020education}.} does not have the necessary education to understand the privacy policies that apply to them.

\begin{figure}
	\centering
	\includegraphics[width=.9\linewidth]{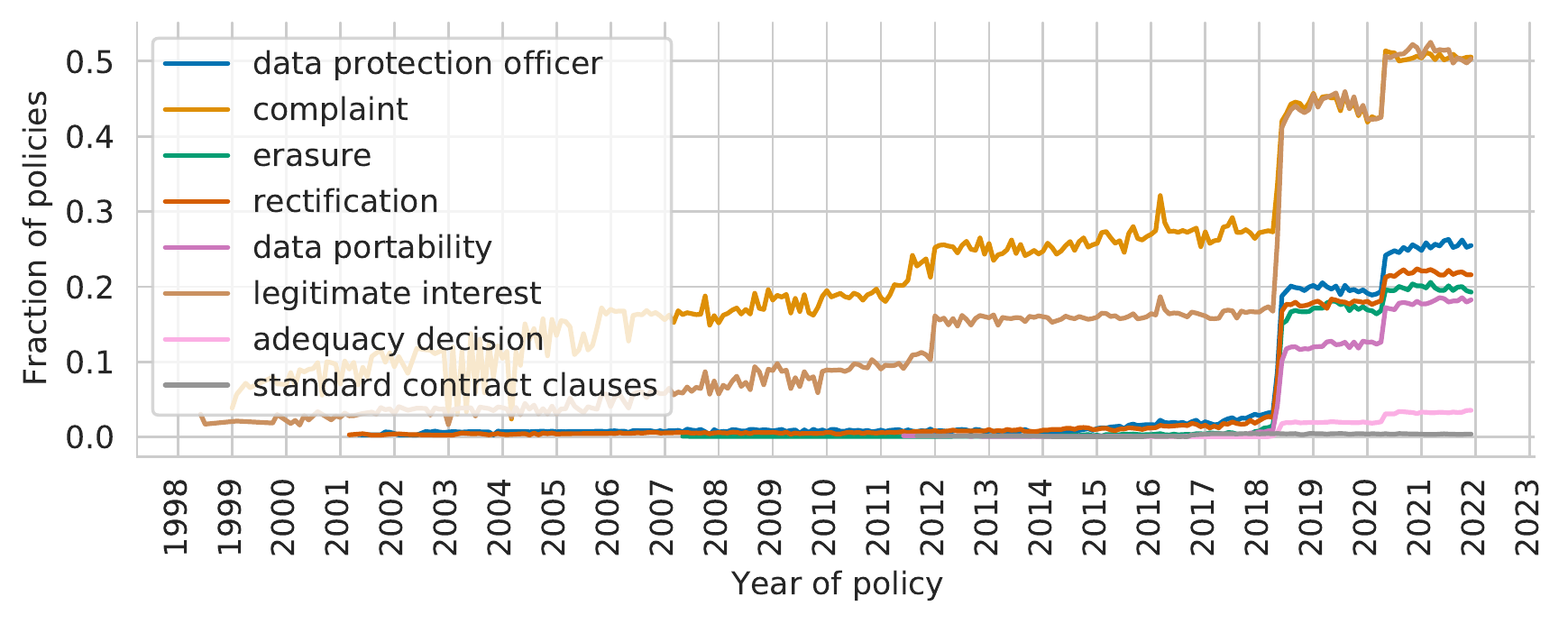}
	\caption{Fraction of policies that mention eight key terms from privacy regulations.}
	\label{fig:gdpr-terms-by-month}
\end{figure}

\subsubsection{Use of key terms}
To evaluate how the content of privacy policies changed with the introductions of the GDPR and CCPA and the court rulings concerning Safe Harbor and Privacy Shield, Figure \ref{fig:gdpr-terms-by-month} shows the fraction of policies that mention eight key terms.
Five GDPR terms concerning user rights -- data protection officer, complaint, erasure, rectification, and data portability \cite{degeling2019we} -- each increase in mentions by 10--20\% in mid-2018, with an additional 5--10\% increase in early 2020.
Mentions of legitimate interest  increased even more (+25\%), which indicates that website owners are attempting to find new legal bases to justify their data practices.

The final two terms, adequacy decision and standard contract clauses, describe two mechanisms that can justify transatlantic data transfers after the invalidation of Privacy Shield.
An adequacy decision, made by the European Commission, means that a non-EU country offers adequate data protection levels \cite{europeancommission2021adequacy}. 
Standard contract clauses are standard texts, published by the EU, that non-EU data processors can use to ensure adequate data protection levels for their EU users \cite{europeancommission2010standard}.
However, Figure \ref{fig:gdpr-terms-by-month} shows only a small increase in mentions of these terms.
This may be caused by our text-matching method, but it may also indicate that website owners refrain from transatlantic data transfers, for example through the use of EU cloud services, that they rely on GDPR article 49 (explicit informed consent), or that they are in breach of European data protection regulations.

\section{Content of privacy policies}
\label{sec:content-analysis}
To analyze the semantic content of privacy policies, we are interested in how many policies each year address each privacy practice, and in what way, e.g., whether they assert collection or sharing of a specific information type.
To this end, we label each privacy policy segment with the top-level category classifier and with each attribute classifier that is relevant to its category labels.
We then eliminate segments with duplicate labels, i.e., we remove segments that belong to the same unique policy if they have the same labels, regardless of the segment text or the policy's timestamp. 
To analyze top-level categories, we retain only the first mention of each category per policy, at its earliest instance.

In addition, we exclude categories and attributes for which the classifiers did not perform well.
Specifically, we retain only labels that have a precision of at least 75\%. This ensures that the predicted labels are most likely correct (low false positives), while accepting the possibility that the classifiers miss some labels (low recall corresponding to high false negatives).
As a result, the reported results may underestimate the true prevalence of privacy practices in privacy policies.
Specifically, for top-level categories, we exclude the \textit{data retention} and \textit{practice not covered} categories. 
For attribute classifiers, we exclude a total of nine labels across the 21 classifiers (marked with an asterisk in Appendix \ref{appendix:attribute-results}).

In the subsequent figures, we present the fraction of policies labeled with specific attributes or combinations of attributes as bar plots, where each bar represents policies from one year. The height of each bar indicates the fraction of policies for which the classifier's confidence was above 0.5 (where combinations of attributes are plotted, both confidences are above 0.5).
We compute 95\% prediction intervals for all results, shown as grey error bars.
Our calculation of prediction intervals relies on the insight that the number of policies with a given label is a random variable with a Poisson Binomial probability distribution \cite{hong2013computing}. The number of policies corresponds to the mean of this distribution, and our prediction intervals are computed based on the 0.975 and 0.025 quantiles. Importantly, this approach takes into account all estimated probabilities, not just the ones above the classification threshold of 0.5 \cite{dechant2015assessment}.

\begin{figure}
	\centering
	\includegraphics[width=\linewidth]{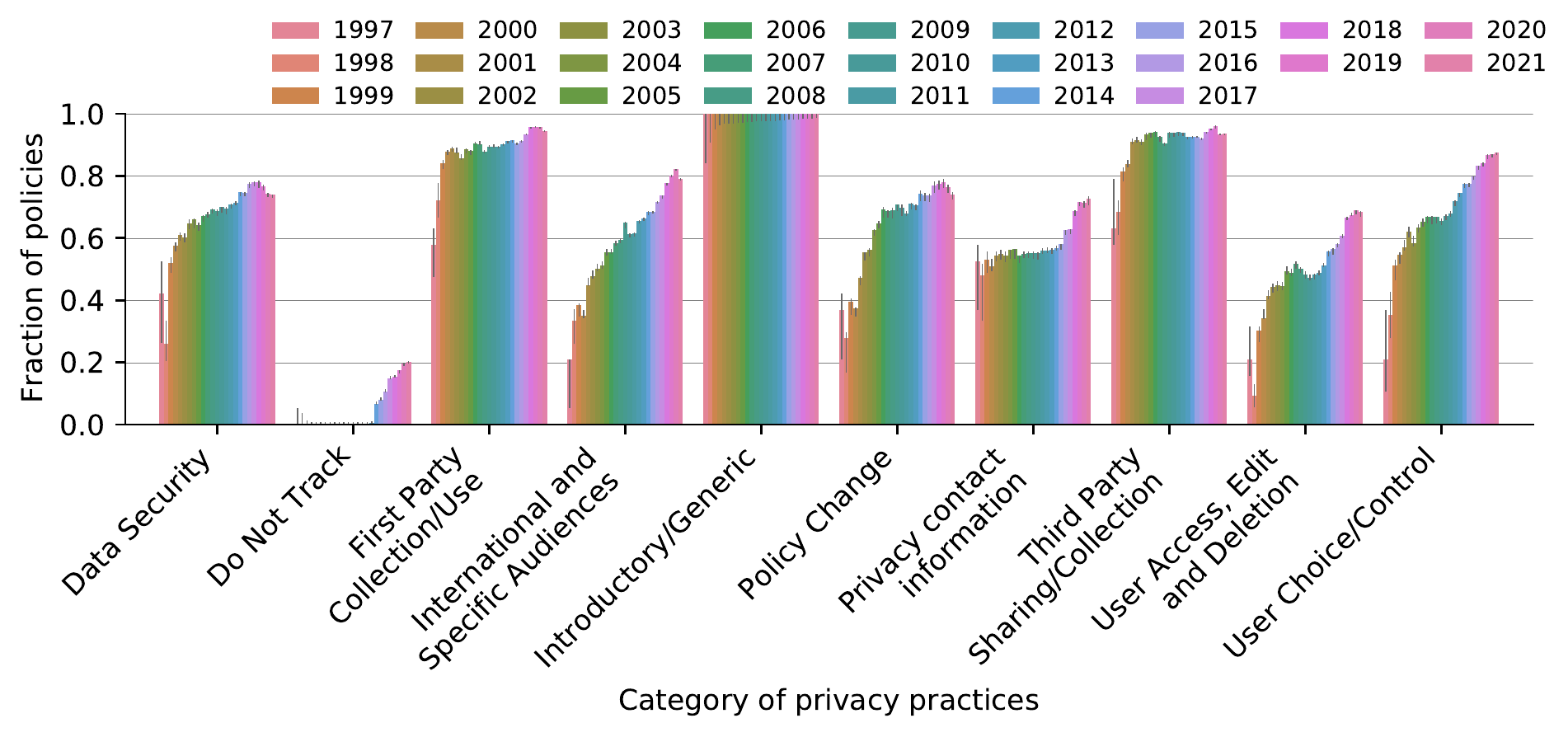}
	\caption{Fraction of privacy policies each year that address each category of data practices.}
	\label{fig:content-categories-by-year}
\end{figure}

\begin{figure*}
	\centering
	\includegraphics[width=\linewidth]{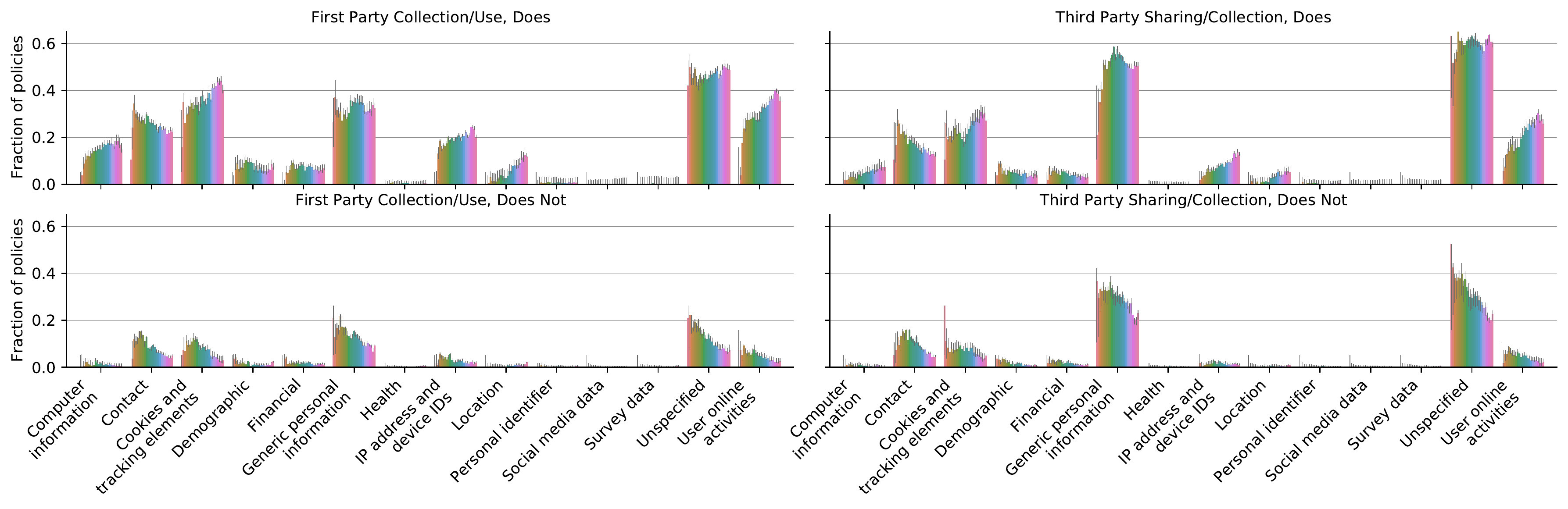}
	\caption{Personal information types collected by first vs. third parties, split by whether the policy \textit{does} or \textit{does not} assert the data collection.}
	\label{fig:content-personal-information-type-does-doesnot}
\end{figure*}

Figure \ref{fig:content-categories-by-year} shows the fraction of policies that address each data practice over the past 25 years.
Overall, privacy policies address more data practices each year, i.e., they have become more comprehensive. Starting in 2018, however, there is a slight decline for some data practices, including statements about \textit{data security} and \textit{policy change}. Almost all policies contain \textit{introductory/generic} statements which we disregard in the remainder of the analysis.
In 2021, \textit{first-party collection and use} is the most commonly addressed data practice (94\% of privacy policies), followed by \textit{third-party sharing and collection} (93\%).
However, from a regulatory viewpoint many privacy policies are still lacking: only 67\% of policies explain users' access, edit, and deletion rights, and only 71\% give contact details for privacy-related queries (both required by the GDPR). The do not track header is mentioned in 20\% of policies, which all assert that they do not respect the header.

Figure \ref{fig:content-personal-information-type-does-doesnot} shows which personal information types are mentioned in privacy policies, split by first- or third-party data collection and by whether the policy \textit{does} or \textit{does not} assert the data collection. 
In most cases, especially for first-party collection, the policies assert that data is indeed collected.
For third-party collection, assertions of non-collection are more frequent, most notably for sharing of \textit{generic personal information} with third parties, where 21\% of 2021 polices state that data is not collected. However, many more policies (51\%) assert third-party sharing of \textit{generic personal information}.

In the remainder of this section, we present results for specific data practices, filtering the results to only include policy statements that \textit{assert} data collection.
\subsection{First-party data collection/use}
The most salient attributes for first-party data collection/use are the collection mode, purpose, personal information types, choice and controls offered to users, and identifiability of collected data.

\subsubsection{Collection mode vs. personal information type, purpose, and choice/control}
\begin{figure*}
	\centering
	\subfloat[Personal information type]{\label{fig:content-pitype-mode} 
	{\includegraphics[width=.48\linewidth]{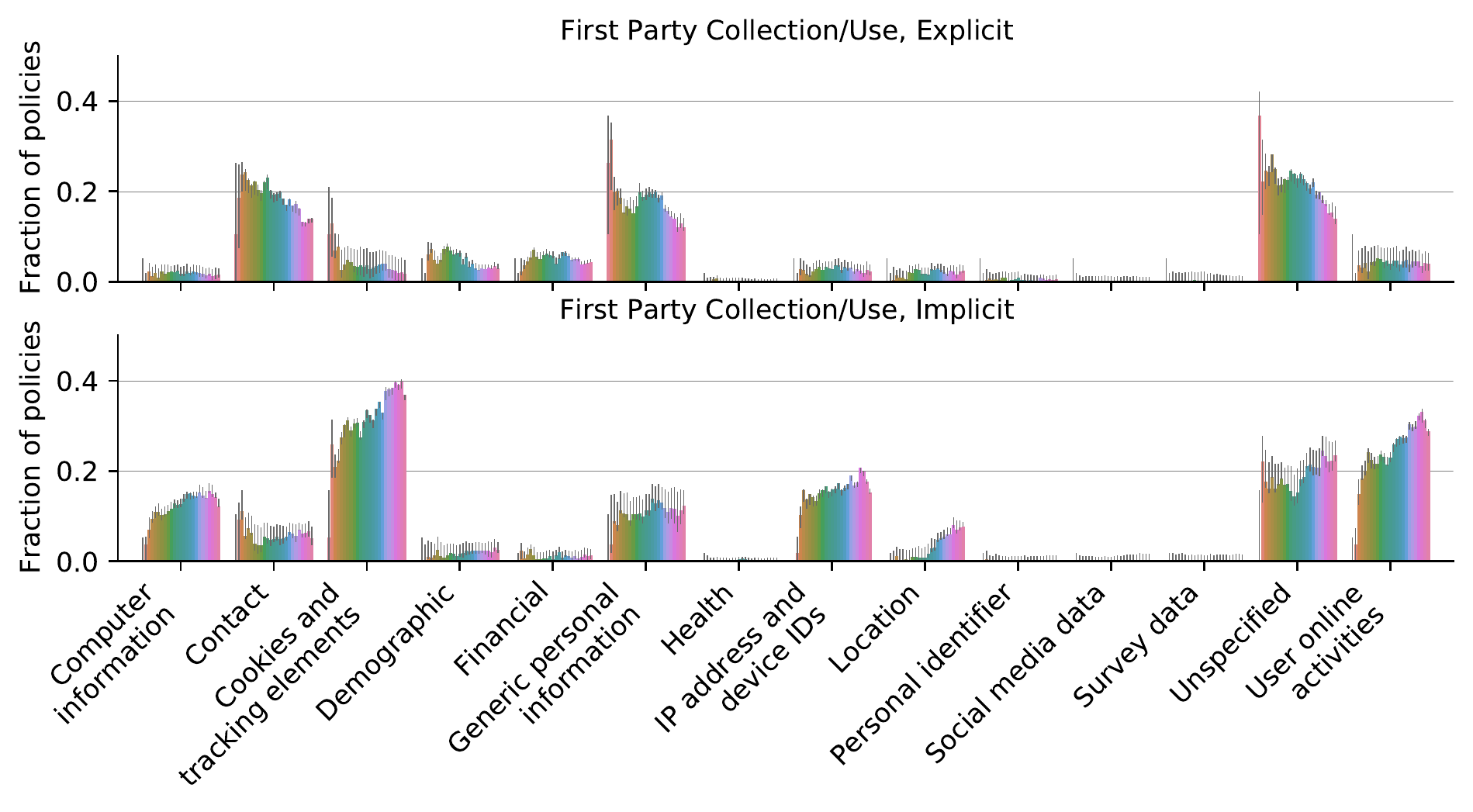}}}
\subfloat[Purpose]{\label{fig:content-purpose-mode} 
	{\includegraphics[width=.48\linewidth]{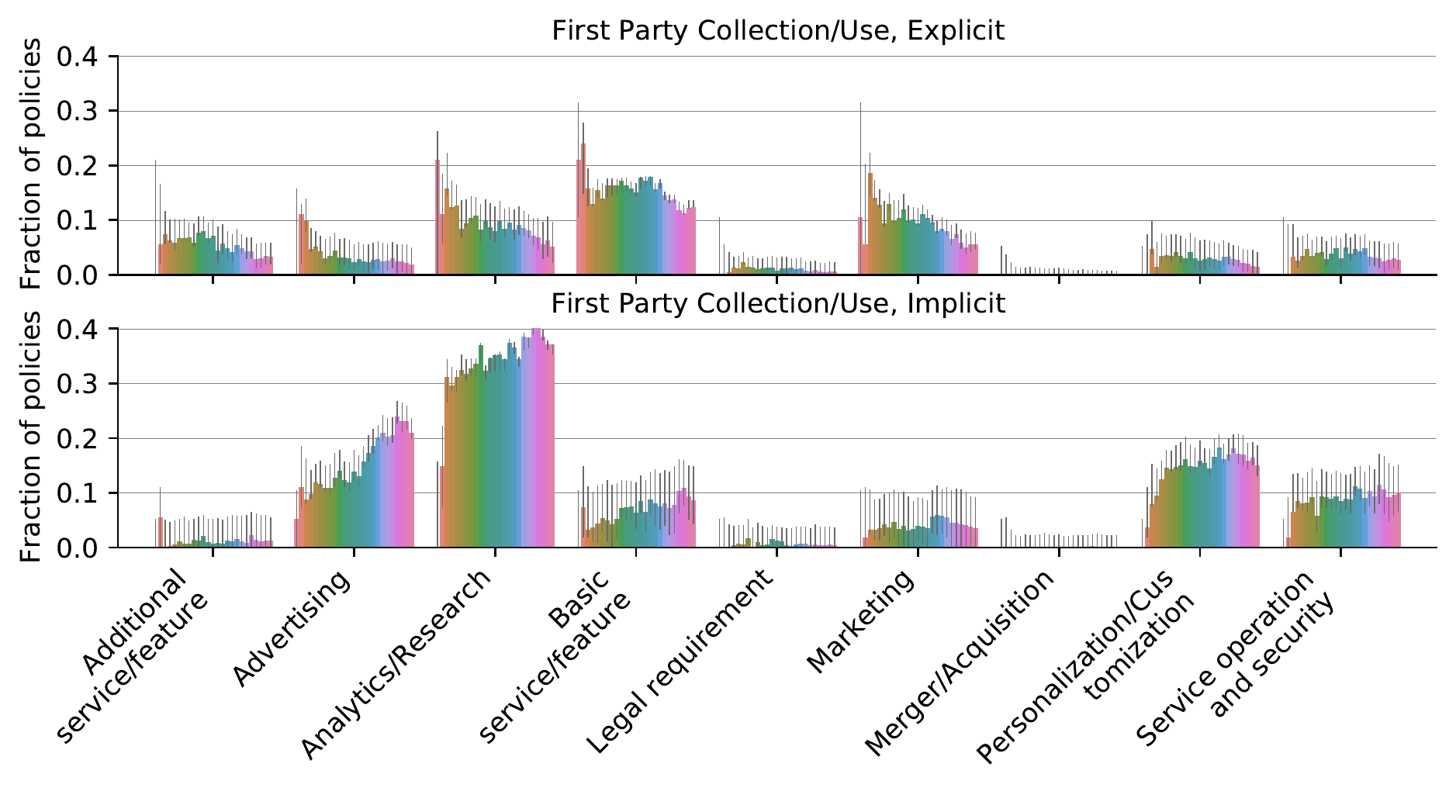}}}\\
\subfloat[Type of choice and controls offered to users]{\label{fig:content-choice-mode} 
	{\includegraphics[width=.48\linewidth]{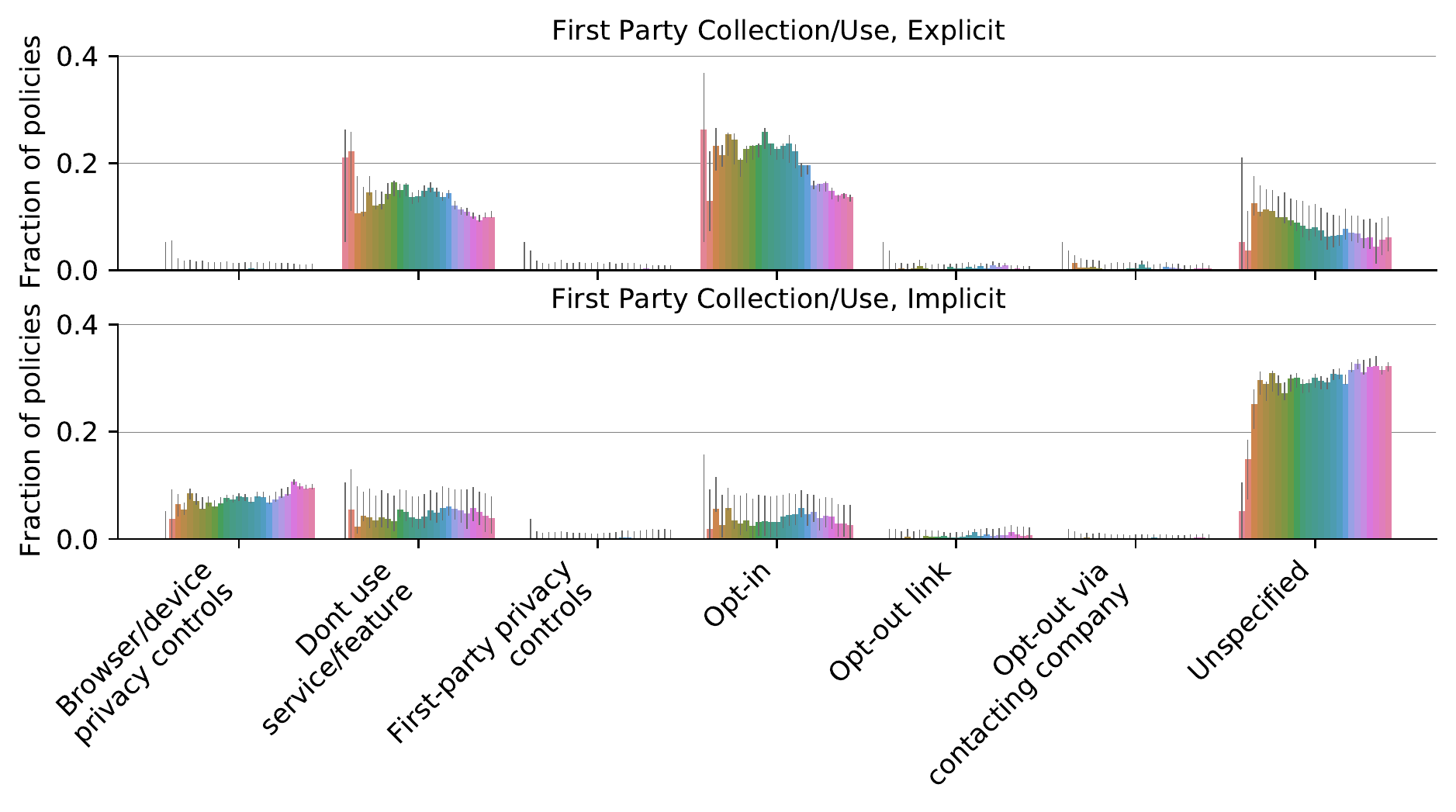}}}
\subfloat[Identifiability of the collected data.]{\label{fig:content-mode-identifiability} 
	{\includegraphics[width=.48\linewidth]{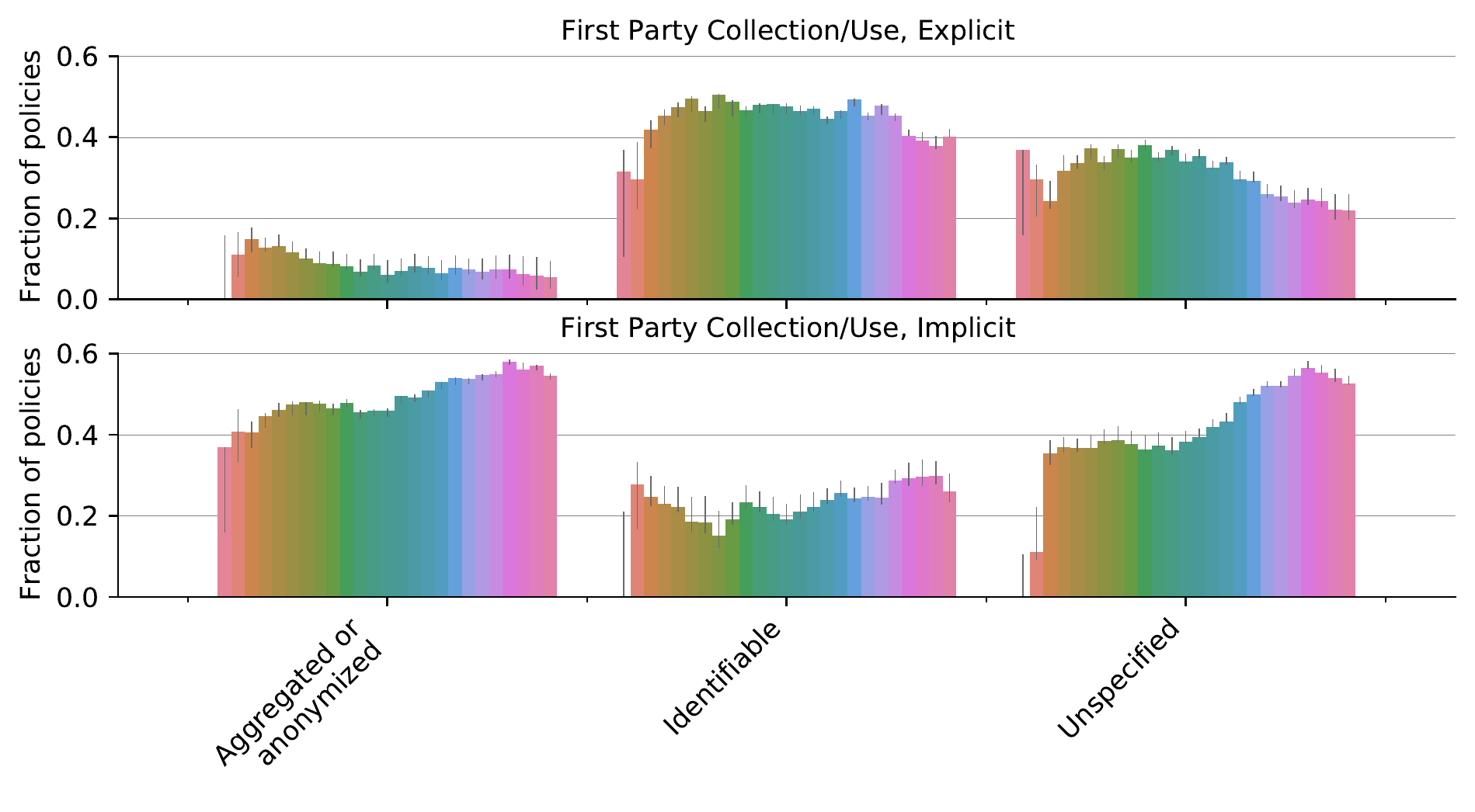}}}
	\caption{Collection mode for first-party data collection.}
	\label{fig:content-mode}
\end{figure*}

The \textit{collection mode} describes whether information is explicitly provided by the user (e.g., data entered in a form), or collected implicitly, e.g., in the background, possibly without the user's knowledge.
Figure \ref{fig:content-pitype-mode} shows personal information types by collection mode. \textit{Generic personal information} and \textit{contact} data is most often collected explicitly, whereas \textit{cookies and tracking} data, data about \textit{user online activities}, and device identifiers (\textit{computer information}, \textit{IP address and device IDs}) are mostly collected implicitly. 29\% of 2021 policies assert that they implicitly record \textit{user online activities}.
\textit{Location} data is collected implicitly at almost triple the rate than explicitly (8\% vs. 3\% in 2021), which is concerning due to the steep rise in location data collection and the sensitivity of location data.

Figure \ref{fig:content-purpose-mode} shows the the purpose of data collection by collection mode.
Explicitly collected data is most often used for \textit{basic service features} (e.g., logins), whereas implicitly collected data is used most for \textit{analytics}, \textit{advertising}, and \textit{personalization}.
Over time, the use of explicit collection for advertising, analytics, or marketing shows a steady decrease, with a corresponding increase in implicit collection.

Figure \ref{fig:content-choice-mode} shows the types of choice and control mechanisms offered to users for each collection mode.
For explicitly collected data, the rate at which opt-ins are offered has decreased over the last decade. In 2021, opt-ins are offered at almost the same rate as the ``choice'' to stop using the service or feature.
For implicitly collected data, most policies leave user choices unspecified. Opt-ins are offered less frequently than asking users to rely on their browser's privacy controls or to stop using the service.
\subsubsection{Identifiability vs. collection mode and purpose}

Figure \ref{fig:content-mode-identifiability} shows that implicitly collected data is more often aggregated or anonymized than explicitly collected data.
This is positive because it shows an effort to protect data that may have been collected without the user's knowledge. However, because anonymizing data is notoriously difficult, most websites are likely to use simple aggregation.
In addition, many policy segments leave the identifiability of collected data unspecified, with a high rate especially for implicit collection.
This indicates that policies are often vague.

\begin{figure}
	\centering
	\includegraphics[width=.9\linewidth]{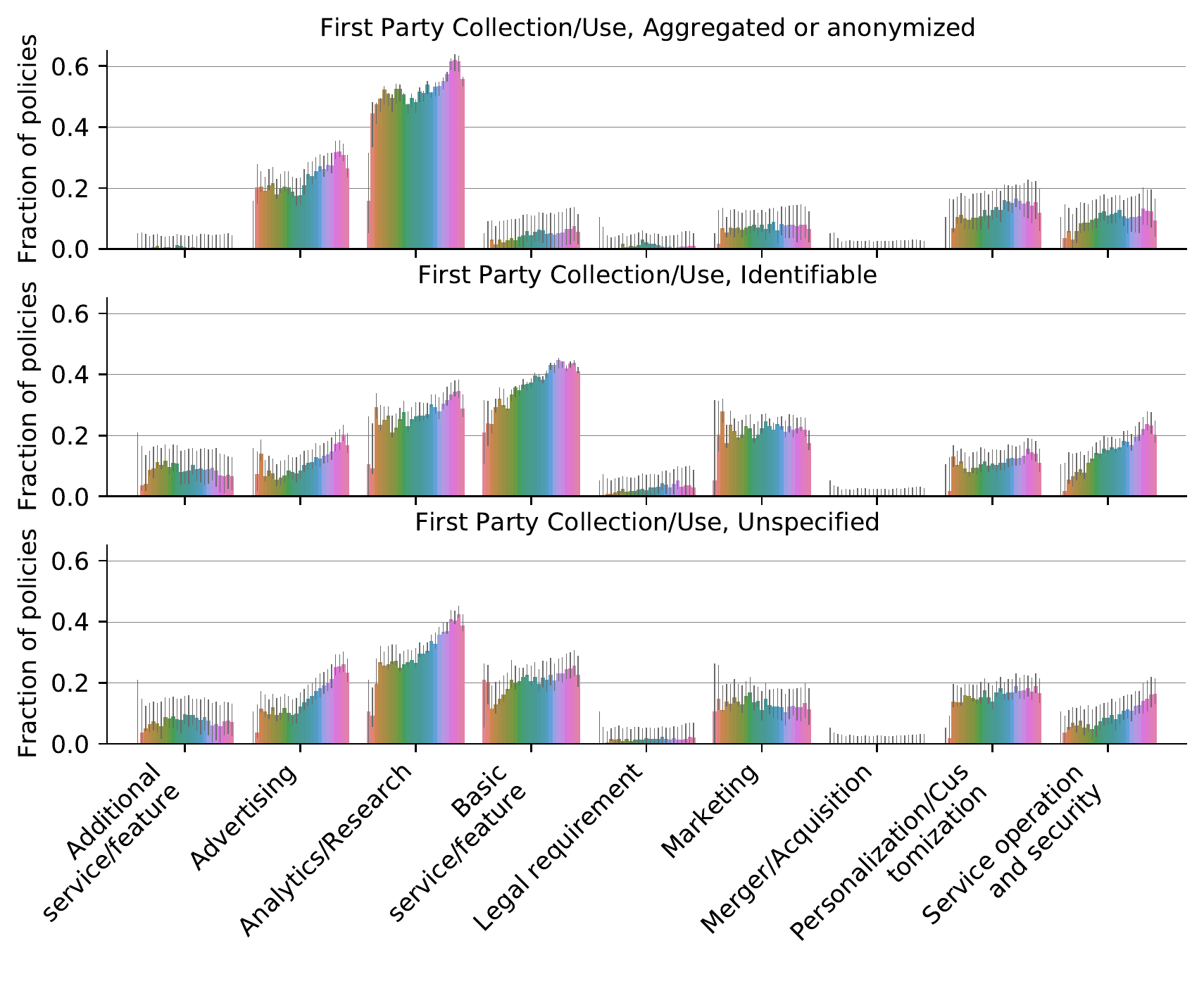}
	\caption{Purpose of data collection, by identifiability of the collected data.}
	\label{fig:content-purpose-identifiability}
\end{figure}

Figure \ref{fig:content-purpose-identifiability} shows the purpose of data collection, by identifiability of the data.
Aggregated or anonymized data is most commonly used for \textit{analytics}, followed by \textit{advertising}.
Identifiable data is used to provide \textit{basic service features}, but also for analytics, advertising, marketing, and service operation. 
However, it is concerning that the number of data collection purposes where identifiability is left unspecified is similar to the other two groups.
Over the years, an increasing number of policies asserts data collection for \textit{analytics}, \textit{advertising}, \textit{service provision}, and \textit{service operation}. The first two purposes reflect web business models driven by advertising revenue, but the last two may indicate an increasing use of \textit{legitimate interest} as a lawful basis for data processing instead of user consent, as for example allowed by the GDPR \cite{kretschmer2021cookie}.

\subsubsection{Personal information type vs. purpose}
\begin{figure*}
	\centering
	\includegraphics[width=\linewidth]{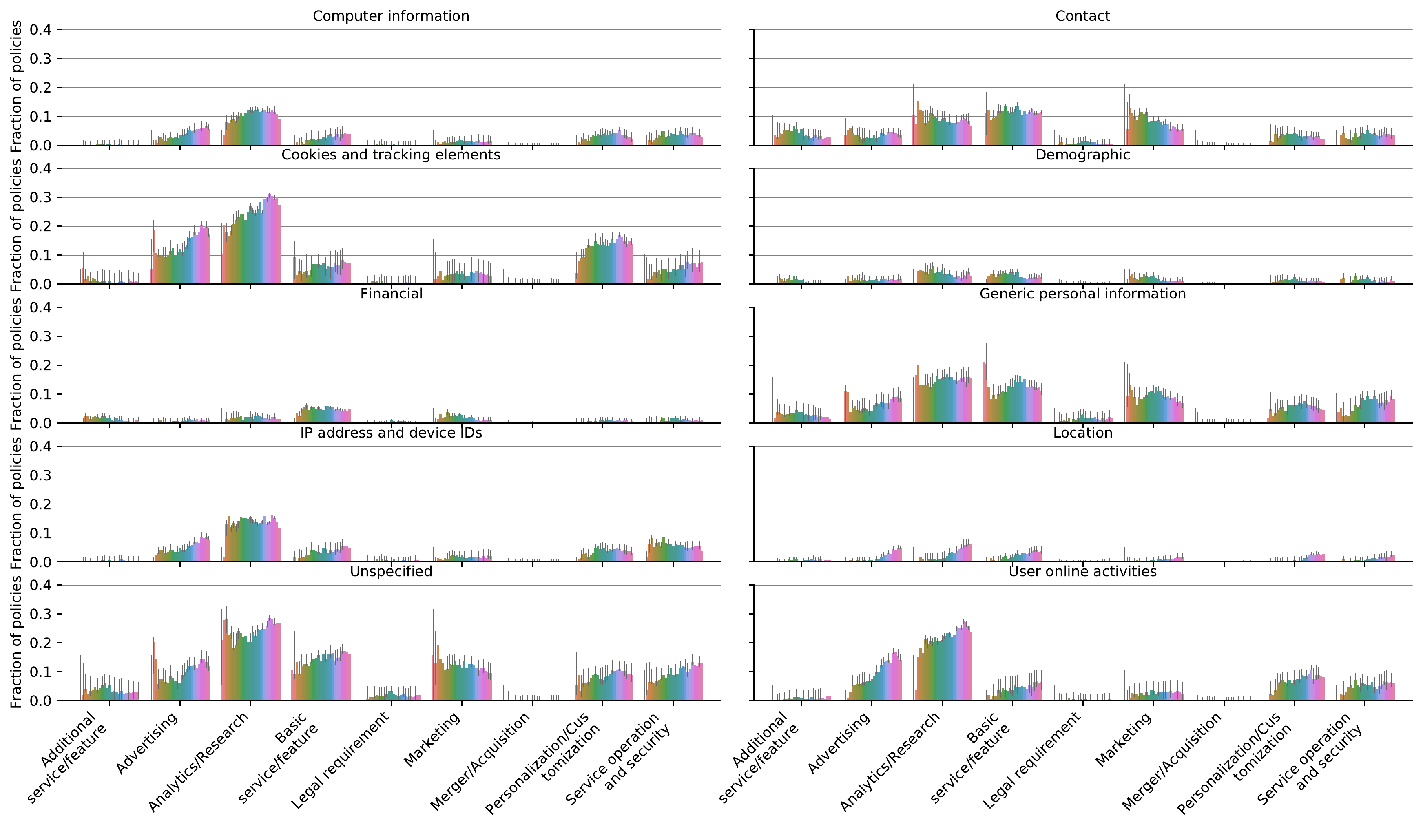}
	\caption{Purpose of data collection, by personal information types collected by first parties.}
	\label{fig:content-purpose-pitype}
\end{figure*}

\begin{figure*}[!ht]
	\centering
	\includegraphics[width=\linewidth]{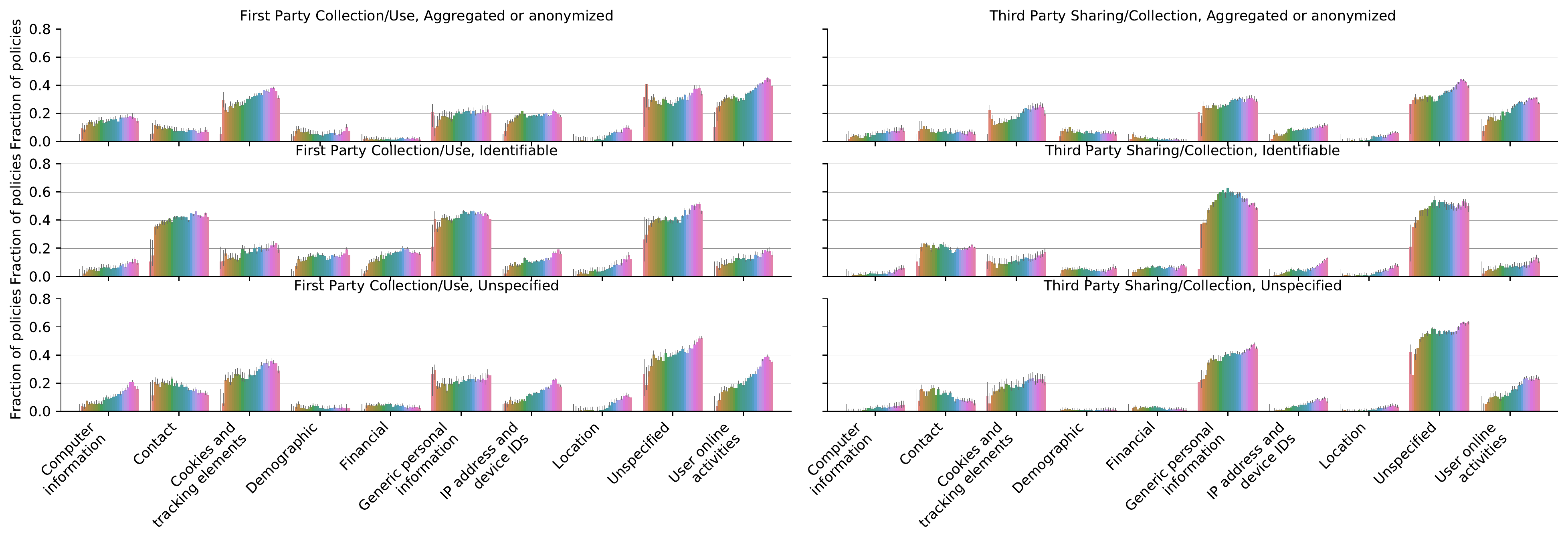}
	\caption{Personal information types collected by first- vs. third-parties, by identifiability of the collected data.}
	\label{fig:content-pitype-identifiability}
\end{figure*}

Figure \ref{fig:content-purpose-pitype} shows the personal information types collected by first parties, depending on the purpose of the data collection (personal information types with very low occurrence rates have been omitted).
Analytics and advertising are the most common purposes for several personal information types: computer information, cookies, generic personal information, IP addresses, and user online activities.
\textit{Demographic} data, \textit{financial} data, and \textit{location} data are used for almost all purposes at relatively low rates.
However, while the collection rates of demographic and financial data have remained stable over the years, location data is collected at an increasing rate. This indicates that pervasive collection of user locations and the associated profiling may become a major concern in the near future.

\subsection{Third-party data sharing/collection}

\subsubsection{Identifiability vs. personal information type}

Figure \ref{fig:content-pitype-identifiability} shows the personal information types collected by first- and third-parties, by  identifiability of collected data.
In most cases, policies assert collection of more personal information types for first-party collection than for third-party sharing, with the exception of \textit{generic personal information}, which is shared with third parties more often than it is used by first parties, regardless of identifiability.
Even though the percentage of policies sharing identifiable PI with third parties has decreased since 2009 (from 61\% to 48\%), this is partially compensated by an increase in policies that do not specify identifiability (from 42\% to 45\%).
Instead of hinting at a reduction in data sharing, this finding therefore suggests that policies are becoming more vague.

\subsubsection{Third-party entity}
\begin{figure}
	\centering
	\includegraphics[width=.9\linewidth]{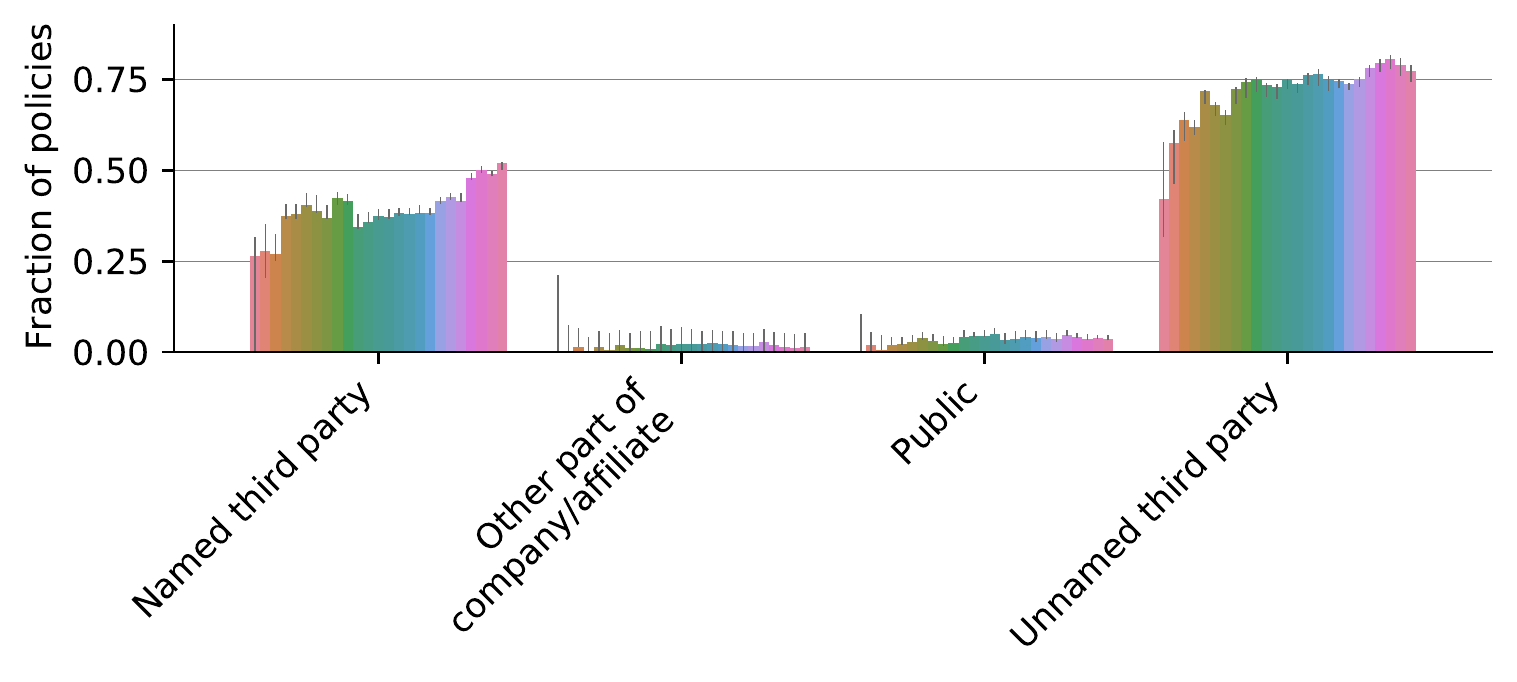}
	\caption{Third-party entities described in privacy policies.}
	\label{fig:content-third-party-entity}
\end{figure}

Figure \ref{fig:content-third-party-entity} shows which types of third-party entities are mentioned in privacy policies.
Over the last decade, there was an increase in \textit{named} third parties, which is slightly positive. However, \textit{named} third parties include categorized third parties that are not identified by name, such as \textit{advertisers}.
In addition, the majority of third parties is unnamed.

\subsubsection{Purpose}
\begin{figure*}
	\centering
	\includegraphics[width=\linewidth]{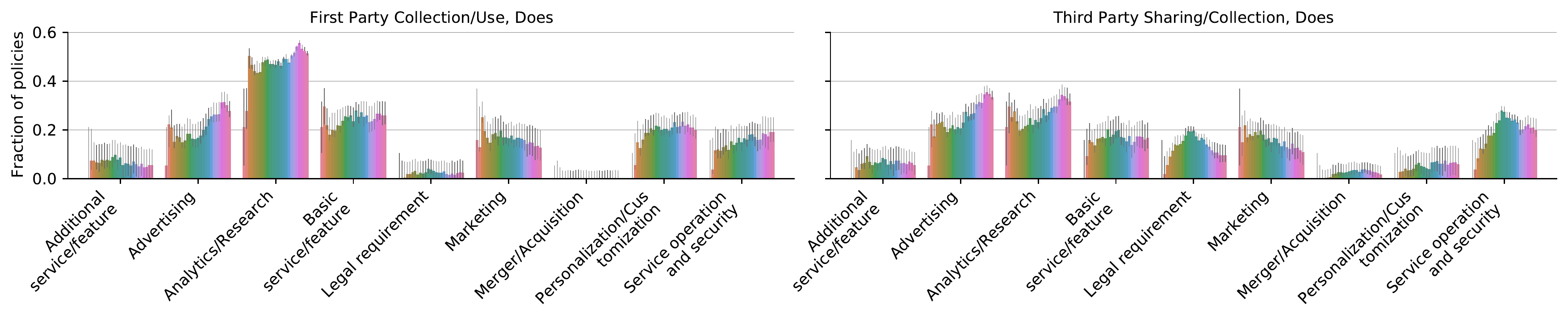}
	\caption{Purposes for data collection for first- vs. third-parties.}
	\label{fig:content-purpose}
\end{figure*}

Figure \ref{fig:content-purpose} compares the purposes of data collection for first- and third-parties.
Overall, slightly fewer purposes are given for third-party sharing, however, some purposes are more prevalent for third-party than first-party collection, including \textit{advertising} and \textit{legal requirements}.
For both first and third parties, data collection for \textit{advertising} and \textit{analytics} has decreased slightly post-GDPR, but in both cases collection rates are still much higher than ten years ago.

\subsection{Policy change}

\begin{figure}
	\centering
	\includegraphics[width=\linewidth]{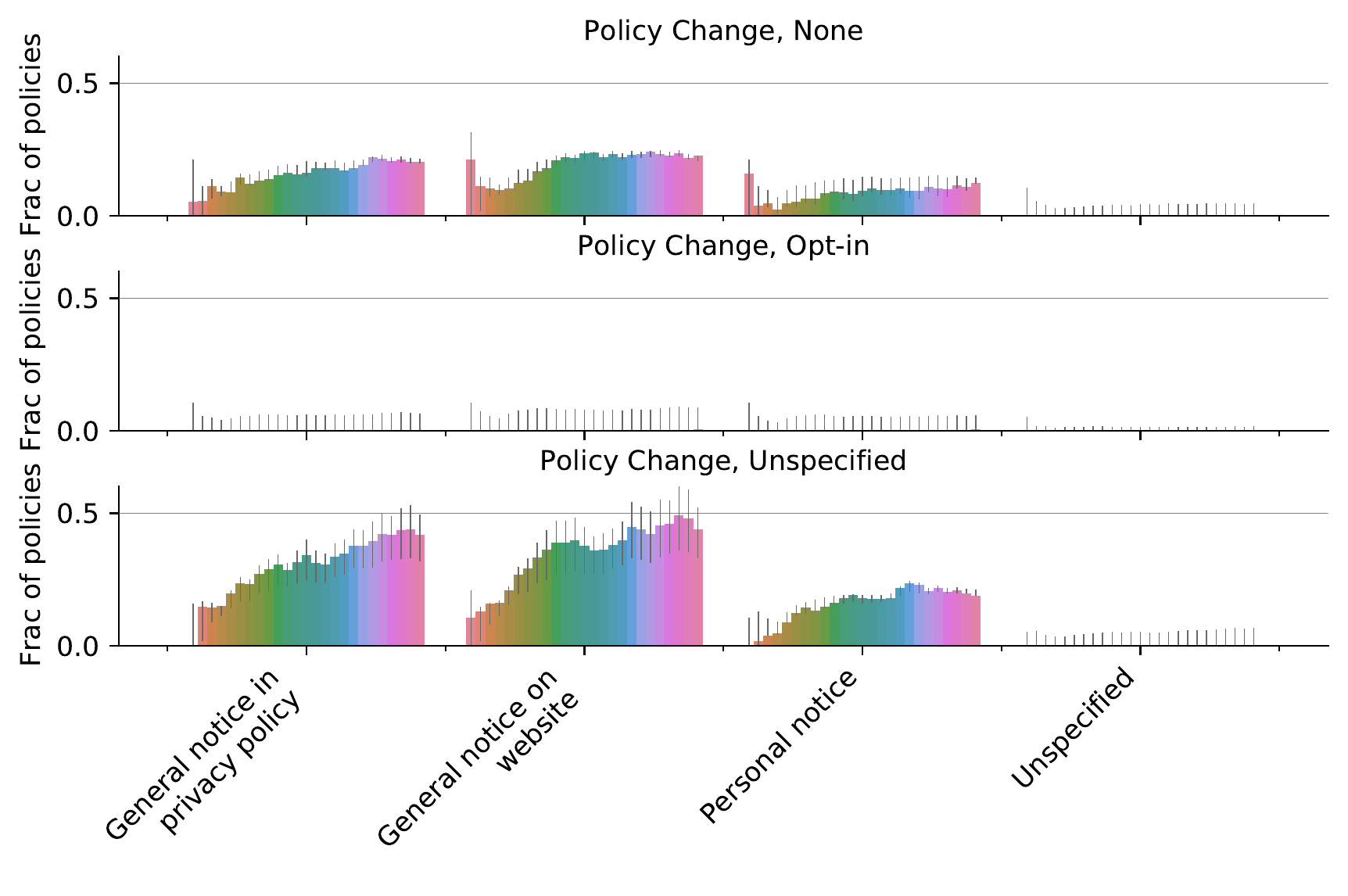}
	\caption{How users are notified of privacy policy changes, by choices they are given.}
	\label{fig:content-change-choice-notification}
\end{figure}

Figure \ref{fig:content-change-choice-notification} shows how users are notified of changes to privacy policies and what choices they are given when this happens.
In 2021, 73\% of policies include a statement about policy change. Of these, 34\% state that changes will be announced by a notice in the privacy policy, 37\% will post a notice on the website, and 22\% will send a personal notice (the remaining policies leave the notification type unspecified).
As a result, most users are unlikely to become aware of changes in privacy policies.
In addition, users are offered almost no meaningful choice when policies change. Of the policies that notify the user of changes, only 12\% offer a new opt-in, whereas 34\% give no choice and 54\% leave it unspecified.

\subsection{User choice/control}

\begin{figure}
	\centering
	\includegraphics[width=\linewidth]{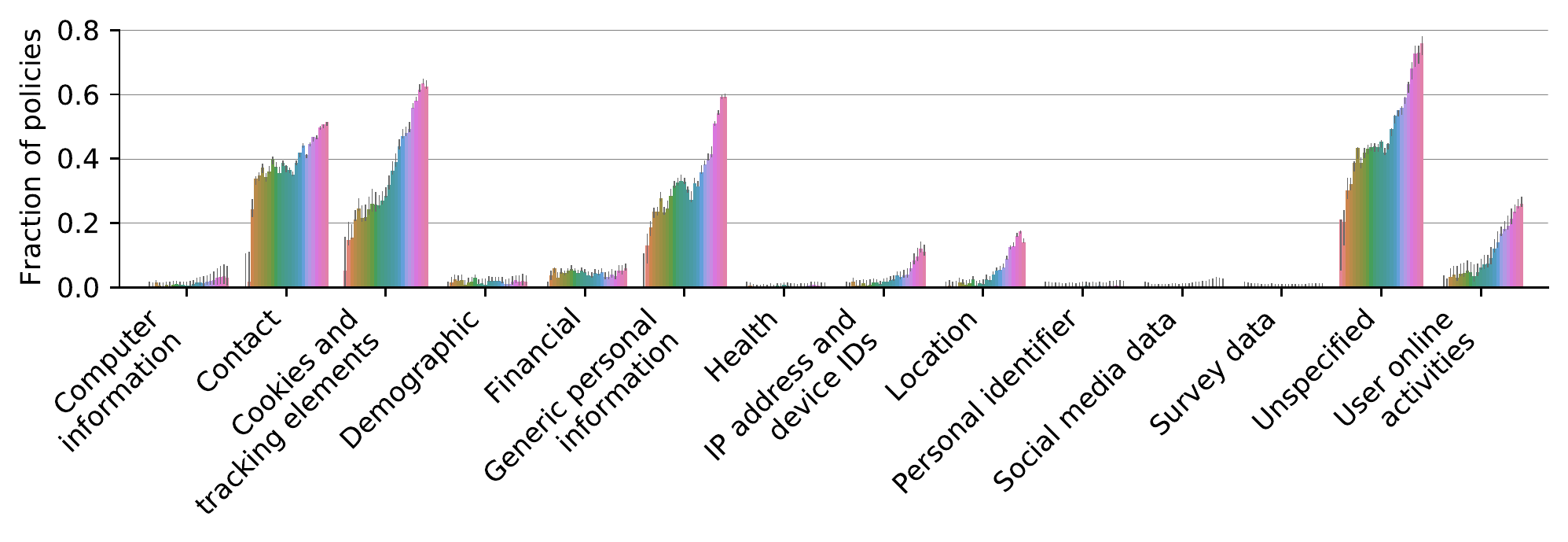}
	\caption{Personal information types for which users are offered choice or control mechanisms.}
	\label{fig:content-user-choicecontrol-personal-information-type}
\end{figure}

Figure \ref{fig:content-user-choicecontrol-personal-information-type} shows the personal information types for which users are offered choice/control mechanisms in general statements.
General choice/control mechanisms are offered most frequently for cookies and tracking elements, but generally at a lower rate than for specific first- or third-party data collection.
In addition, the choices regarding cookies are insufficient to protect users from all tracking because choice or control mechanisms are rarely offered for \textit{computer information}, \textit{device identifiers}, and \textit{personal identifiers}, which allow tracking users via fingerprinting \cite{pugliese2020longterm}.

\subsection{User access, edit, deletion}

\begin{figure}
	\begin{minipage}[b]{0.47\textwidth}
		\centering
		\includegraphics[width=\linewidth]{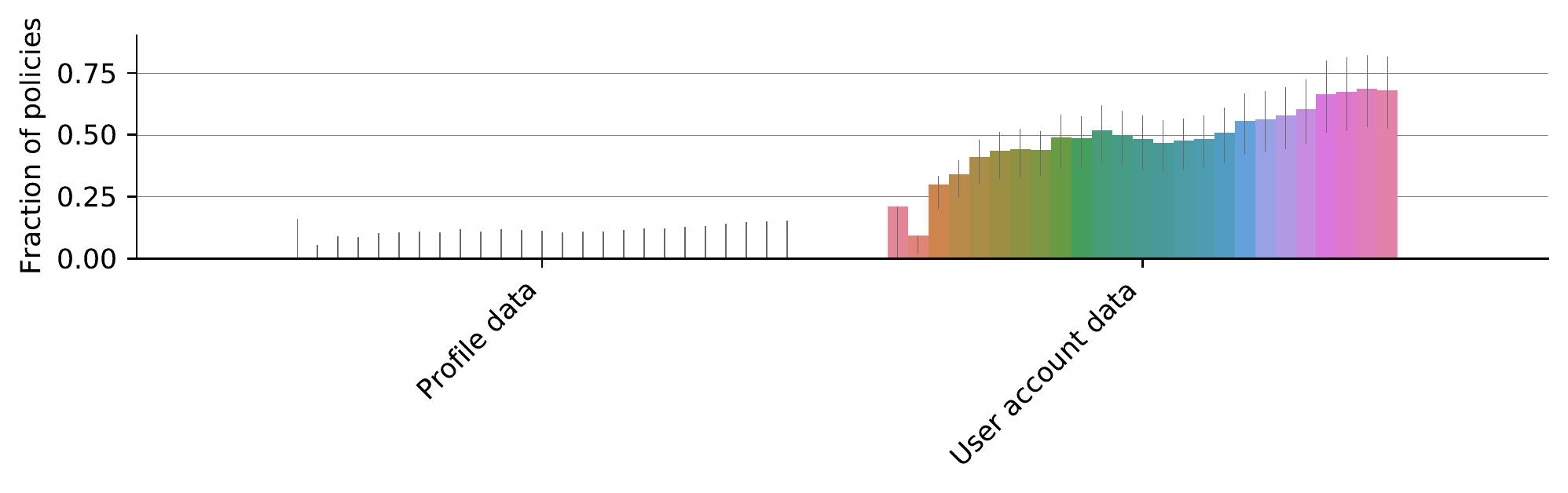}
		\caption{Scope of data for which users are offered access, edit, and deletion rights.}
		\label{fig:content-user-access-edit-and-deletion-access-scope}
	\end{minipage}\hfill
	\begin{minipage}[b]{0.47\textwidth}
		\centering
		\includegraphics[width=\linewidth]{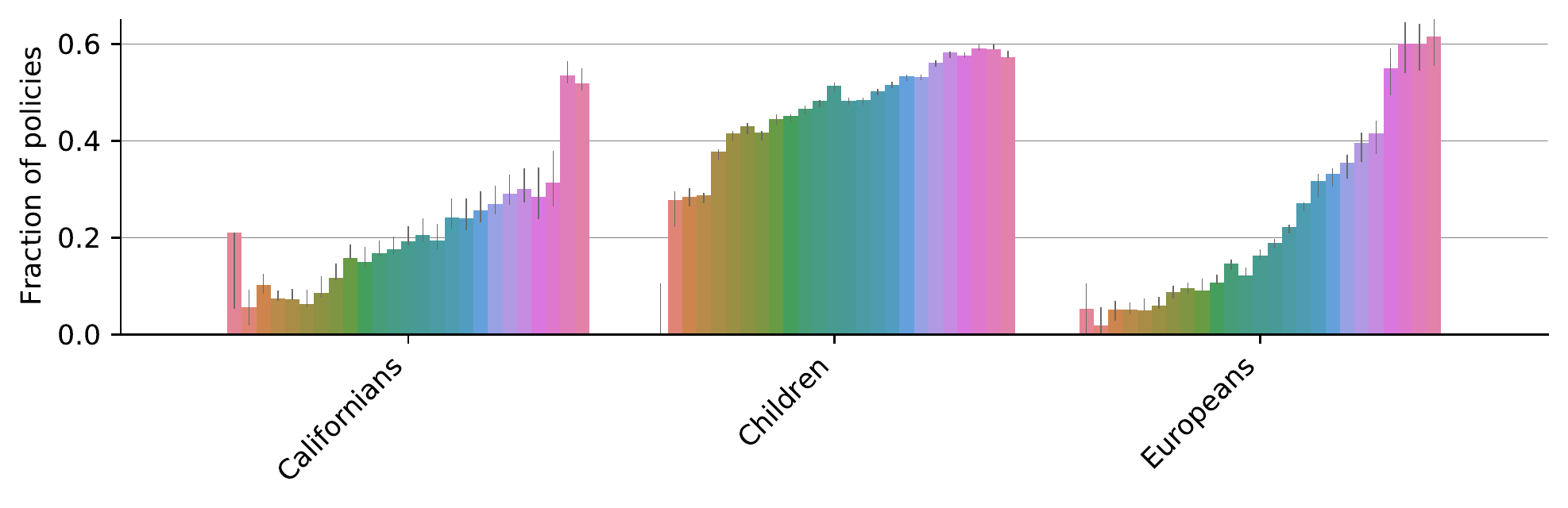}
		\caption{Audience types described in privacy policies.}
		\label{fig:content-international-and-specific-audiences-audience-type}
	\end{minipage}
\end{figure}

Figure \ref{fig:content-user-access-edit-and-deletion-access-scope} shows the scope of data for which privacy policies offer access, edit, and deletion rights to users.
User access is mostly offered for account data, i.e., data explicitly specified by users, but very rarely for profile data which is collected implicitly.

\subsection{International/specific audiences}

Figure \ref{fig:content-international-and-specific-audiences-audience-type} shows the audience types that are singled-out in privacy policies.
Children are most frequently mentioned, due to longstanding legislation in many countries that require differential treatment of minors.
There was a 20\% increase in mention of Europeans (from 37\% to 56\%) after the introduction of the GDPR, and a similar 20\% increase in mention of Californians (from 34\% to 54\%) after the introduction of the CCPA. This indicates an increasing tailoring of privacy policies to specific audiences, and it may mean other audiences do not benefit from the increased protections afforded to Europeans and Californians with the respective regulations.

\subsection{Data security}

\begin{figure}
	\centering
	\includegraphics[width=\linewidth]{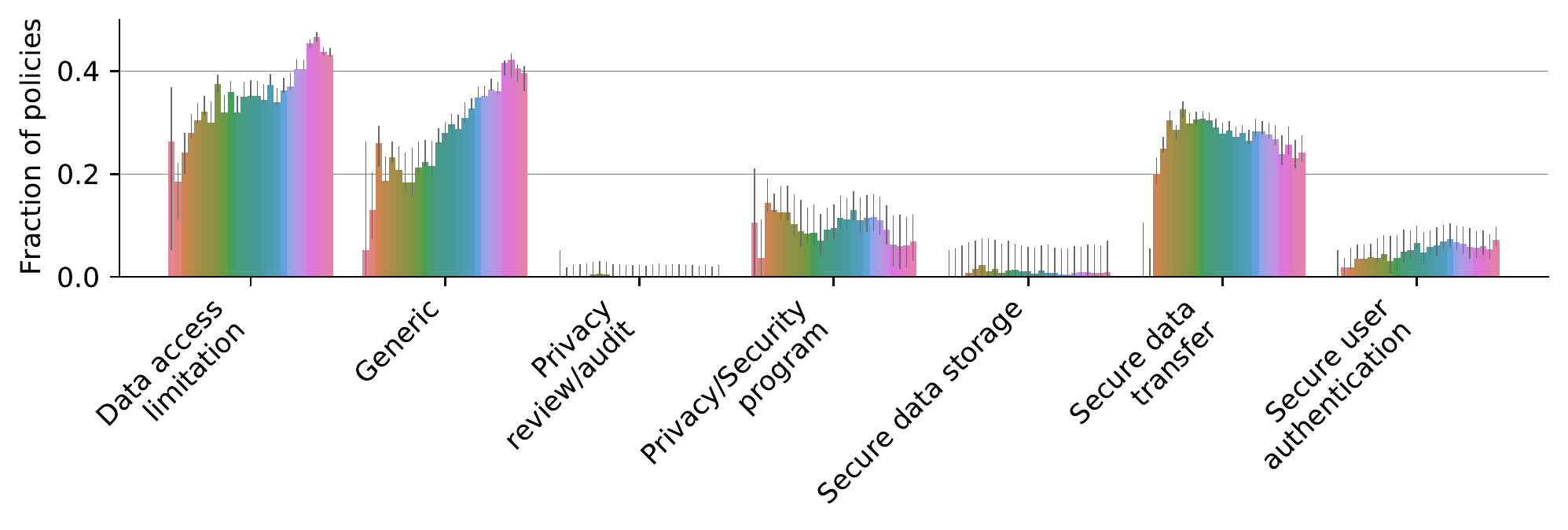}
	\caption{Security measures described in privacy policies.}
	\label{fig:content-data-security-security-measure}
\end{figure}

Figure \ref{fig:content-data-security-security-measure} shows the security measures mentioned in privacy policies.
Most policies mention data access limitations and generic security measures without giving specific details.
Post-GDPR, mentions of privacy/security programs and secure data transfer have decreased, whereas mentions of generic security measures have increased. This again indicates that policies are becoming more vague.

\section{Limitations}
\label{sec:limitations}
We have presented a large-scale, longitudinal study of the readability and contents of privacy policies.
A major but common limitation of our study is our focus on English-language policies.
To lift this limitation in a future study, we would need valid readability formulas as well as a labeled corpus of privacy policies for each target language.
The Flesch Reading Ease formula has already been adapted to other Indo-European and Slavic languages, but this may be a challenge for, e.g., Asian languages.

\subsection{Retrieval of privacy policies}
We focus on a longitudinal evaluation of privacy policies and rely on the Internet Archive's Wayback Machine because it is commonly seen as the most complete and reliable source of archived Internet sources \cite{lerner2016internet,brunelle2015not}.
However, if a site is not archived by the Wayback Machine, we do not have its historical privacy policies available for analysis.
Sites can be excluded from the Wayback Machine for different reasons, including a restriction in the robots.txt file. 

Our crawler is sensitive to variations in how websites link to their privacy policies.
Even though we attempt to find privacy policy content under various names, the process can fail if websites are creative in how they name their links and link titles.
For example, if the link title is \textit{here} (as in ``read our privacy policy \textit{here}''), and the URL includes the word ``policy'' but not ``privacy'', the crawler fails to find the policy.
This limitation could be lifted by adding privacy policy links manually, as was done in \cite{degeling2019we}.
In addition, the crawler fails for some cases of regional differentiation, where the website asks the user to select a language before showing the privacy policy.
If the names and titles of policy links correspond to the chosen language (``English'') instead of indicating presence of a privacy policy, the crawler fails to find the policy.
Finally, some websites choose to make their privacy policies available as a PDF download only.
Our crawler detects these links and downloads the policy, but is not able to analyze the binary file.

\subsection{Content evaluation}
Our use of machine learning to segment and label the contents of privacy policies introduces limitations related to training data, segmentation, and classifier accuracy.

The OPP-115 corpus was published in 2016 and is based on privacy policies from that time. It is not clear whether classifiers trained based on this data are applicable to a longitudinal policy corpus from 1996 to 2021. For example, new regulations may have introduced new terms or new ways of phrasing data practices which are not present in the training data.
In addition, the corpus only consists of 115 policies and as a result some data practices, e.g., in the \textit{data retention} or \textit{do not track} categories, occur infrequently. This makes it difficult to train accurate classifiers.

The policy segments created by GraphSeg are sometimes longer than segments we would have created manually. This may negatively influence labeling accuracy.

\section{Conclusion}
\label{sec:conclusion}

In this paper, we have presented a longitudinal corpus of over 50,000 privacy policies from 1996 to 2021 and analyzed the policies in terms of their length, readability, and content.
We found evidence for increasing policy bloat and decreasing readability, in particular after the introduction of recent privacy regulations.
In addition, we found concerning developments in the data practices described in policies, such as increased collection and sharing of sensitive data and lack of choice.

It is especially concerning that these data practices are obscured in lengthy policies that require university education to understand, and that would take users more than one hour per day to read.
Websites have shown that they can adopt standards for machine-readable formats quickly. For example, the ads.txt standard has reached 60\% adoption rate within two years \cite{bashir2019longitudinal}.
It is therefore not unreasonable to expect that privacy policies could be treated similarly.
However, as the lack of adoption of P3P and the lack of respect for the DNT header show, it does not appear to be in the industry's interest to respect user privacy.
The quantitative evidence presented in this paper shows that privacy policies are a mechanism that fails users and serves website owners.

As a result, we believe that three different approaches may together form a way forward:
first, technical measures on the user-side that automatically classify privacy practices, match them against user preferences, and block unwanted data collection--in essence realizing P3P on the client-side;
second, regulatory measures that mandate specific formats and locations for privacy policies and respect for specific privacy standards such as DNT;
and third, measurement approaches that verify compliance of policies with actual data flows, e.g., building on work on flow-to-policy consistency \cite{andow2020actions}.

\bibliographystyle{IEEEtran}
\bibliography{IEEEabrv,main}

\appendices

\section{Evaluation of readability formula implementations}
\label{sec:readability-implementations}

In our initial data analysis, we observed inconsistencies between readability values computed by different implementations, for example differences in ranking the readability of policy texts. 
We therefore evaluated five open source Python implementations of readability formulas, some of which have been used in prior work \cite{shipp2020how,amos2021privacy}:
textstat (TS)\footnote{\url{https://pypi.org/project/textstat/}}, 
spacy\_readability (SR)\footnote{\url{https://pypi.org/project/spacy-readability/}},
py-readability-metrics (PR)\footnote{\url{https://pypi.org/project/py-readability-metrics/}},
readability (AV)\footnote{\url{https://pypi.org/project/readability/}}, and
readability-score (WM)\footnote{\url{https://github.com/wimmuskee/readability-score}}.
Table \ref{tab:input-diffs-readability} summarizes how the five libraries compute the input values to readability formulas: the numbers of words, sentences, syllables, characters, complex words, difficult words, and polysyllabic words.
The last line, denoted SR*, shows changes we made to the SR library.

\newcolumntype{L}[1]{>{\raggedright\arraybackslash\hspace{0pt}}p{#1}}  %
\begin{table*}
	\centering
	\footnotesize
	\caption{Methods used by five Python readability libraries to compute inputs for readability formulas.}
	\label{tab:input-diffs-readability}
	\begin{tabular}{p{0.6cm}p{0.7cm}L{1.3cm}L{1.4cm}L{1.2cm}L{1.75cm}L{1.4cm}L{1.4cm}L{1.4cm}}
		Library & Version & Words & Sentences & Syllables & Characters & Complex words & Difficult words & Polysyllabic words \\
		\toprule
		TS & 0.6.2 & split by whitespace & regex, ignore short sentences & pyphen & length of text, no punctuation, no spaces & 3+ syllables and not in DC list & 2+ syllables and not in DC list\footnote{DC list refers to the 1948 Dale-Chall word list} & words with 3+ syllables \\
		
		SR & 1.4.1 & spaCy tokenizer & spaCy tokenizer & syllapy & length of each word, no punctuation, no digits & n/a & word and lemmatized word not in DC list & syllables in words with 3+ syllables\\
		
		PR & 1.4.4 & NLTK TweetTokenizer & NLTK sentence tokenizer & count vowels & length of each word, no punctuation & 3+ syllables, not uppercase, no dash & stemmed word not in stemmed DC list & words with 3+ syllables\\
		
		AV & 0.3.1 & split by whitespace & split by newline & count vowels + dictionary & length of each word & 3+ syllables, not uppercase & word not in DC list & words with 3+ syllables, not uppercase \\
		
		WM & 2.1 & split by whitespace & NLTK sentence tokenizer & pyphen & length of each word & n/a & word not in DC list & words with 3+ syllables\\
		
		SR* & -- & spaCy tokenizer, filter spaces & NLTK sentence tokenizer & syllapy & length of each word, no punctuation, no digits, no spaces & n/a & word not in updated DC list & words with 3+ syllables\\
		\bottomrule
	\end{tabular}
\end{table*}

\subsection{Evaluation against ground truth}
We first evaluate these libraries against a corpus of text passages with known ground-truth values.
We construct our corpus based on four sources. 
First, we use short excerpts from privacy policies and manually count their characters, words, syllables, polysyllabic words, and sentences. 
Second, we use published text samples with ground-truth counts for words, sentences, and difficult words as well as Dale-Chall scores, based on the 1995 word list and formula \cite{chall1995readability}.
Third, we use published text samples with ground-truth readability scores for DC, FRE, FKG, GF, and SMOG \cite{chall1996qualitative}.
We augmented these samples with a manual count of words and sentences.
Fourth, we use text samples from \cite{dubay2007smart} and add manual counts of words, sentences, syllables, polysyllabic words, and characters.
All manual counts for words and characters were verified against counts from LibreOffice.
In total, our corpus consists of 70 text passages.

We omit the Gunning-Fog formula from this evaluation because it was only implemented in three of the libraries, and because we could not obtain samples with ground truth values from the literature.
\begin{figure*}
	\centering
	\subfloat[Words]{\label{fig:deviation-count-words} {\includegraphics[width=.3\linewidth]{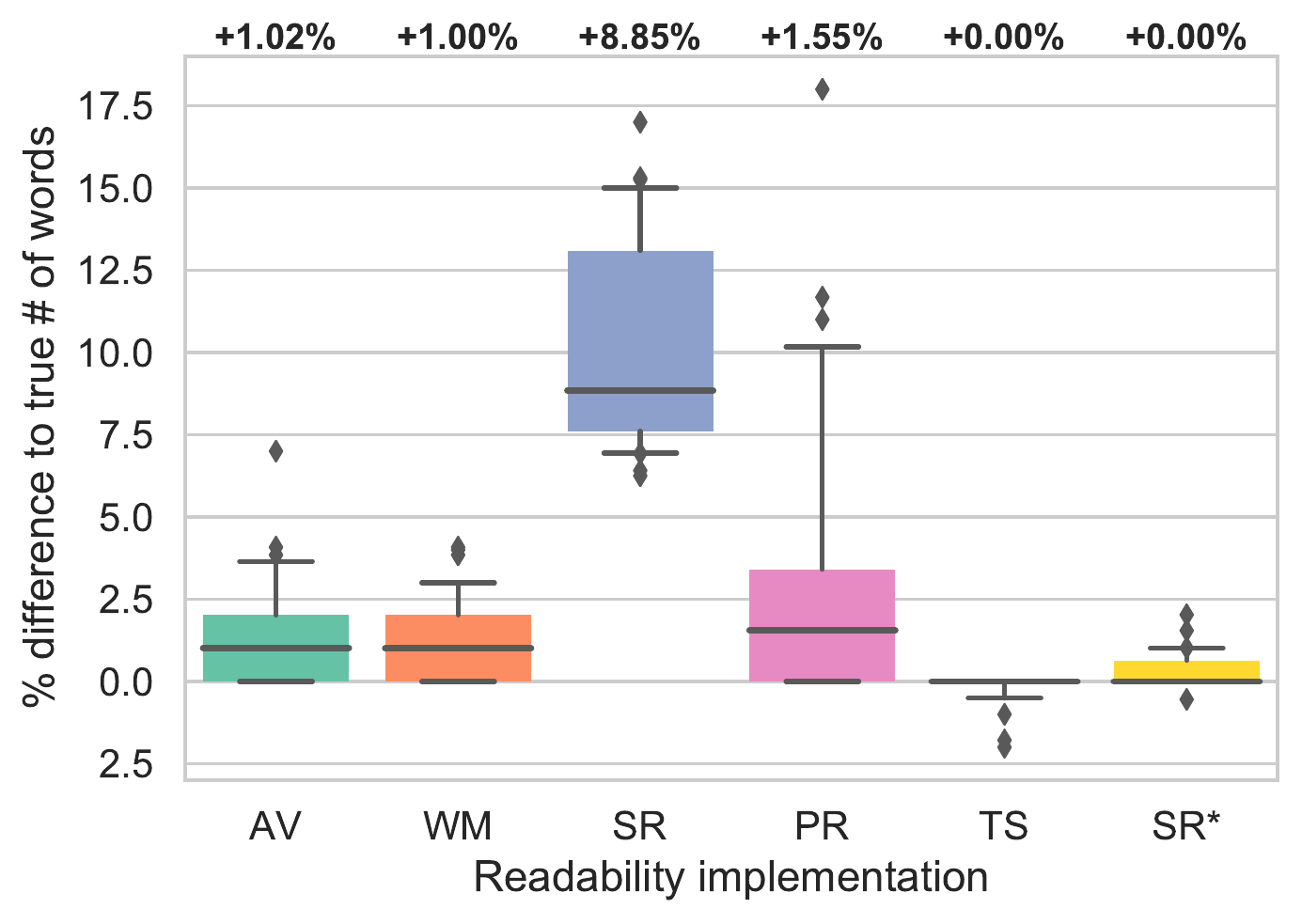} }}
	\hspace{.05cm}
	\subfloat[Sentences]{\label{fig:deviation-count-sentences} {\includegraphics[width=.3\linewidth]{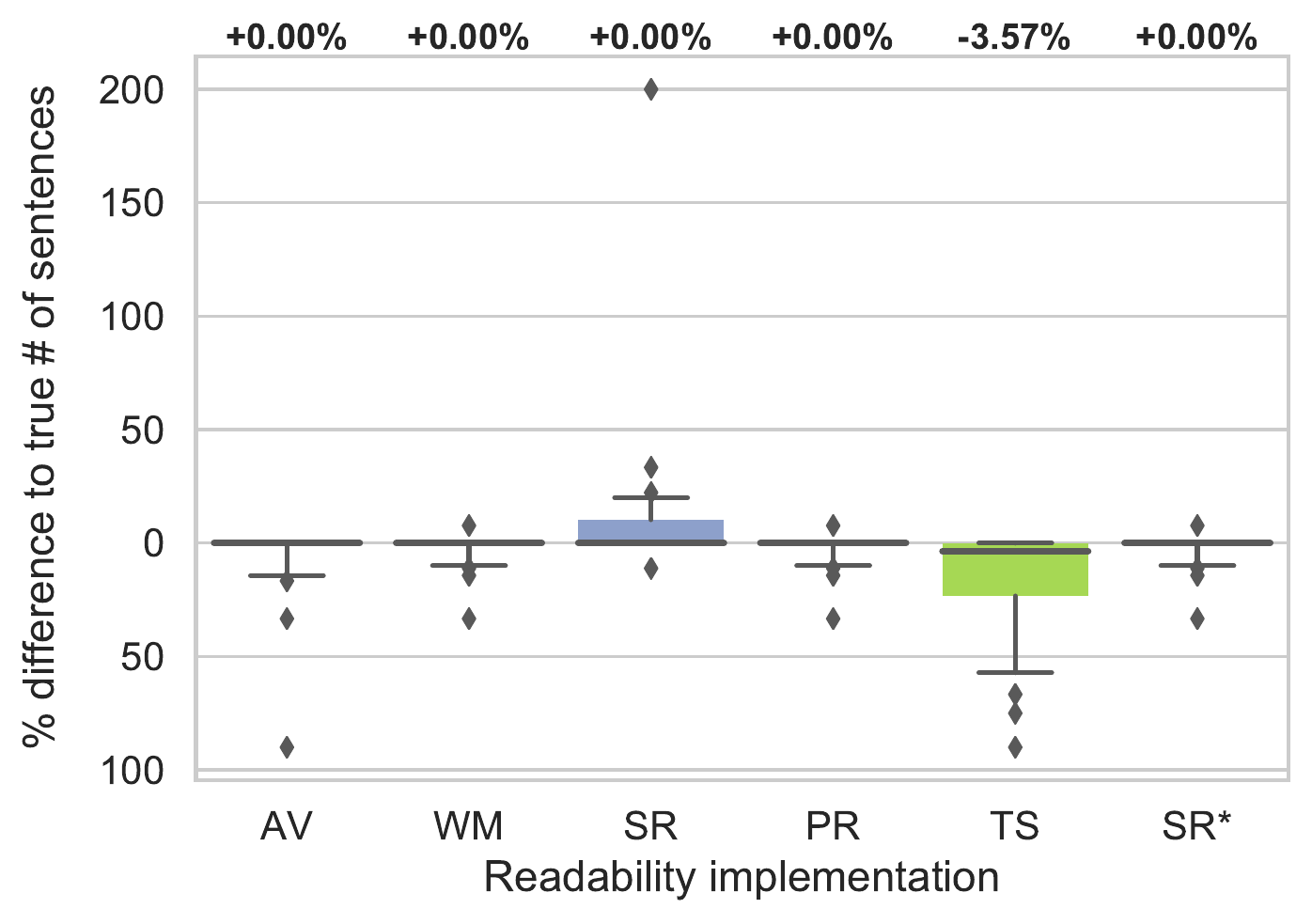} }}
	\hspace{.05cm}
	\subfloat[Syllables]{\label{fig:deviation-count-syllables} {\includegraphics[width=.3\linewidth] {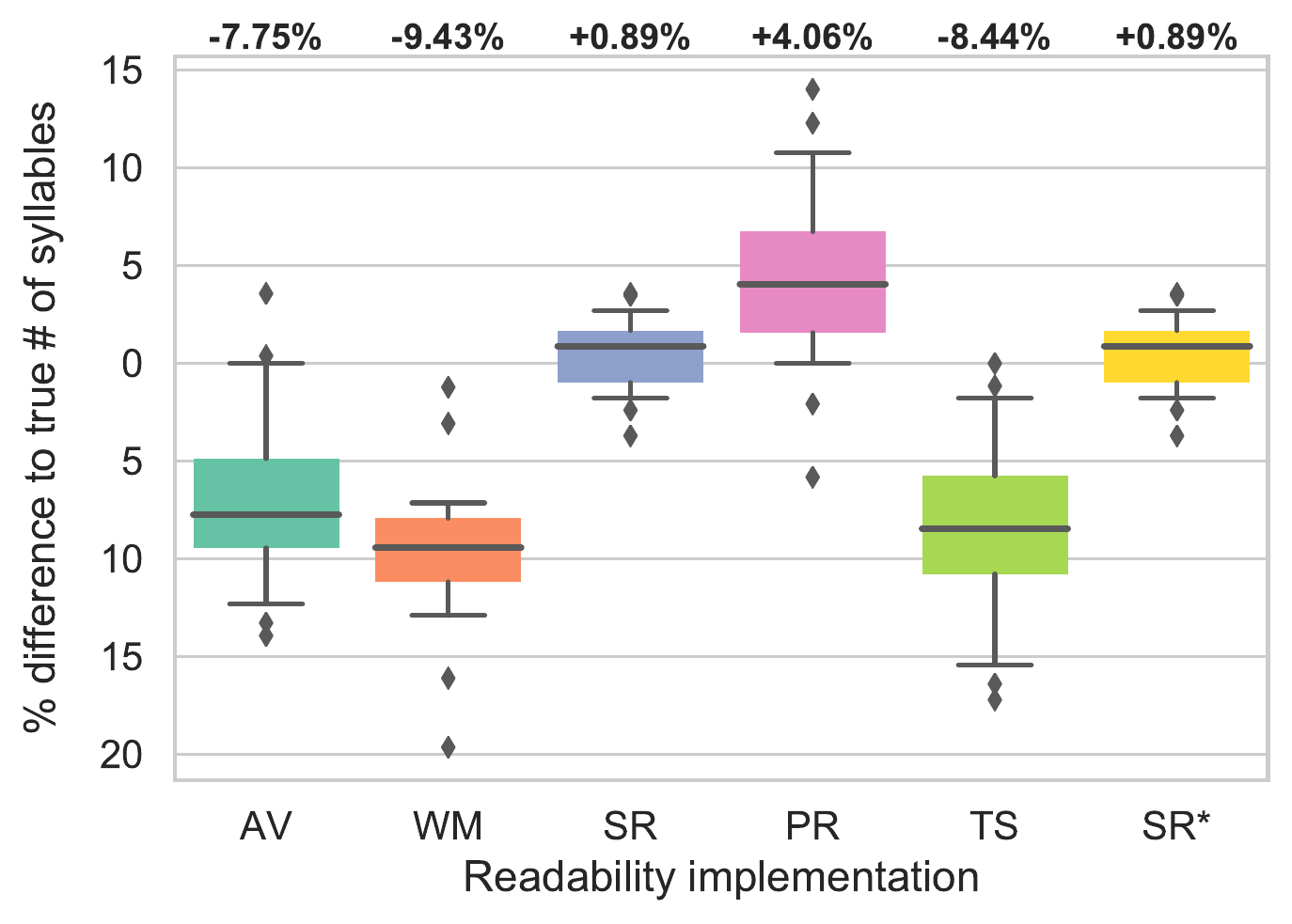} }} \\
	\subfloat[Characters]{\label{fig:deviation-count-characters} {\includegraphics[width=.3\linewidth]{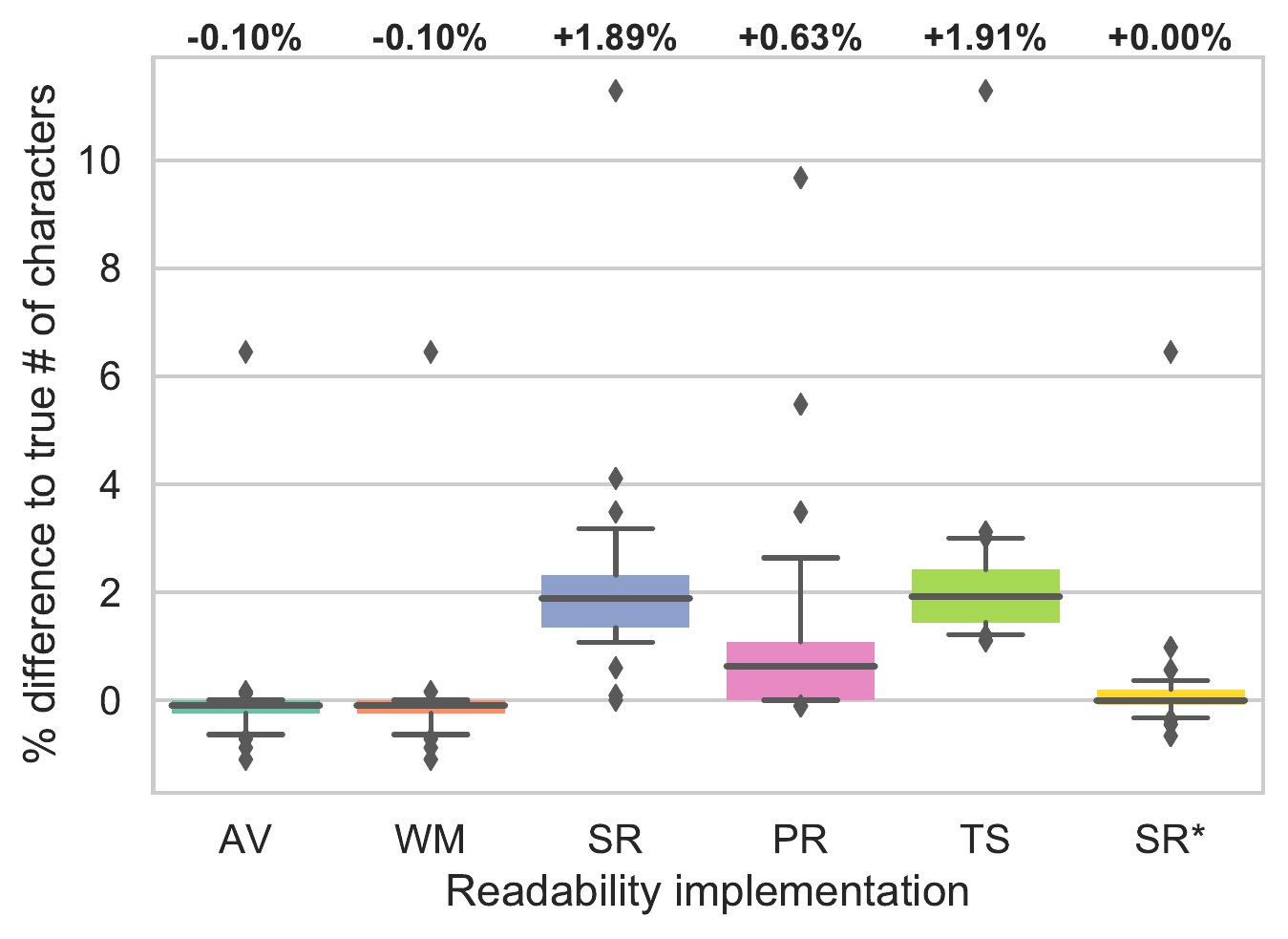} }}
	\hspace{.05cm}
	\subfloat[Difficult words]{\label{fig:deviation-count-diff_words} {\includegraphics[width=.3\linewidth]{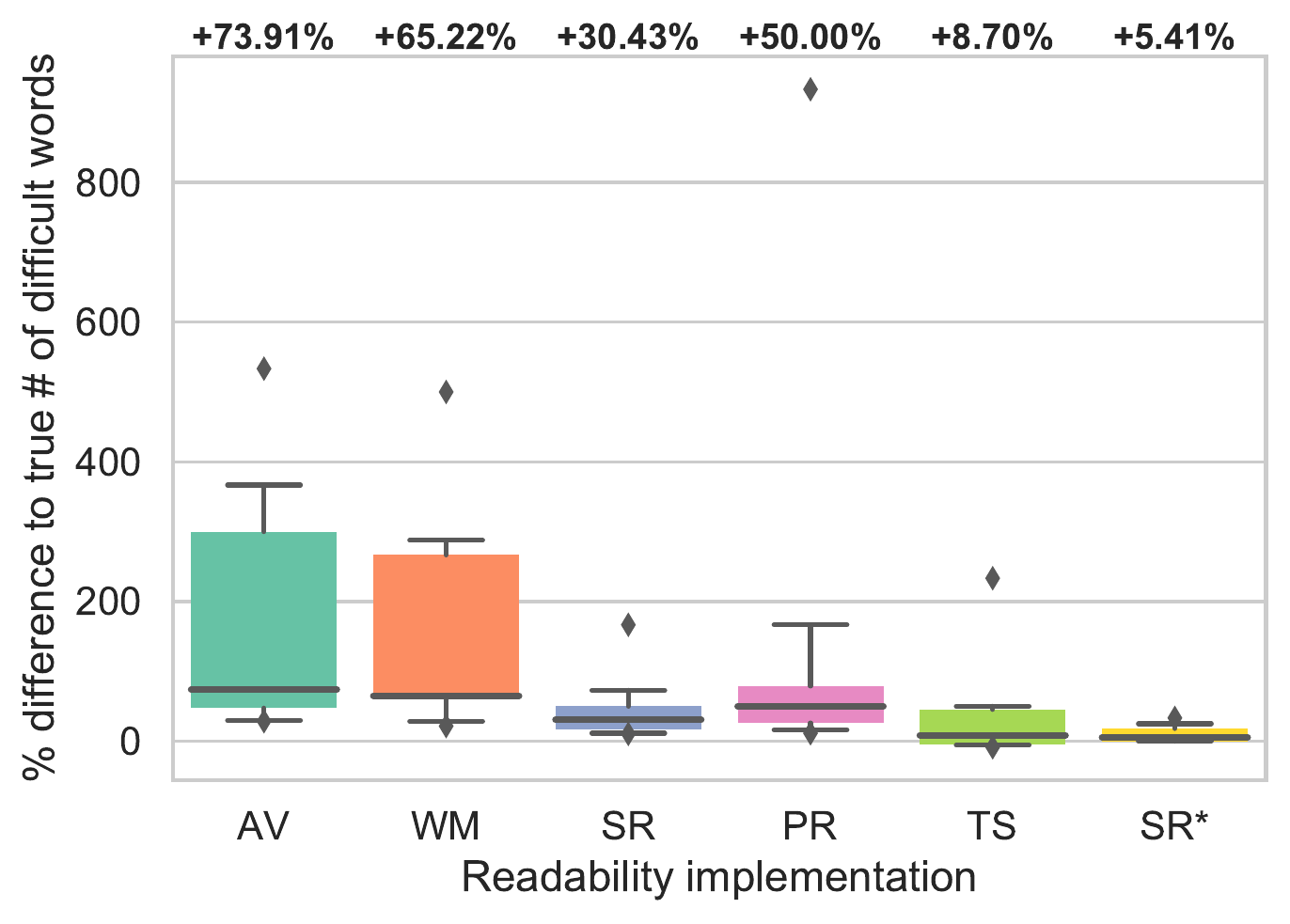} }}
	\hspace{.05cm}
	\subfloat[Polysyllabic words]{\label{fig:deviation-count-polysyllables} {\includegraphics[width=.3\linewidth] {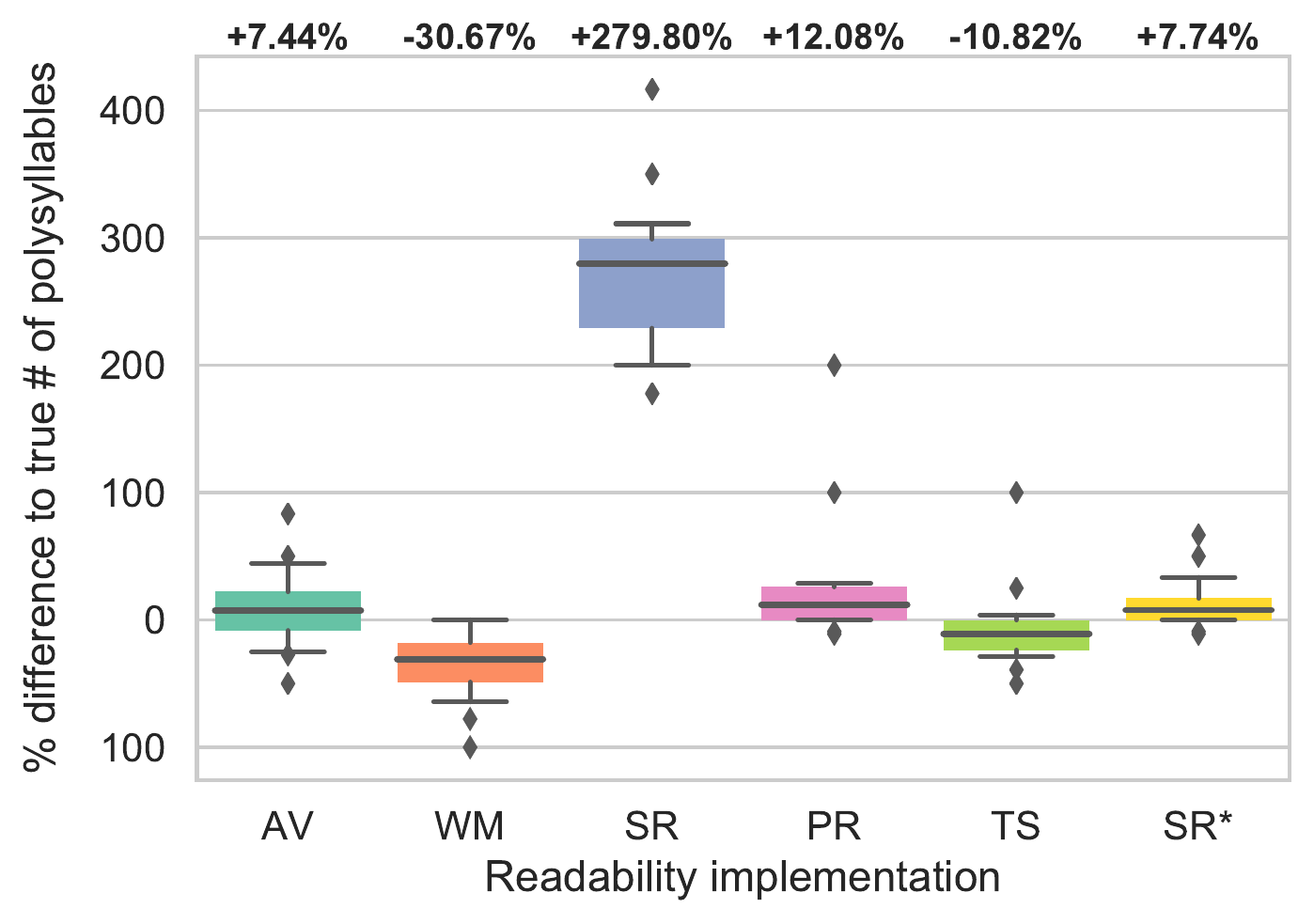} }}
	\caption{Deviation (in \%) from the true count of words, sentences, syllables, characters, difficult words, and polysyllabic words.The boxes range from the 25\% to the 75\% percentile, while whiskers extend to the 10\% and 90\% percentiles.}
	\label{fig:deviation-counts}
\end{figure*}

Figure \ref{fig:deviation-counts} shows by how many percent each readability implementation deviates from the true counts of words, sentences, syllables, characters, difficult words, and polysyllabic words, on all 70 text samples.
It is clear that the differences in counting methods described in Table \ref{tab:input-diffs-readability} lead to differences in how much each count deviates from the true count.
The deviation in word count in SR is caused by SR failing to remove some empty tokens from the count. After fixing this bug in SR*, the median count equals the true count.
The deviation in sentence counts is caused by differences in how partial sentences, such as headings, are counted. The best implementations use the NLTK sentence tokenizer and under-count about 15\% of sentences (those between the 10\% and 25\% percentiles) by less than 10\%, while the spaCy tokenizer over-counts 25\% of sentences by less than 10\%, and 15\% of sentences by around 10\%. For this reason, we chose the NLTK tokenizer in our improved SR* implementation.
For syllable counts, the counts produced by syllapy (SR and SR*) are very close to true counts. The pyphen and vowel counting methods have larger variation in our corpus, and on average produce counts that are further from the true count.
The character counts are quite accurate, with all medians within 2\% of the true count. For SR*, we have improved the count in SR by filtering whitespace characters.

The deviations in the number of difficult words are the largest, on average, of all the counts. There are two main reasons for this. 
First, all five libraries use the outdated 1945 version of the Dale-Chall word list, while the ground truth in our corpus is based on the 1995 version. 
Second, none of the five libraries take into account the detailed rules for classifying words as \textit{difficult} that accompany the word list, whereas we have implemented most of these additional rules in SR*. 

Finally, all implementations slightly under-count the number of polysyllabic words, with the exception of SR, which massively over-counts polysyllabic words because it counts the total number of syllables in polysyllabic words instead of the number of words.

\begin{figure*}
	\centering
	\subfloat[Flesch Reading Ease]{\label{fig:deviation-measure-fre} {\includegraphics[width=.3\linewidth]{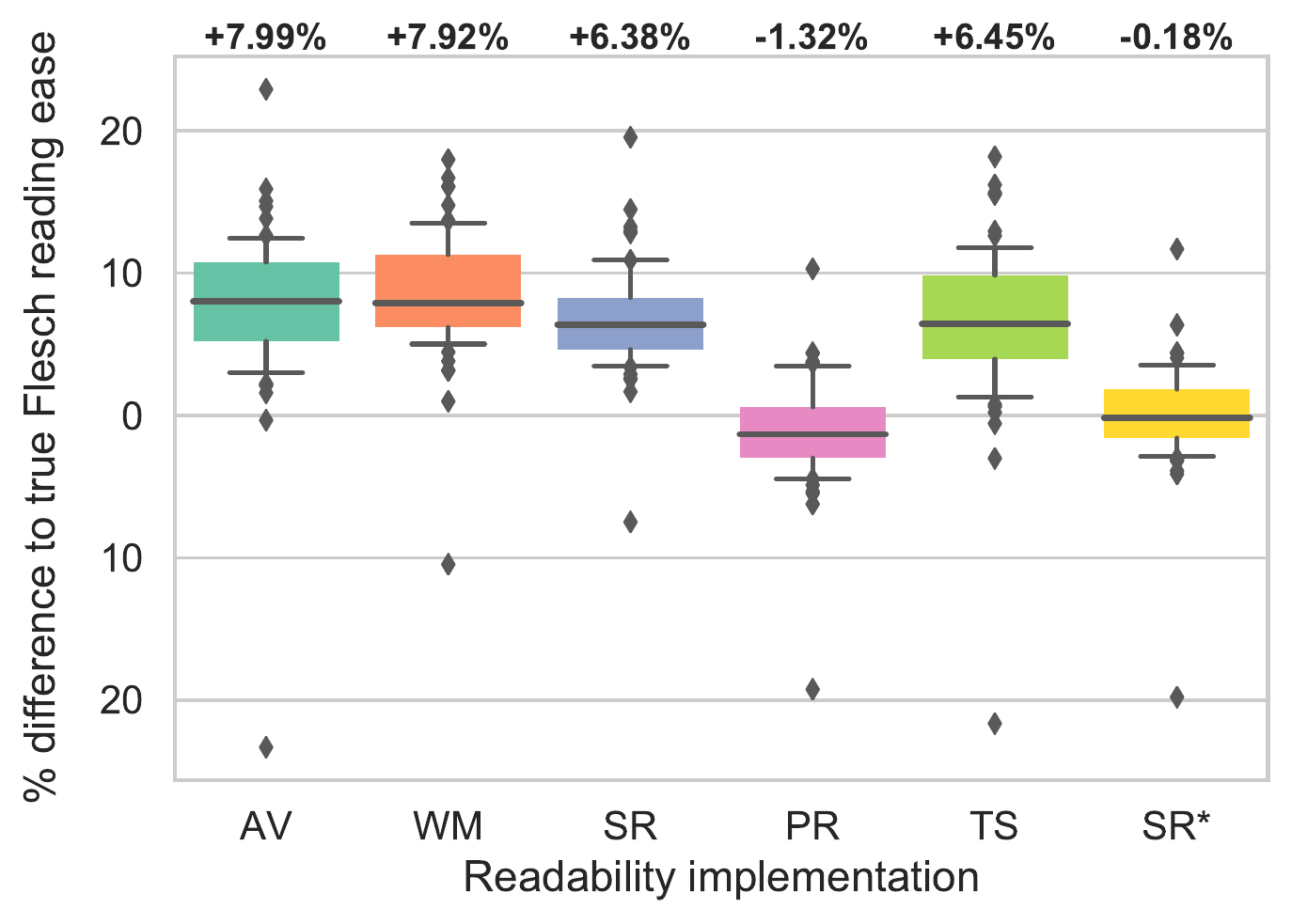} }}
	\hspace{.05cm}
	\subfloat[Flesch-Kincaid Grade Level]{\label{fig:deviation-measure-fkg} {\includegraphics[width=.3\linewidth]{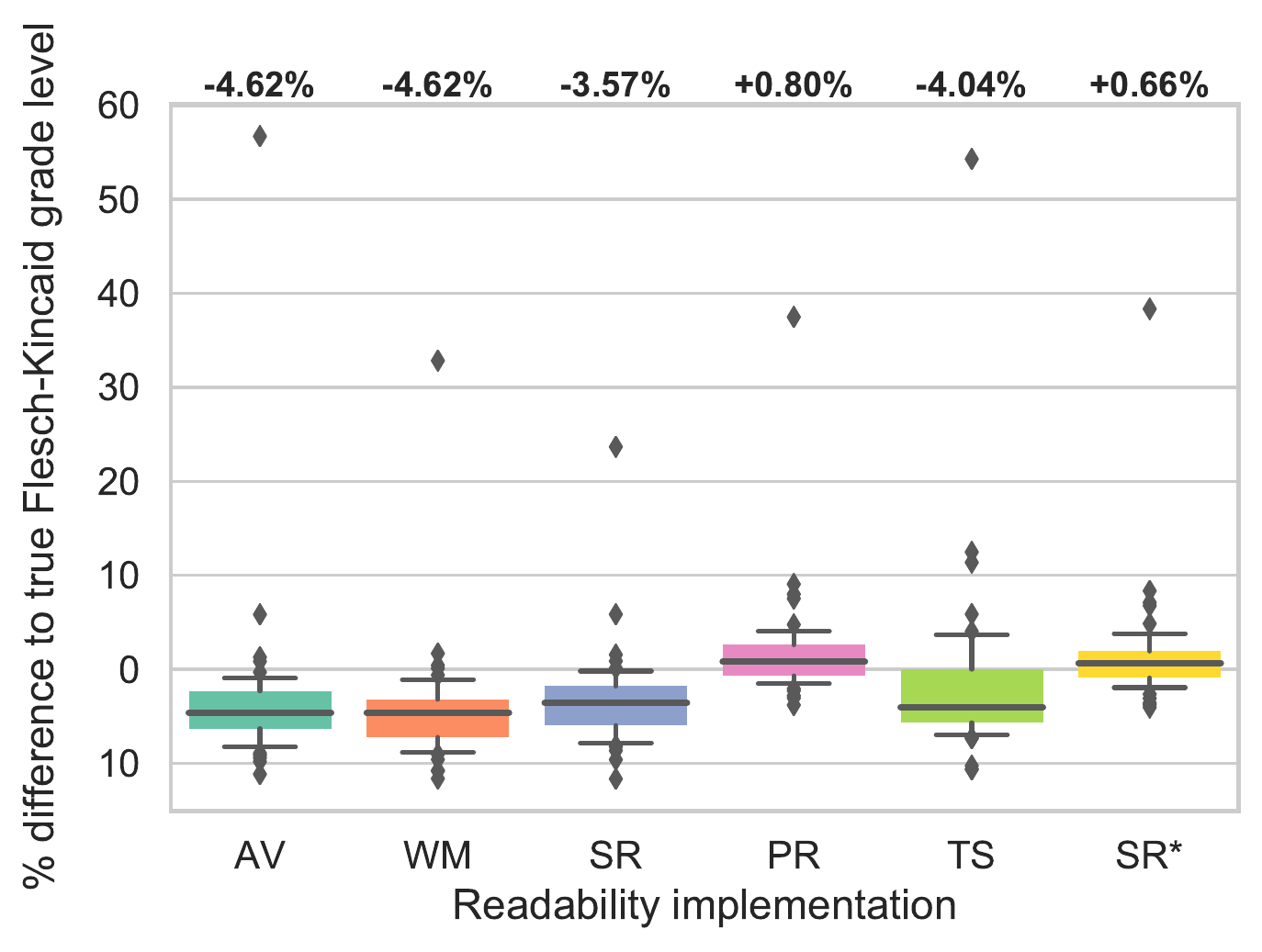} }}
	\hspace{.05cm}
	\subfloat[SMOG]{\label{fig:deviation-measure-smog} {\includegraphics[width=.3\linewidth] {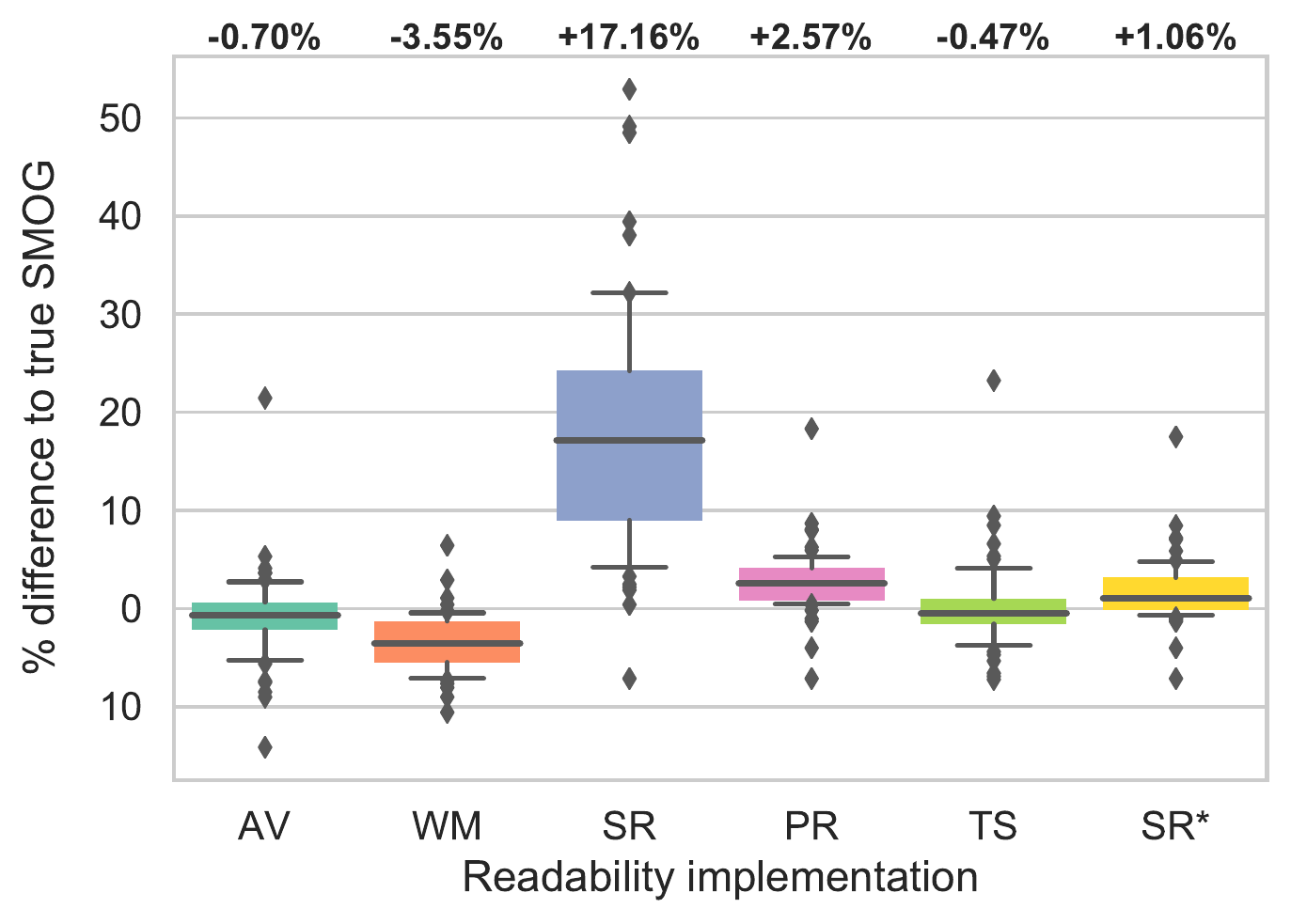} }} \\
	\caption{Deviation (in \%) from the true values of readability measures, on our corpus of sample texts. The boxes range from the 25\% to the 75\% percentile, while whiskers extend to the 10\% and 90\% percentiles.}
	\label{fig:deviation-measures}
\end{figure*}

Figure \ref{fig:deviation-measures} shows the effect these deviations in counts have on readability values.
We compute readability scores for all measures that we have ground truth values for (FRE, FKG, and SMOG), but omit Dale-Chall due to its steep computation requirement.
Figure \ref{fig:deviation-measures} shows that our improved implementation SR* is very close to the true values for all three measures.
We also observe that PR is close to true values.
The other libraries--AV, WM, SR, and TS--have similar deviation patterns for FRE and FKG, overestimating the median FRE score by more than 5\% and underestimating the median FKG by 5\%.

\subsection{Evaluation of ranking consistency}
\begin{figure*}
	\centering
	\subfloat[Flesch Reading Ease]{\label{fig:sample-corr-fre} {\includegraphics[width=.3\linewidth]{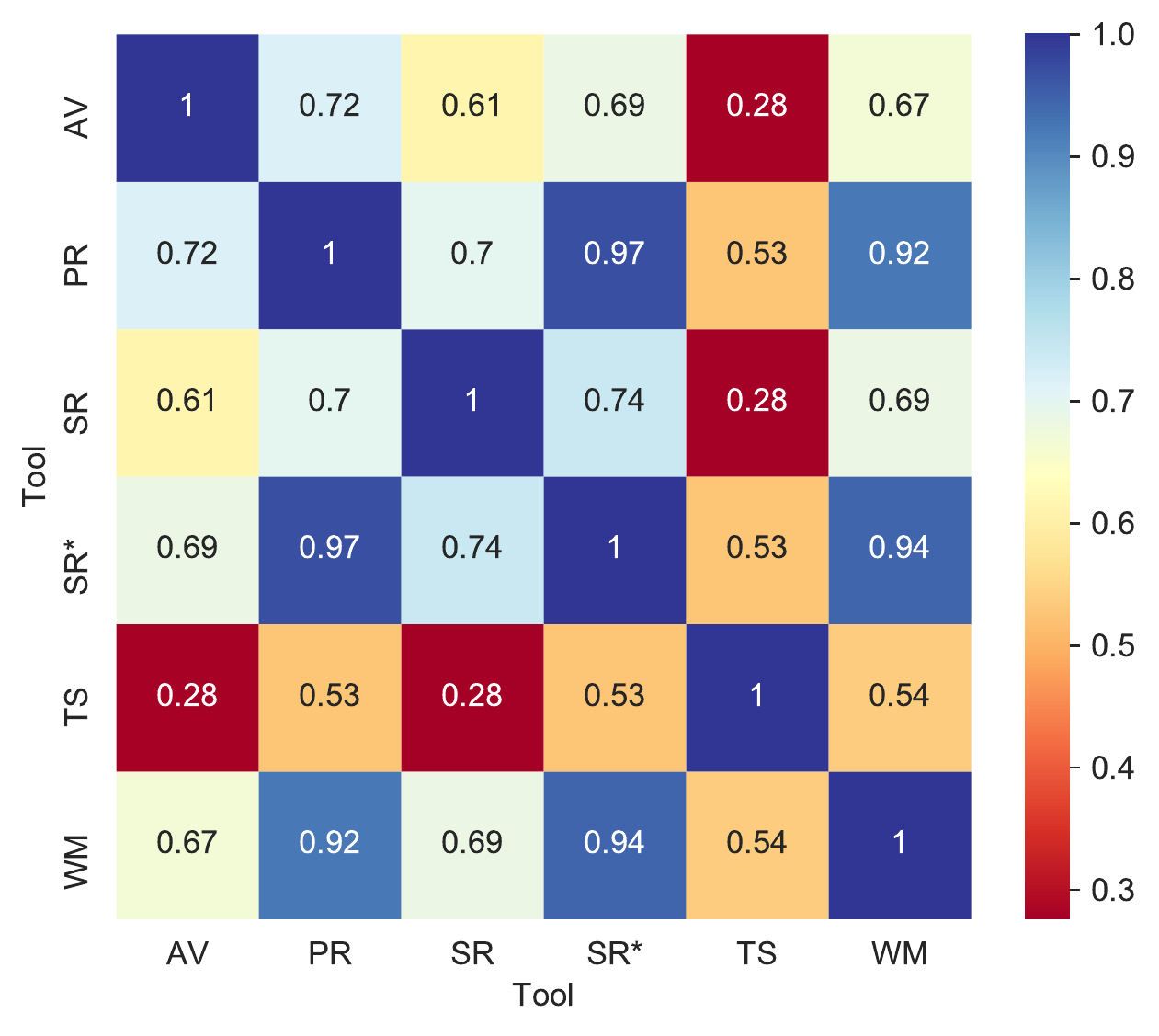} }}
	\hspace{.05cm}
	\subfloat[Coleman-Liau]{\label{fig:sample-corr-cl} {\includegraphics[width=.3\linewidth]{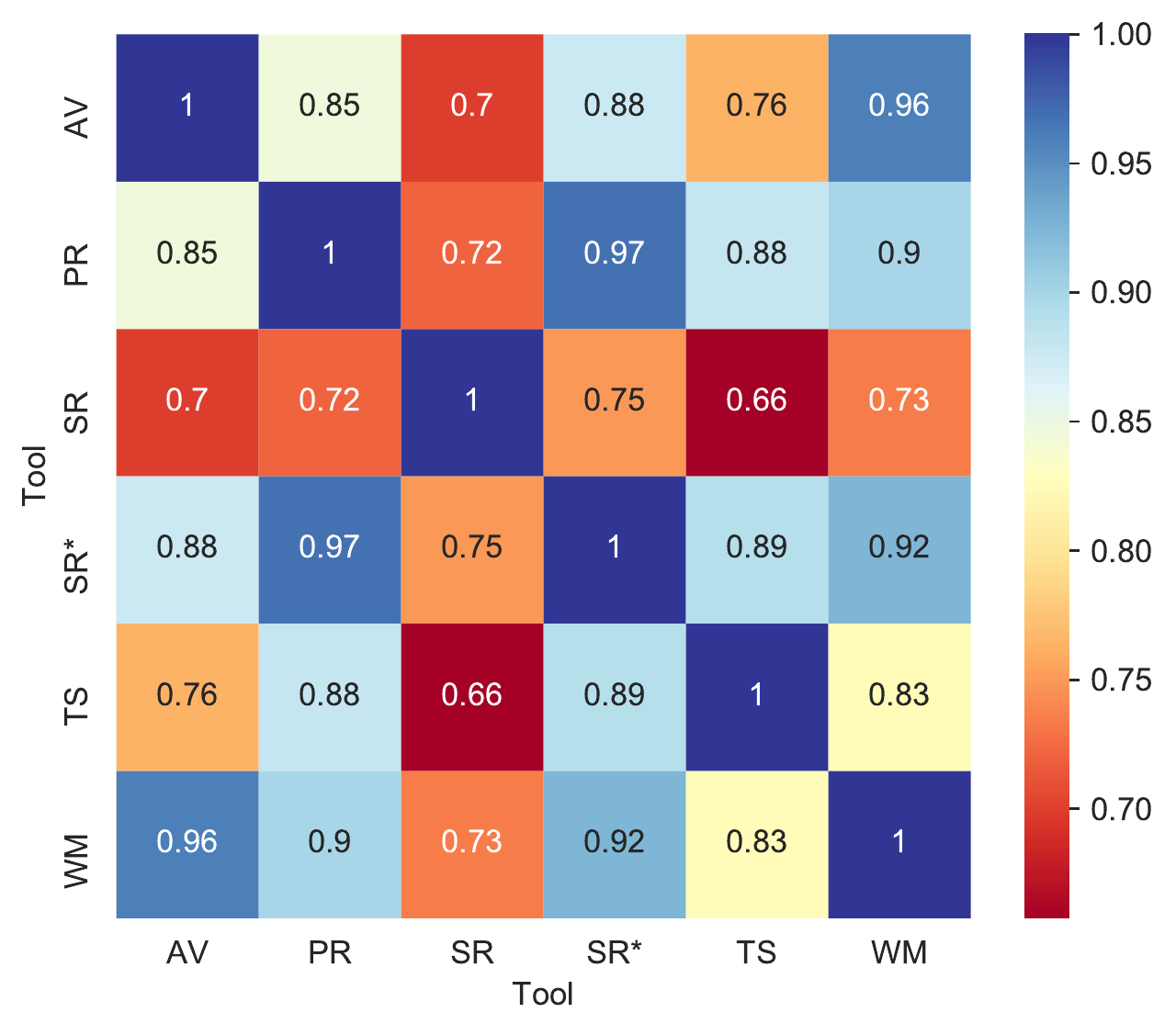} }}
	\hspace{.05cm}
	\subfloat[SMOG]{\label{fig:sample-corr-smog} {\includegraphics[width=.3\linewidth]{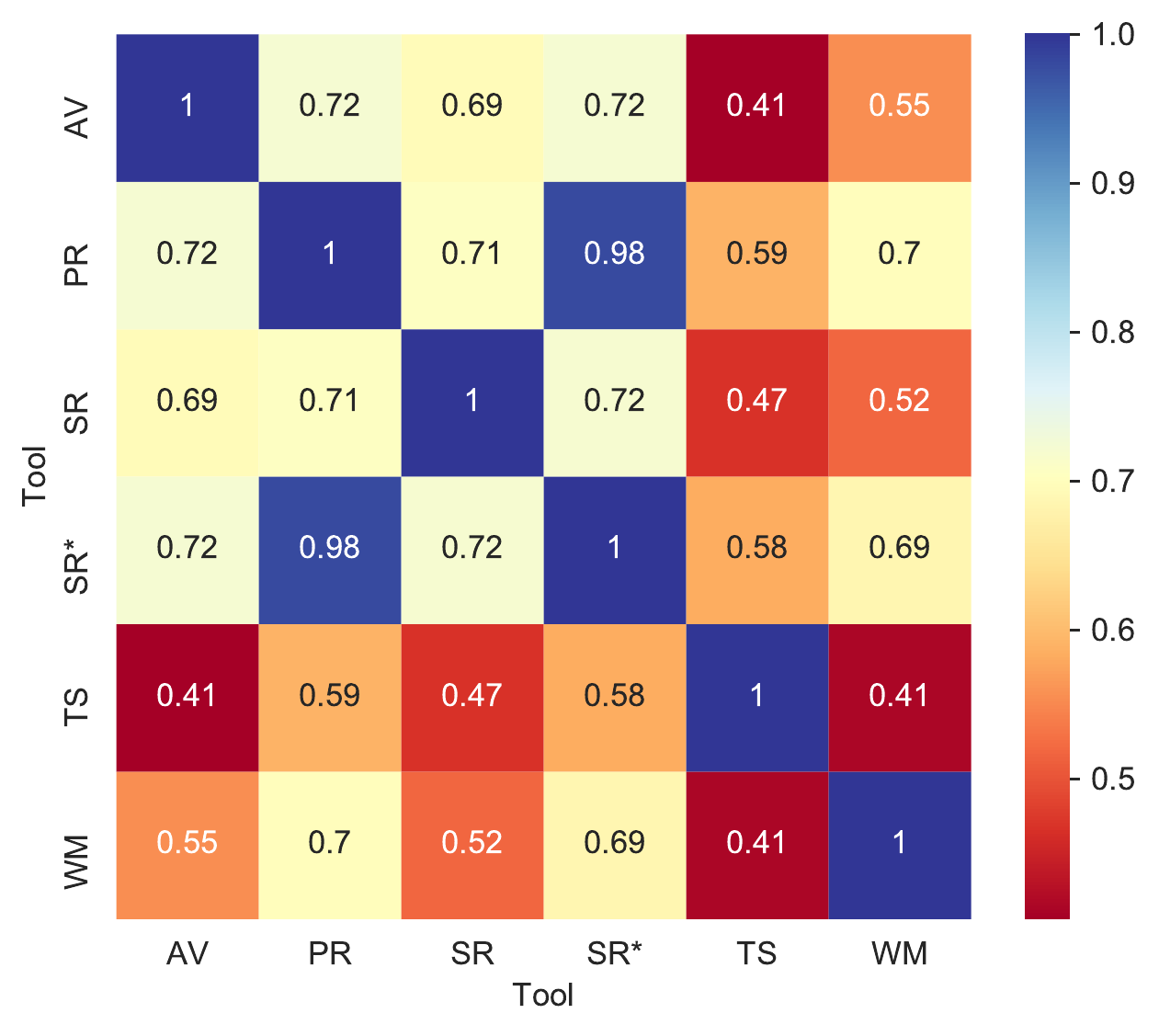} }}
	\caption{Spearman rank-correlation coefficient between the six different implementations of three readability measures, based on a sample of 1,500 privacy policies from our corpus}
	\label{fig:sample-corr}
\end{figure*}

Our evaluation has shown that the numeric values for readability measures differ between implementations.
This may not be a problem in itself, if the implementations are consistent in the relative ranks of different texts, i.e., if they can reliably indicate which texts are relatively more readable than others.
To evaluate this ranking consistency, Figure \ref{fig:sample-corr} shows the Spearman rank-correlation coefficients for three readability measures, each comparing the six implementations, based on 1,500 privacy policies from our corpus.
The figure shows that correlations vary between the three readability measures, but in most cases there is substantial disagreement between the implementations. The only exceptions are PR and SR*, which produce nearly identical ranks. These are the same two implementations whose numeric readability values were close to the ground truth values.
As a result, PR and SR* can be recommended to evaluate readability.

\begin{figure}
	\centering
	\includegraphics[width=.6\linewidth]{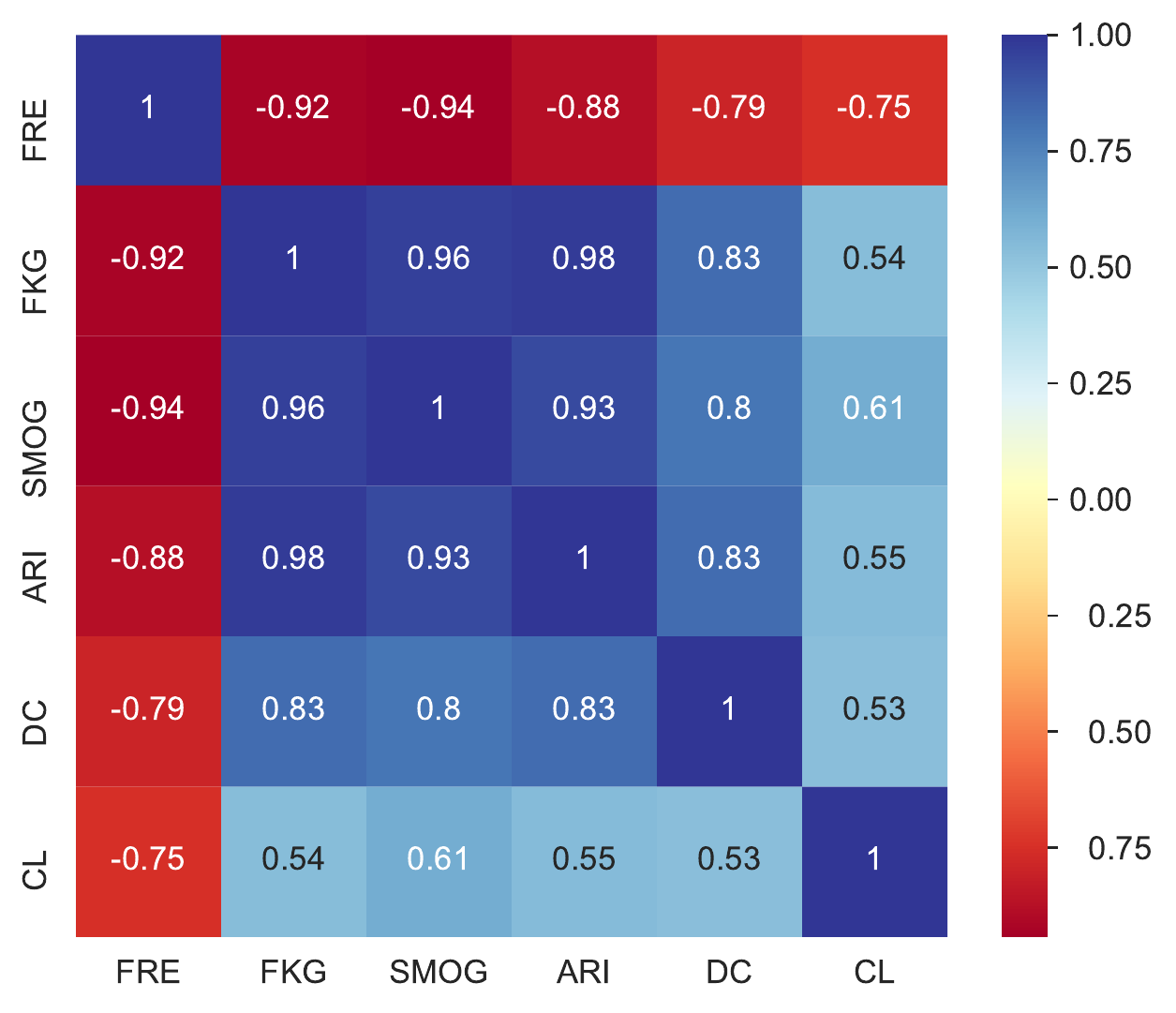}
	\caption{Spearman rank-correlation between readability measures, as computed by SR*, on a sample of 1500 privacy policies.}
	\label{fig:correlation}
\end{figure}
To evaluate to what extent the different readability measures are consistent with each other, Figure \ref{fig:correlation} shows the Spearman rank-correlation coefficient for readability values generated by our SR* implementation.
We can see that most readability measures rank texts in a similar way, except for FKG which (as expected) ranks them inversely, and CL. The low correlation with CL is most likely because most readability measures are computed based on sentence length, whereas CL uses inverse sentence length.

\subsection{Evaluation of computation efficiency}

Aside from consistency between implementations and between readability measures, efficient computation is important when applying readability measures to large amounts of text (as in our case).
The Dale-Chall score in particular requires the identification of proper names and places, for which we have used the spaCy entity type recognition, which comes at a cost in terms of computation time.
Table \ref{tab:dale-chall-computation} shows the computation times for the DC score. Our implementation is orders of magnitude slower than the others, but, according to Figure \ref{fig:deviation-count-diff_words}, the most accurate. For this reason, we have excluded the DC score from our study.

\begin{table}
	\centering
	\caption{Mean times to compute the Dale-Chall score per privacy policy}
	\label{tab:dale-chall-computation}
	\begin{tabular}{ll}
		library & computation time [s] \\
		\midrule
		AV & $1.97 \times 10^{-6}$ \\
		WM & $0.14$ \\
		TS & $0.007$ \\
		PR & $1.06 \times 10^{-5}$ \\
		SR & $0.18$ \\
		SR* & $24.9$ \\
	\end{tabular}
\end{table}

\subsection{Concluding remarks on selection of readability formulas and implementations}

Our evaluation of readability formulas shows that implementations often make subtle mistakes in counting sentences, words, and characters, which can alter the intended semantics of the readability measure.
In addition, not all concepts used in readability formulas can be implemented efficiently, for example the recognition of proper names and places, and the recognition of \textit{complex} or \textit{difficult} words.
Finally, readability formulas that rely on a word list (such as Dale-Chall) are inconvenient because their exceptions and rules are computationally expensive to implement. Due to the ever-changing nature of human language, they also require up-to-date word lists to produce accurate estimates of readability.

Based on our evaluation, we have selected three readability measures for our study of privacy policies, one based on syllables (FRE), one based on characters (CL), and one based on word complexity (SMOG).
We prefer FRE over FKG to improve comparability with prior work such as \cite{libert2018automated,shipp2020how}.
We select CL instead of ARI  because CL correlates least with FRE and because its sentence complexity measure is the inverse of the others.
We select SMOG over GF and DC because SMOG is the only measure that uses a computationally efficient measure of sentence complexity, i.e., polysyllabic words per sentence instead of \textit{complex} or \textit{difficult} words per sentence.
We do not select GF because determining which words are \textit{complex} according to the GF definition is not feasible and based on heuristics that may not hold for privacy policies.
We do not select DC because of weaknesses in its word-list based approach and its computational complexity.

\section{Results for attribute-level classifiers}
\label{appendix:attribute-results}

Tables \ref{tab:attr-access-scope-performance} to \ref{tab:attr-user-type} show classification results for attribute-level classifiers. Labels marked with an asterisk have been excluded from the analysis due to their low precision.

\begin{table}[!h]
	\footnotesize
	\caption{F1 for attribute-level classifier: Access Scope (85 epochs)}
	\label{tab:attr-access-scope-performance}
	\centering
	\begin{tabular}{p{2.8cm}llll}
		\toprule
		& precision    &recall  &f1-score   &support\\
		\midrule
     Profile data  &    1.000  &   0.167 &    0.286    &     6\\
*Unspecified  &    0.600  &   1.000 &    0.750  &       3\\
User account data&  0.950   &  1.000 &    0.974 &       19\\
\textit{micro avg}   &   0.885   &  0.821  &   0.852 &       28\\
\textit{macro avg}   &   0.850  &   0.722&     0.670  &      28\\
		\bottomrule
	\end{tabular}
\end{table}

\begin{table}[!h]
	\footnotesize
	\caption{F1 for attribute-level classifier: Access Type (165 epochs)}
	\label{tab:attr-access-type-performance}
	\centering
	\begin{tabular}{p{2.8cm}llll}
		\toprule
		&  precision    &recall  &f1-score   &support\\
		\midrule
Edit information & 0.957 &    1.000   &  0.978   &     22\\
Unspecified   &   1.000  &   1.000  &   1.000   &      1\\
View    &  1.000  &   0.556   &  0.714    &     9\\
\textit{micro avg}  &    0.966  &   0.875  &   0.918   &     32\\
\textit{macro avg}  &    0.986  &   0.852  &   0.897   &     32\\
		\bottomrule
	\end{tabular}
\end{table}

\begin{table}[!h]
	\footnotesize
	\caption{F1 for attribute-level classifier: Action First-Party (40 epochs, data augmentation: segments labeled with different classes were combined into a new segment labeled with the union of the classes)}
	\label{tab:attr-action-first-party-performance}
	\centering
	\begin{tabular}{p{2.8cm}llll}
		\toprule
		&precision    &recall  &f1-score   &support\\
		\midrule
Collect in mobile app   &   0.975  &   0.928  &   0.951    &    83\\
Collect on mobile website & 1.000  &   0.429  &   0.600   &      7\\
Collect on website    &  0.970  &   0.989    & 0.979   &    360\\
Unspecified    &  0.975  &   0.909 &   0.941   &    298\\
\textit{micro avg}  &    0.972   &  0.945   &  0.959   &    748\\
\textit{macro avg}   &   0.980  &   0.814  &   0.868 &      748\\
		\bottomrule
	\end{tabular}
\end{table}

\begin{table}[!h]
	\footnotesize
	\caption{F1 for attribute-level classifier: Action Third Party (62 epochs)}
	\label{tab:attr-action-third-party-performance}
	\centering
	\begin{tabular}{p{2.8cm}llll}
		\toprule
		&precision    &recall  &f1-score   &support\\
		\midrule
*Collect on first party website/app& 0.688   &  0.500 &    0.579   &     22\\
Receive/Shared with   &   0.971  &   0.937 &    0.953    &   142\\
See   &   1.000  &   0.857  &   0.923    &     7\\
Track on first party website/app&  0.923 &    0.923  &   0.923    &    26\\
Unspecified   &   1.000   &  0.200   &  0.333   &      5\\
\textit{micro avg}    &  0.941  &   0.866   &  0.902   &    202\\
\textit{macro avg}    &  0.916 &    0.683  &   0.742   &    202\\
		\bottomrule
	\end{tabular}
\end{table}

\begin{table}[!h]
	\footnotesize
	\caption{F1 for attribute-level classifier: Audience Type (117 epochs)}
	\label{tab:attr-audience-type}
	\centering
	\begin{tabular}{p{2.8cm}llll}
		\toprule
		&precision    &recall  &f1-score   &support\\
		\midrule
Californians  & 0.941   &  0.941  &   0.941   &     17\\
Children      & 0.968   &  0.968 &    0.968   &     31\\
Europeans     & 1.000   &  1.000  &   1.000   &      3\\
\textit{micro avg}     & 0.961   &  0.961  &   0.961   &     51\\
\textit{macro avg}     & 0.970   &  0.970  &   0.970   &     51\\
		\bottomrule
	\end{tabular}
\end{table}
\begin{table}[!h]
	\footnotesize
	\caption{F1 for attribute-level classifier: Change Type (91 epochs)}
	\label{tab:attr-change-type}
	\centering
	\begin{tabular}{p{2.8cm}llll}
		\toprule
		&precision    &recall  &f1-score   &support\\
		\midrule
Privacy relevant change & 1.000    & 0.714  &  0.833     &    7\\
Unspecified   &   0.917  &   1.000  &   0.957   &     11\\
\textit{micro avg}    &  0.941 &    0.889   &  0.914   &     18\\
\textit{macro avg}   &   0.958   &  0.857   &  0.895    &    18\\
		\bottomrule
	\end{tabular}
\end{table}
\begin{table}[!h]
	\footnotesize
	\caption{F1 for attribute-level classifier: Choice Scope (104 epochs)}
	\label{tab:attr-choice-scope}
	\centering
	\begin{tabular}{p{2.8cm}llll}
		\toprule
		&precision    &recall  &f1-score   &support\\
		\midrule
                      Both   &   1.000 &    0.200  &   0.333    &    10\\
Collection  &    0.882   &  0.857   &  0.870   &     70\\
First party collection  &    0.900   &  0.692  &   0.783    &    13\\
First party use   &   0.848   &  0.812  &   0.830     &   48\\
Third party sharing/collection & 0.864  &   0.679   &  0.760  &      28\\
*Third party use &     0.000  &   0.000 &    0.000     &    9\\
*Unspecified    &  0.692   &  0.882 &   0.776    &    51\\
Use   &   1.000  &   0.535  &   0.697    &    43\\
\textit{micro avg}    &  0.835&     0.724   &  0.776 &      272\\
\textit{macro avg}    &  0.773  &   0.582  &   0.631  &     272\\
		\bottomrule
	\end{tabular}
\end{table}

\begin{table}[!h]
	\footnotesize
	\caption{F1 for attribute-level classifier: Choice Type (90 epochs)}
	\label{tab:attr-choice-type}
	\centering
	\begin{tabular}{p{2.8cm}llll}
		\toprule
		&precision    &recall  &f1-score   &support\\
		\midrule
Browser/device privacy controls & 0.900  &   0.923 &    0.911   &     39\\
Dont use service/feature  &    0.811  &   0.750   &  0.779     &   40\\
First-party privacy controls & 0.857  &   0.400  &   0.545    &    15\\
Opt-in    &  0.909   &  0.811  &   0.857    &    74\\
Opt-out link   &   0.970  &   0.800   &  0.877  &      40\\
Opt-out via contacting company&  0.923 &    0.828&     0.873 &       29\\
*Third-party privacy controls&  0.733   &  0.458 &    0.564 &       24\\
Unspecified    &  0.831   &  0.844   &  0.837    &    64\\
\textit{micro avg}   &   0.875  &   0.778&    0.824   &    325\\
\textit{macro avg}   &   0.867   &  0.727  &   0.780   &    325\\
		\bottomrule
	\end{tabular}
\end{table}

\begin{table}[!h]
	\footnotesize
	\caption{F1 for attribute-level classifier: Collection Mode (50 epochs)}
	\label{tab:attr-collection-mode}
	\centering
	\begin{tabular}{p{2.8cm}llll}
		\toprule
		&precision    &recall  &f1-score   &support\\
		\midrule
Explicit    &  0.938 &    0.882 &    0.909  &      68\\
Implicit    &  0.920 &    0.920 &    0.920  &     100\\
*Unspecified &  0.696 &    0.762 &    0.727  &      21\\
\textit{micro avg}   &  0.898 &    0.889 &    0.894  &     189\\
\textit{macro avg}   &  0.851 &    0.855 &    0.852  &     189\\
		\bottomrule
	\end{tabular}
\end{table}

\begin{table}[!h]
	\footnotesize
	\caption{F1 for attribute-level classifier: Do Not Track policy (400 epochs)}
	\label{tab:attr-do-not-track}
	\centering
	\begin{tabular}{p{2.8cm}llll}
		\toprule
		&precision    &recall  &f1-score   &support\\
		\midrule
 Honored    &  1.000 &    1.000  &   1.000    &     1\\
Not honored & 1.000  &   1.000    & 1.000     &    4\\
\textit{micro avg}   &  1.000 &    1.000  &   1.000    &     5\\
\textit{macro avg}   &  1.000 &    1.000  &   1.000    &     5\\
		\bottomrule
	\end{tabular}
\end{table}

\begin{table}[!h]
	\footnotesize
	\caption{F1 for attribute-level classifier: Does/Does Not (24 epochs)}
	\label{tab:attr-does-does-not}
	\centering
	\begin{tabular}{p{2.8cm}llll}
		\toprule
		&precision    &recall  &f1-score   &support\\
		\midrule
    Does  &    0.984  &   0.978  &   0.981     &  323\\
Does Not  &    0.944 &    0.829  &   0.883     &   41\\
\textit{micro avg} &     0.980 &    0.962 &    0.971    &   364\\
\textit{macro avg} &     0.964 &    0.904  &   0.932    &   364\\
		\bottomrule
	\end{tabular}
\end{table}

\begin{table}[!h]
	\footnotesize
	\caption{F1 for attribute-level classifier: Identifiability (100 epochs)}
	\label{tab:attr-identifiability}
	\centering
	\begin{tabular}{p{2.8cm}llll}
		\toprule
		&precision    &recall  &f1-score   &support\\
		\midrule
Aggregated or anonymized&  0.912  &  0.963     &0.937  &      54\\
Identifiable  &    0.976&     0.910  &   0.942    &   134\\
Unspecified  &    0.767    & 0.958   &  0.852  &     48\\
\textit{micro avg}    &  0.909  &   0.932 &    0.921    &   236\\
\textit{macro avg}   &   0.885  &   0.944 &    0.910    &   236\\
		\bottomrule
	\end{tabular}
\end{table}

\begin{table}[!h]
	\footnotesize
	\caption{F1 for attribute-level classifier: Notification Type (150 epochs, data augmentation: segments labeled with different classes were combined into a new segment labeled with the union of the classes)}
	\label{tab:attr-notification-type}
	\centering
	\begin{tabular}{p{2.8cm}llll}
		\toprule
		&precision    &recall  &f1-score   &support\\
		\midrule
General notice in privacy policy&  0.931  &   1.000  &   0.964   &     27\\
General notice on website   &   0.963 &    1.000   &  0.981    &    26\\
Personal notice    &  1.000  &   0.900 &    0.947   &     20\\
Unspecified   &   0.875  &   0.875   &  0.875     &    8\\
\textit{micro avg}   &   0.951  &   0.963  &   0.957   &     81\\
\textit{macro avg}   &   0.942   &  0.944  &   0.942   &     81\\
		\bottomrule
	\end{tabular}
\end{table}

\begin{table}[!h]
	\footnotesize
	\caption{F1 for attribute-level classifier: Personal Information Type (50 epochs, data augmentation: segments labeled with different classes were combined into a new segment labeled with the union of the classes)}
	\label{tab:attr-personal-information-type}
	\centering
	\begin{tabular}{p{2.8cm}llll}
		\toprule
		&precision    &recall  &f1-score   &support\\
		\midrule
     Computer information   &   0.954 &    0.926 &    0.940   &    135\\
Contact &     0.978   &  0.952  &   0.965     &  330\\
Cookies and tracking elements&  0.985   &  0.997   &  0.991    &   339\\
Demographic   &   0.963 &    0.895 &    0.928   &     86\\
Financial   &   0.991 &    0.973  &   0.982   &    112\\
Generic personal information&  0.953   &  0.950   &  0.951      & 577\\
Health   &   1.000  &   0.852&     0.920  &      27\\
IP address and device IDs      &1.000   &  0.960&     0.980  &     176\\
Location    &  0.991 &    0.924   &  0.957   &    119\\
Personal identifier    &  1.000   &  0.548  &   0.708    &    31\\
Social media data    &  1.000 &    0.074   &  0.138     &   27\\
Survey data  &    1.000  &   0.200  &   0.333    &    15\\
Unspecified   &   0.882 &    0.848  &   0.865     &  395\\
User online activities  &    0.959  &   0.924  &   0.941  &     277\\
\textit{micro avg}    &  0.958   &  0.917  &   0.937 &    2646\\
\textit{macro avg}    &  0.975   &  0.787  &   0.828 &     2646\\
		\bottomrule
	\end{tabular}
\end{table}

\begin{table}[!h]
	\footnotesize
	\caption{F1 for attribute-level classifier: Purpose (65 epochs)}
	\label{tab:attr-purpose}
	\centering
	\begin{tabular}{p{2.8cm}llll}
		\toprule
		&precision    &recall  &f1-score   &support\\
		\midrule
Additional service/feature    &  0.881  &   0.552  &   0.679    &    67\\
Advertising   &   0.941   &  0.909 &    0.925   &     88\\
Analytics/Research   &   0.887    & 0.910   &  0.899    &    78\\
Basic service/feature  &    0.909&     0.738 &    0.814   &    122\\
Legal requirement  &    0.969 &    0.838  &   0.899     &   37\\
Marketing   &   0.924  &   0.839   &  0.880     &   87\\
Merger/Acquisition  &    1.000  &   0.895  &   0.944     &   19\\
Personalization/ Customization &  0.933  &   0.764  &   0.840   &     55\\
Service operation and security &0.879   &  0.797  &   0.836     &   64\\
*Unspecified     & 0.589   &  0.825 &    0.688   &     40\\
\textit{micro avg}    &  0.885  &   0.799  &   0.840    &   657\\
\textit{macro avg}    &  0.891   &  0.807  &   0.840  &     657\\
		\bottomrule
	\end{tabular}
\end{table}

\begin{table}[!h]
	\footnotesize
	\caption{F1 for attribute-level classifier: Retention Period (161 epochs)}
	\label{tab:attr-retention-period}
	\centering
	\begin{tabular}{p{2.8cm}llll}
		\toprule
		&precision    &recall  &f1-score   &support\\
		\midrule
Indefinitely &  1.000  &   1.000  &   1.000   &      3\\
*Limited      &  0.667  &   0.857  &   0.750   &      7\\
Unspecified  &  0.833  &   1.000  &   0.909   &      5\\
\textit{micro avg}    &  0.778  &   0.933  &   0.848    &    15\\
\textit{macro avg}    &  0.833  &   0.952  &   0.886   &     15\\
		\bottomrule
	\end{tabular}
\end{table}

\begin{table}[!h]
	\footnotesize
	\caption{F1 for attribute-level classifier: Retention Purpose (600 epochs)}
	\label{tab:attr-retention-purpose}
	\centering
	\begin{tabular}{p{2.8cm}llll}
		\toprule
		&precision    &recall  &f1-score   &support\\
		\midrule
         Legal requirement   &   1.000 &    0.800  &   0.889   &      5\\
Perform service  &    0.800 &    0.800  &   0.800       &  5\\
Service operation and security & 1.000 &    0.500 &   0.667    &     2\\
Unspecified    &  1.000   &  1.000   &  1.000      &   1\\
\textit{micro avg}   &   0.909   &  0.769    & 0.833  &      13\\
\textit{macro avg}   &   0.950   &  0.775    & 0.839   &     13\\
		\bottomrule
	\end{tabular}
\end{table}

\begin{table}[!h]
	\footnotesize
	\caption{F1 for attribute-level classifier: Security Measure (300 epochs, data augmentation: segments labeled with different classes were combined into a new segment labeled with the union of the classes)}
	\label{tab:attr-security-measure}
	\centering
	\begin{tabular}{p{2.8cm}llll}
		\toprule
		&precision    &recall  &f1-score   &support\\
		\midrule
Data access limitation    &  1.000   &  0.879 &    0.935    &    33\\
Generic     & 1.000 &    1.000 &    1.000   &     64\\
Privacy review/audit  &   1.000  &   0.333 &    0.500   &      3\\
Privacy/Security program  &  1.000 &    1.000   &  1.000   &      3\\
Secure data storage   &   1.000    & 0.429  &   0.600    &     7\\
Secure data transfer     & 1.000   &  1.000 &    1.000     &   26\\
Secure user authentication & 1.000   &  0.500   &  0.667    &     4\\
\textit{micro avg} &     1.000  &   0.914   &  0.955 &      140\\
\textit{macro avg} &     1.000  &   0.734   &  0.815 &      140\\
		\bottomrule
	\end{tabular}
\end{table}

\begin{table}[!h]
	\footnotesize
	\caption{F1 for attribute-level classifier: Third Party Entity (90 epochs)}
	\label{tab:attr-third-party-entity}
	\centering
	\begin{tabular}{p{2.8cm}llll}
		\toprule
		&precision    &recall  &f1-score   &support\\
		\midrule
          Named third party   &  0.868   &  0.787    & 0.825    &    75\\
Other part of company/affiliate & 0.923  &   0.800 &    0.857  &      15\\
Public    &  0.857  &   0.667   &  0.750     &    9\\
Unnamed third party  &    0.885 &    0.959 &    0.921     &  121\\
Unspecified    &  0.625  &   0.625 &    0.625      &   8\\
\textit{micro avg}   &   0.872   &  0.868  &   0.870   &    228\\
\textit{macro avg}   &   0.832   &  0.767  &  0.796    &   228\\
		\bottomrule
	\end{tabular}
\end{table}

\begin{table}[!h]
	\footnotesize
	\caption{F1 for attribute-level classifier: User Choice (199 epochs)}
	\label{tab:attr-user-choice}
	\centering
	\begin{tabular}{p{2.8cm}llll}
		\toprule
		&precision    &recall  &f1-score   &support\\
		\midrule
    None   &   0.875   &  0.875 &    0.875    &     8\\
Opt-in     & 1.000   &  0.500  &   0.667       &  2\\
Unspecified & 0.800  &   1.000  &   0.889     &    4\\
\textit{micro avg}   &  0.857  &   0.857   &  0.857    &    14\\
\textit{macro avg}   &  0.892  &   0.792   &  0.810    &    14\\
		\bottomrule
	\end{tabular}
\end{table}

\begin{table}[!h]
	\footnotesize
	\caption{F1 for attribute-level classifier: User Type (19 epochs)}
	\label{tab:attr-user-type}
	\centering
	\begin{tabular}{p{2.8cm}llll}
		\toprule
		&precision    &recall  &f1-score   &support\\
		\midrule
  Unspecified   &   0.939   &  0.939   &  0.939 &      132\\
User with account & 0.941 &    0.877 &    0.908 &       73\\
\textit{micro avg}   &   0.940   &  0.917  &   0.928    &   205\\
\textit{macro avg}   &   0.940   &  0.908  &   0.924    &   205\\
		\bottomrule
	\end{tabular}
\end{table}

\end{document}